\documentclass[journal,twocolumn,10pt]{IEEEtran}
\usepackage{amsmath,epsfig}
\usepackage{amssymb}
\usepackage{booktabs}
\usepackage{amsthm}
\usepackage{graphicx}
\usepackage{caption}
\usepackage{subcaption}
\usepackage{rotating}
\usepackage{floatflt} % text flows around figures
\usepackage{paralist} % compact and in-paragraph lists
\usepackage{algorithm} % package for pseudo code
\usepackage{xcolor}
\usepackage{color}
\usepackage{epstopdf}
\usepackage[normalem]{ulem}
\usepackage{float}
\usepackage{todonotes}
\usepackage{mathtools}
\newcommand{\mypar}[1]{{\bf #1.}}
\theoremstyle{definition}

\newcommand{\R}{\ensuremath{\mathbb{R}}}

\newcommand{\Cc}{\ensuremath{\mathcal{C}}}
\DeclareMathOperator{\Id}{I}

\DeclareMathOperator{\Dd}{D}

\DeclareMathOperator{\Nei}{Nei}

\def\x{\mathbf{x}}
\def\t{\mathbf{t}}

\def\y{\mathbf{y}}
\def\z{\mathbf{z}}

\def\N{\mathcal{N}}

\def\V{\mathcal{V}}

\def\U{\mathcal{U}}

\def\E{\mathcal{E}}

\def\one{{\bf 1}}

\DeclareMathOperator{\LL}{L}

\DeclareMathOperator{\Adj}{A}

\DeclareMathOperator{\Pj}{P}
\DeclareMathOperator{\Mm}{M}

\DeclareMathOperator{\Ee}{E}
\DeclareMathOperator{\X}{X}
\DeclareMathOperator{\Y}{Y}
\DeclareMathOperator{\W}{W}
\DeclareMathOperator{\Z}{Z}
\DeclareMathOperator{\Rr}{R}

\usepackage{xr}
\begin{document}
\title{ Localization, Decomposition, and Dictionary Learning of Piecewise-Constant Signals on
  Graphs } \author{Siheng~Chen,~\IEEEmembership{Student~Member,~IEEE},
  Yaoqing~Yang,~\IEEEmembership{Student~Member,~IEEE},
  Jos{\'e} M. F. Moura,~\IEEEmembership{Fellow,~IEEE},
  Jelena~Kova\v{c}evi\'c,~\IEEEmembership{Fellow,~IEEE}% <-this % stops a space
  \thanks{ S. Chen, Y. Yang, J. M. F. Moura and J. Kova\v{c}evi\'c are
    with the Dept. of Electrical and Computer Engineering. Moura and
    Kova\v{c}evi\'c are also with the Dept. of Biomedical Engineering
    (by courtesy), Carnegie Mellon University, Pittsburgh, PA, 15213
    USA. Emails: \{sihengc,yaoqingy,moura,jelenak\}@andrew.cmu.edu.
  }% <-this % stops a space
  \thanks{The authors gratefully acknowledge support from the NSF
    through awards CCF 1421919, 1563918 and
    1513936.}% <-this % stops a space
}
%% The paper headers
%\markboth{IEEE Trans. Signal Process., March 2016. In preparation.} {Chen \MakeLowercase{\textit{et al.}}: Representations}
 \maketitle

%\tableofcontents

\begin{abstract}
  Motivated by the need to extract meaning from large amounts of
  complex data that can be best modeled by graphs, we consider three
  critical problems on graphs: localization, decomposition, and
  dictionary learning of piecewise-constant signals. These graph-based
  problems are related to many real-world applications, such as
  localizing virus attacks in cyber-physical systems, localizing
  stimulus in brain connectivity networks, and mining traffic events
  in city street networks, where the key issue is to separate
  localized activated patterns and background noise; in other words,
  we aim to find the supports of localized activated patterns.
  Counterparts of these problems in classical signal/image processing,
  such as impulse detection, foreground detection, and wavelet
  construction, have been intensely studied over the past few decades.
  We use piecewise-constant graph signals to model localized patterns,
  where each piece indicates a localized pattern that exhibits
  homogeneous internal behavior and the number of pieces indicates the
  number of localized patterns.  For such signals, we show that
  decomposition and dictionary learning are natural extensions of
  localization, the goal of which is not only to efficiently
  approximate graph signals, but also to accurately find supports of
  localized patterns. For each of the three problems, i.e.,
  localization, decomposition, and dictionary learning, we propose a
  specific graph signal model, an optimization problem, and a
  computationally efficient solver.  The proposed solvers directly
  find the supports of arbitrary localized activated patterns without
  tuning any thresholds, which is a notorious challenge in many
  localization problems.  We then conduct an extensive empirical study
  to validate the proposed methods on both simulated and real data
  including the analysis of a large volume of spatio-temporal
  Manhattan urban data.  From taxi-pickup activities and using our
  methods, we are able to detect both everyday as well as special
  events and distinguish weekdays from weekends. Our findings validate
  the effectiveness of the approach and suggest that graph signal
  processing tools may aid in urban planning and traffic forecasting.
\end{abstract}


\section{Introduction}

\begin{figure}[h]
  \begin{center}
    \begin{tabular}{cc}
      \includegraphics[width=0.4\columnwidth]{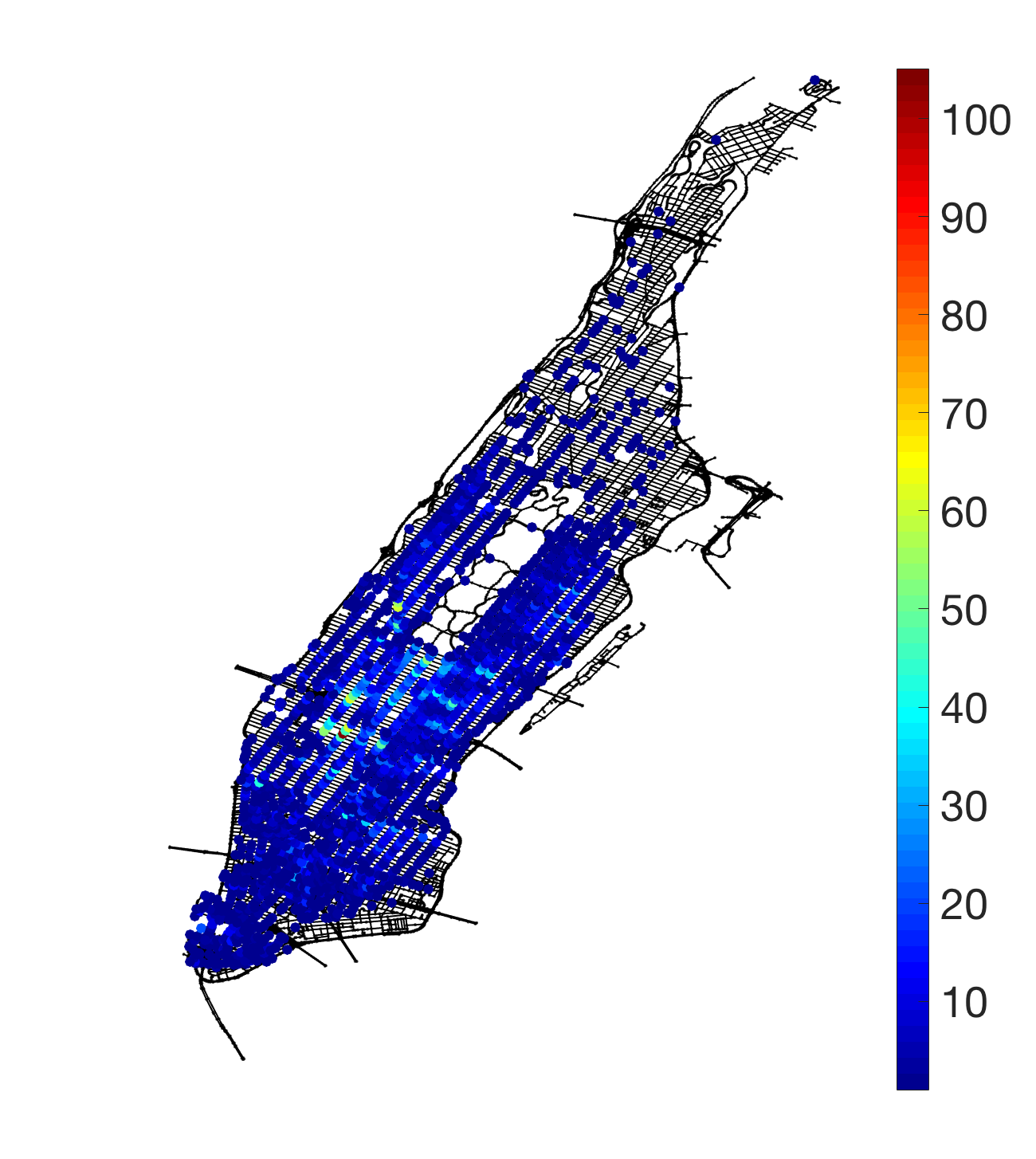}
      &
      \includegraphics[width=0.4\columnwidth]{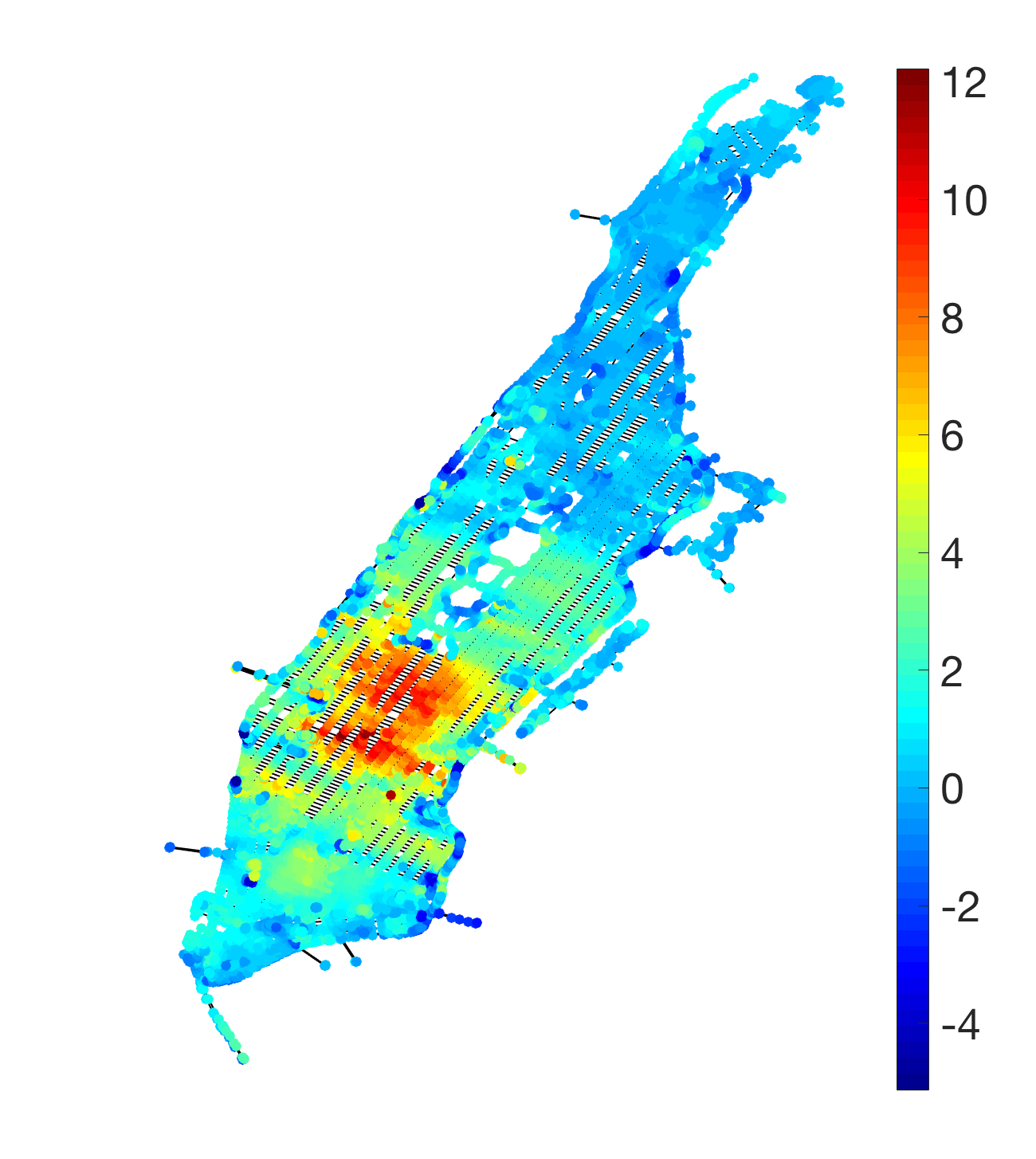}
      \\
      {\small (a) Taxi-pickup distribution.} &  {\small (b) Nonlinear approximation}
      \\
      &  {\small (50 best graph frequencies).}
      \\
      \includegraphics[width=0.4\columnwidth]{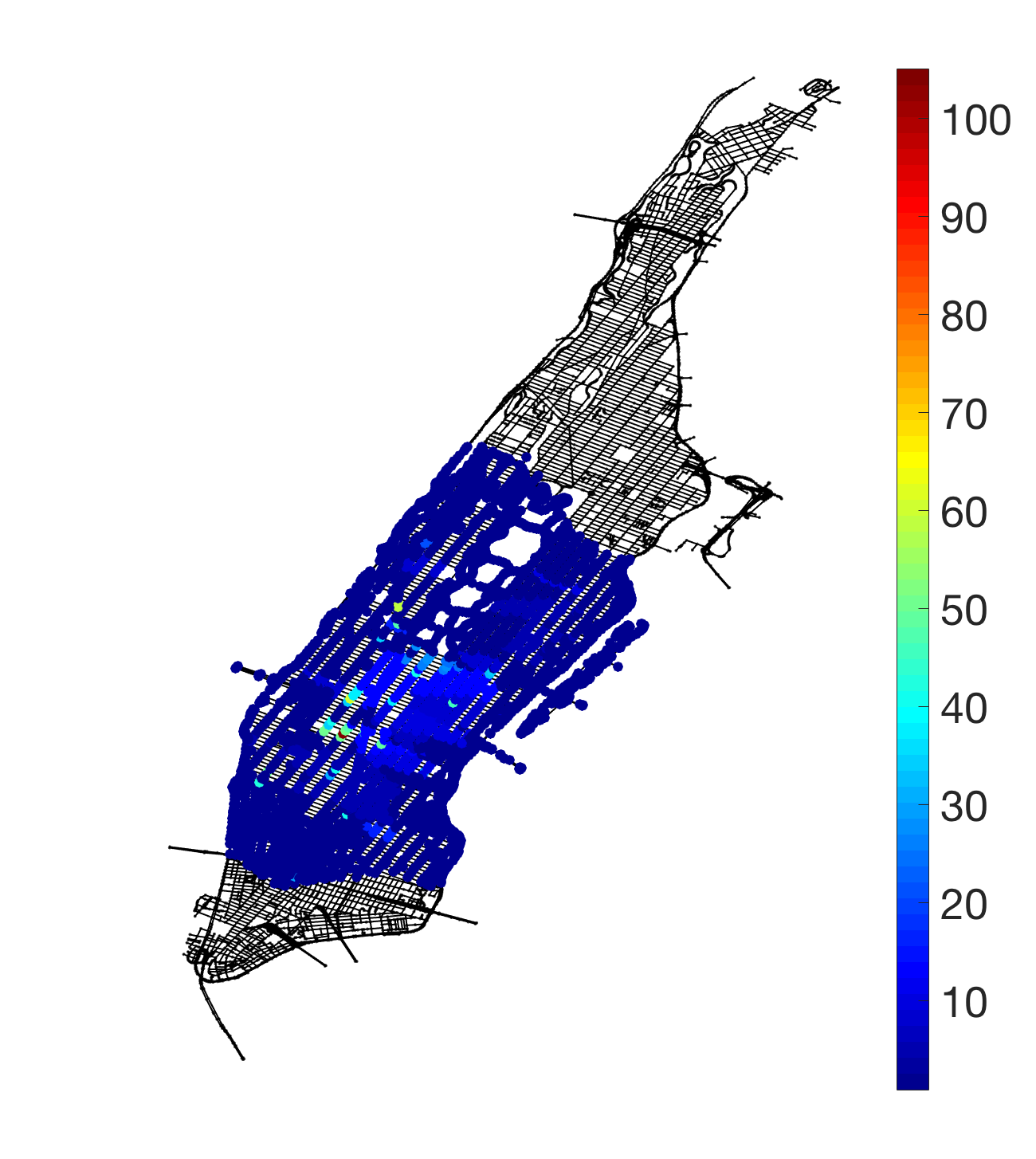}  &
      \includegraphics[width=0.4\columnwidth]{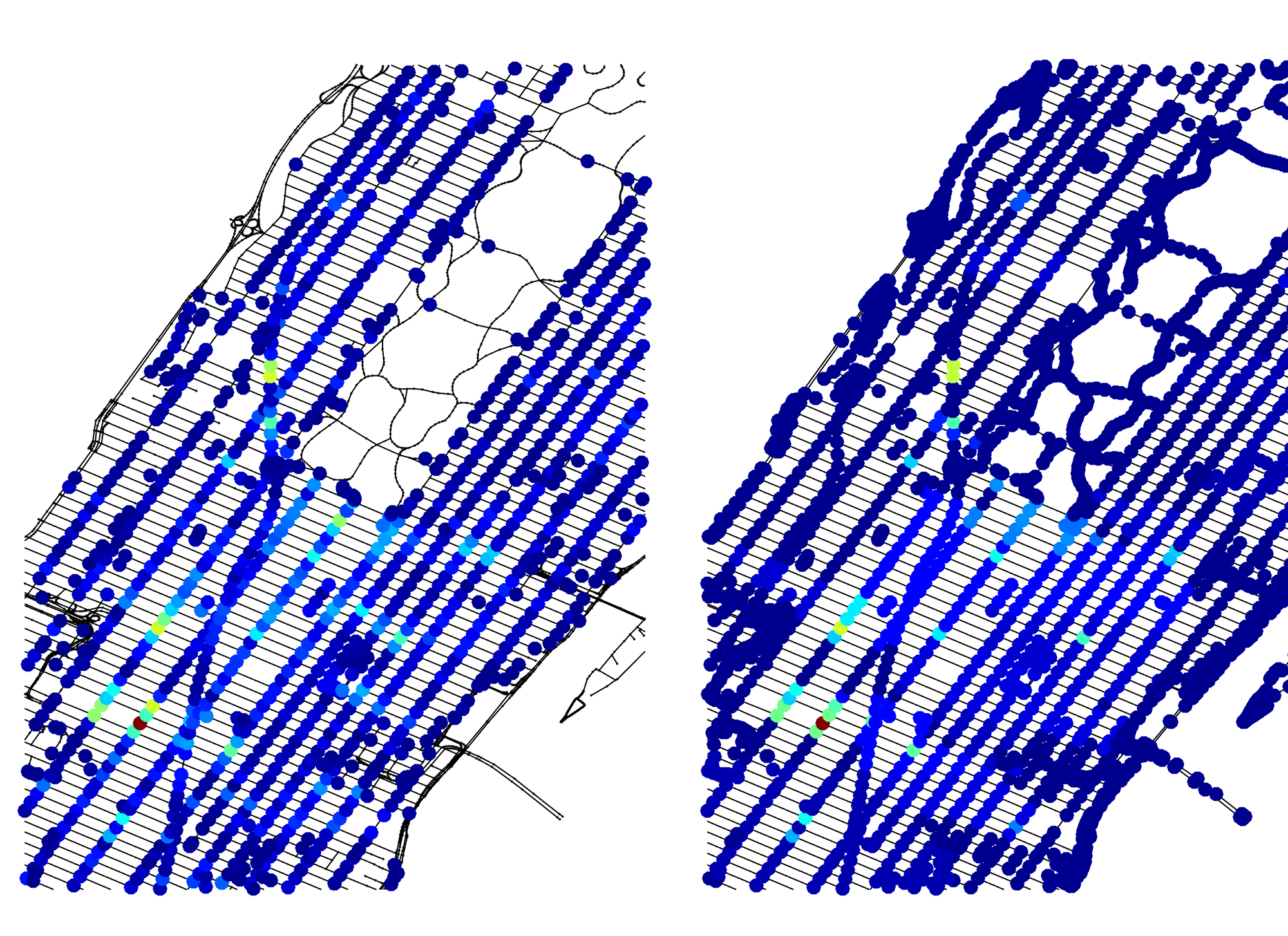}  
      \\
      {\small (c) Piecewise-constant}  & 
      {\small (d) Zoom-in comparison:} 
      \\
      {\small  approximation (50 pieces).}   & {\small real vs. PC approximation.} 
    \end{tabular}
  \end{center}
  \vspace{-4mm}
  \caption{\label{fig:toy_pc} Piecewise-constant approximation well
    represents irregular, nonsmooth graph signals by capturing the
    large variations on the boundary of pieces and ignoring small
    variations inside pieces. Plot (a) shows Taxi-pickup distribution
    at 7 pm on Jan 1st, 2015 in Manhattan. Visually, the distribution
    is well approximated by a piecewise-constant (PC) graph signal
    with 50 pieces in Plot (c). On the other hand, the graph frequency
    based approximation in Plot (b) fails to capture localized
    variations.}
  \vspace{-1mm}
\end{figure}

Today's data is being generated at an unprecedented level from a diversity of
sources, including social networks, biological studies and physical
infrastructure.  The necessity of analyzing such complex data has led
to the birth of \emph{signal processing on
  graphs}~\cite{SandryhailaM:13, SandryhailaM:14, ShumanNFOV:13},
which generalizes classical signal processing tools to data supported
on graphs; the data is the graph signal indexed by the nodes of the
underlying graph. Recent additions to the toolbox include sampling of
graph signals~\cite{AnisGO:15, ChenVSK:15, MarquesSGR:15,
  ChenVSK:15c}, recovery of graph signals~\cite{ChenSMK:14,
  ChenCRBGK:13, NarangGO:13,KotzagiannidisD:16}, representations for
graph signals~\cite{ZhuM:12, ThanouSF:14, TremblayB:16, ChenJVSK:16},
uncertainty principles on graphs~\cite{AgaskarL:13, TsitsveroBL:15},
stationary graph signal processing~\cite{PerraudinV:16,
  MarquesSGR:16}, graph-based filter
banks~\cite{NarangO:12,NarangO:13, EkambaramFAR:15, ZengCO:16},
denoising~\cite{NarangO:12, ChenSMK:14a}, community detection and
clustering on graphs~\cite{Tremblay:14, DongFVN:14, ChenO:14} and
graph-based transforms~\cite{HammondVG:11,NarangSO:10,ShumanFV:16}.

The task of finding activated signal/image supports has been
intensely studied in classical signal/image processing from various
aspects over the past few decades. For example, impulse detection
localizes impulses in a noisy signal~\cite{ZhangK:02}; support
recovery of sparse signals localizes sparse activations with a limited
number of samples~\cite{HauptCN:09}; foreground detection
localizes foreground in a video sequence~\cite{KimCHD:05}; cell
detection and segmentation localize cells in microscopy
images~\cite{BuggenthinMSHHST:13} and matched filtering localizes
radar signals in the presence of additive stochastic
noise~\cite{Turin:60}.

We  consider here the counterpart problem on graphs, that is,
localizing activated supports of signals in a large-scale graph, a
task relevant to many real-world applications from localizing virus
attacks in cyber-physical systems to activity in brain connectivity
networks, to traffic events in city road networks.  Similarly to
classical localization problems, the key issue is to separate
localized activated patterns on graphs from background noise; in other
words, we aim to find the supports of localized activated patterns on
graphs.

Similarly to how piecewise-constant time series are used in classical
signal processing, we model localized activated patterns as
piecewise-constant graph signals since they capture large variations
between pieces and ignore small variations within pieces. In other
words, piecewise-constant graph signals allow us to find supports of
localized patterns in the graph vertex domain.  Each piece in the
signal indicates a localized pattern that exhibits homogeneous
internal behavior while the number of pieces indicates the number of
localized patterns. As an example, Figure~\ref{fig:toy_pc} shows that
a piecewise-constant graph signal can well approximate a real
signal---taxi-pickup distribution in Manhattan.

While most of the work in graph signal processing happens in the graph
frequency domain, we  work here in the graph vertex domain because it
provides better localization and is easier to visualize
(Figure~\ref{fig:toy_pc}(b)). This is similar to classical image
processing, where most methods for edge detection and image
segmentation work in the space domain (instead of the frequency
domain).

Based on the piecewise-constant graph signal model, we consider three
key problems: localization, decomposition, and dictionary learning,
each of which builds on the previous one.  Localization identifies an
activated piece in a \emph{single} graph signal, decomposition identifies \emph{multiple} pieces in a \emph{single} graph signal and dictionary
learning identifies \emph{shared} activated pieces in \emph{multiple}
graph signals.  The aim of studying decomposition and dictionary
learning is not only for efficient approximation, but also for
accurately localizing patterns.

\begin{figure}[h]
  \begin{center}
    \begin{tabular}{ccc}
 \includegraphics[width=0.3\columnwidth]{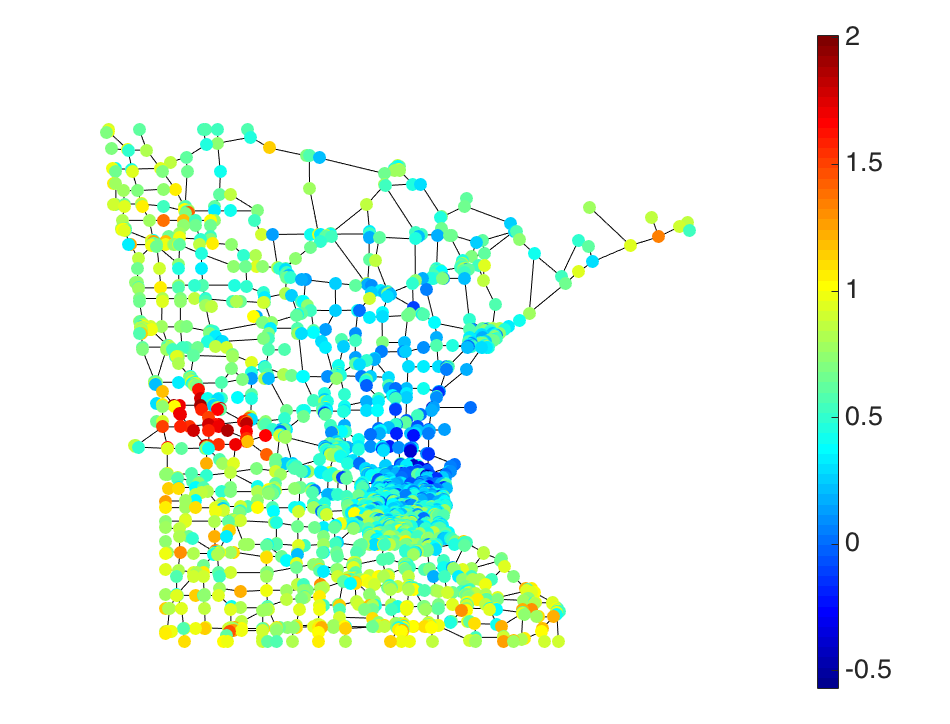}
 &
  \includegraphics[width=0.3\columnwidth]{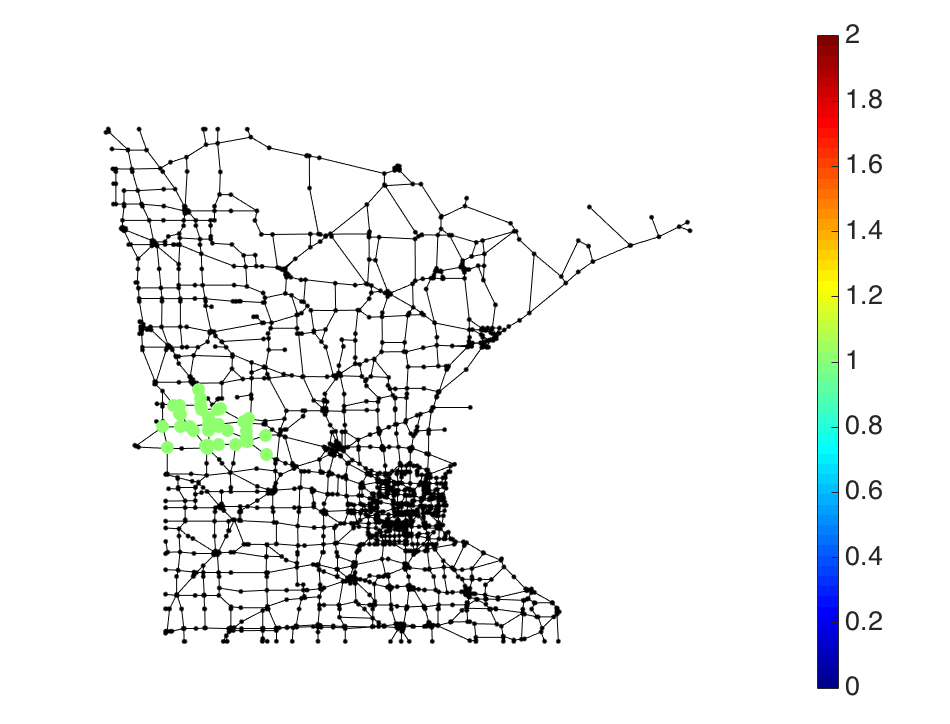}
   &
  \includegraphics[width=0.3\columnwidth]{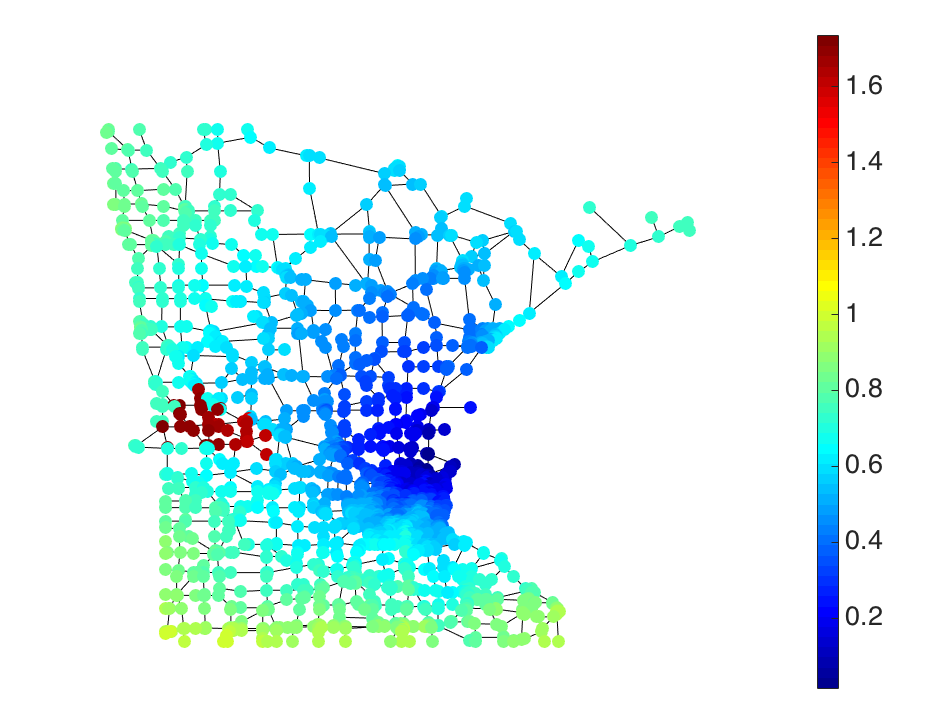}
  \\
    {\small (a) Signal.} &  {\small (b) Activated piece.}  &  {\small (c) Noiseless signal.}
\end{tabular}
  \end{center}
  \caption{\label{fig:toy_local} Signal localization on graphs. Given
    a signal (a), the aim is to identify an activated piece
    (b) while denoising aims to obtain a noiseless  signal
    (c). When a smooth background is ignored, localization is equivalent
    to denoising. }
\end{figure}

\mypar{Localization} This task is the counterpart to matched filtering
in classical signal processing with the aim of identifying a set of
connected nodes where the graph signal switches values---we call this
an \emph{activated piece}. For example, given a graph signal in
Figure~\ref{fig:toy_local}(a), we are looking for an underlying
activated piece as in Figure~\ref{fig:toy_local}(b).  As the original
signal is noisy, this task is related but not equivalent to denoising,
as denoising aims to obtain a noiseless graph signal (see
Figure~\ref{fig:toy_local}(c)), which is not necessarily localized. In
this paper, we focus on localizing activated pieces in noisy
piecewise-constant signals when localization is, in fact, equivalent to
denoising with proper thresholding.

\begin{figure}[h]
  \begin{center}
    \begin{tabular}{ccc}
 \includegraphics[width=0.3\columnwidth]{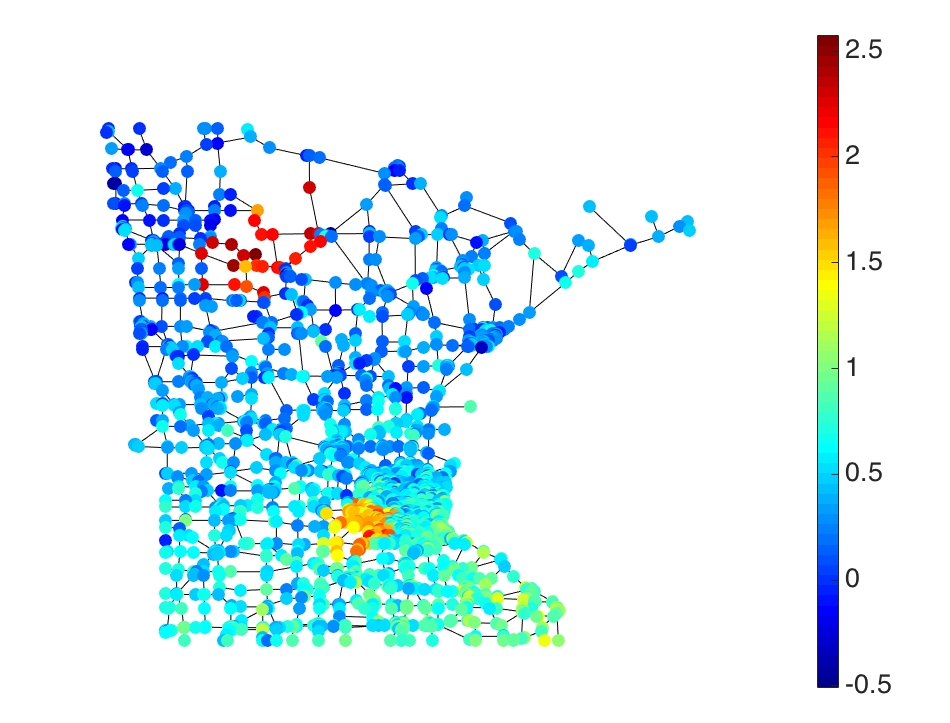}
 &
  \includegraphics[width=0.3\columnwidth]{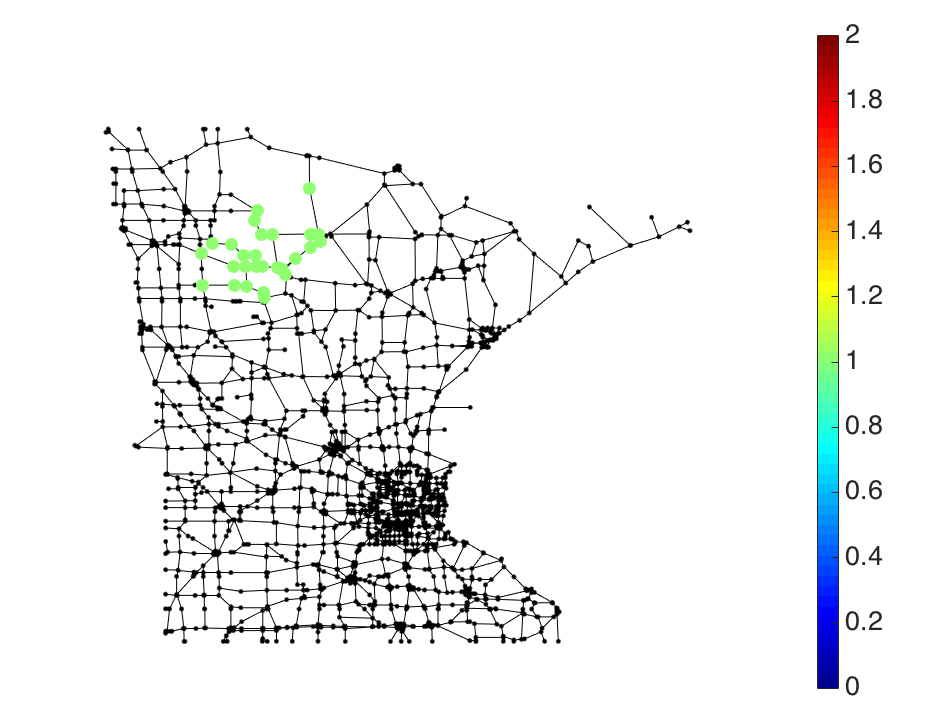}
   &
  \includegraphics[width=0.3\columnwidth]{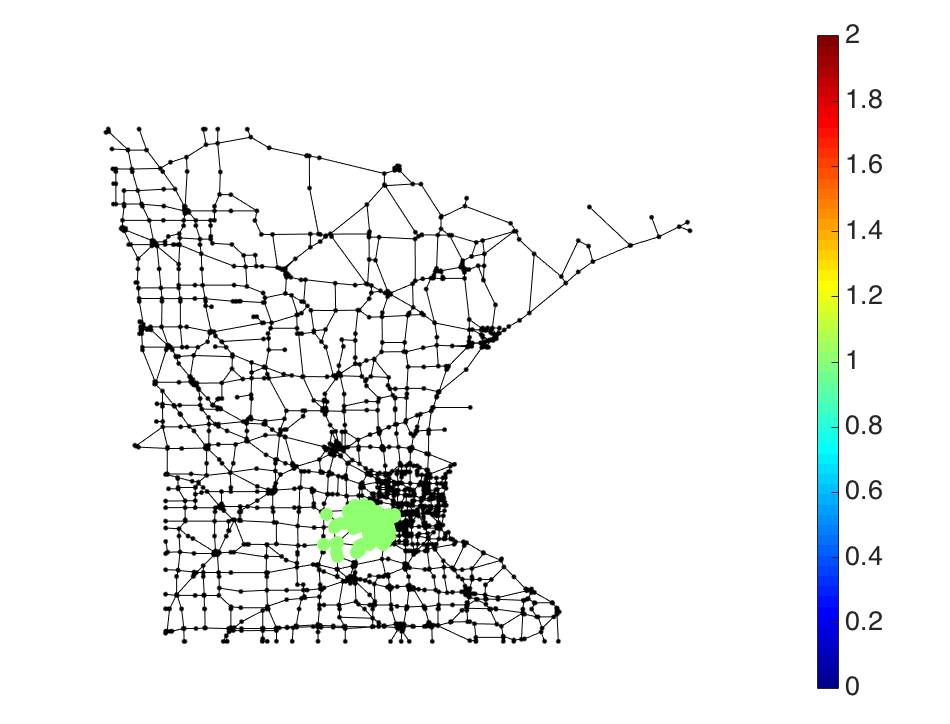}
  \\
    {\small (a) Signal.} &  {\small (b) Piece 1.}  &  {\small (c) Piece 2.}
\end{tabular}
  \end{center}
  \caption{\label{fig:toy_decomposition} Signal decomposition on
    graphs. Given a signal (a), we aim to decompose it into two
    activated pieces (b) and (c). }
\end{figure}

\mypar{Decomposition} This task is the counterpart to independent
component analysis in classical signal processing with the aim of
decomposing or representing a signal as a linear combination of
building blocks---activated pieces.  For example, given a graph signal
in Figure~\ref{fig:toy_decomposition}(a), we decompose it into two
activated pieces, shown in (b) and (c), respectively.

\begin{figure}[h]
  \begin{center}
    \begin{tabular}{ccc}
 \includegraphics[width=0.3\columnwidth]{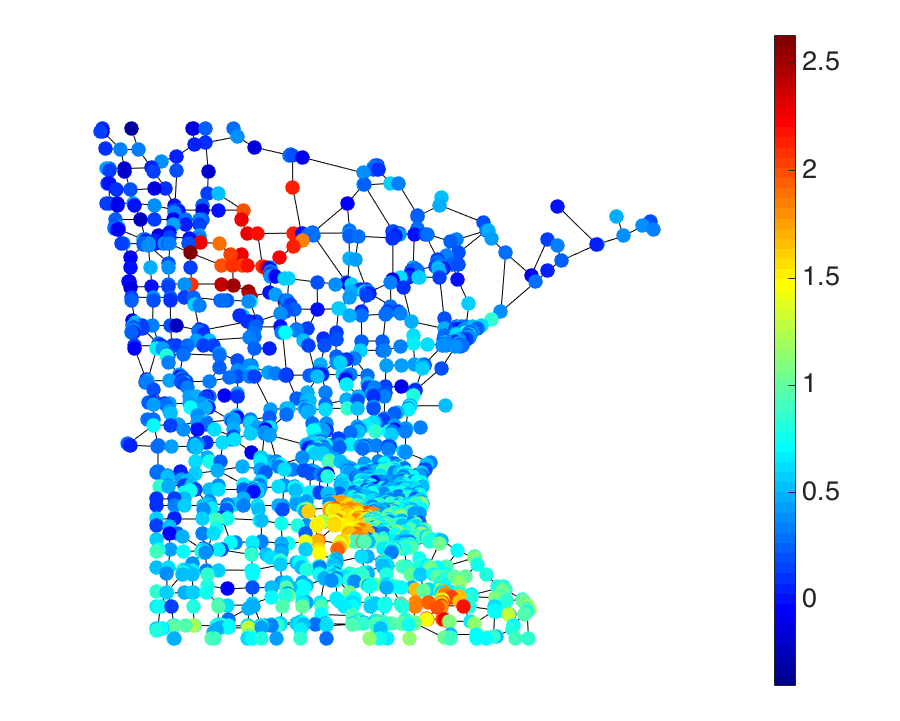}
 &
  \includegraphics[width=0.3\columnwidth]{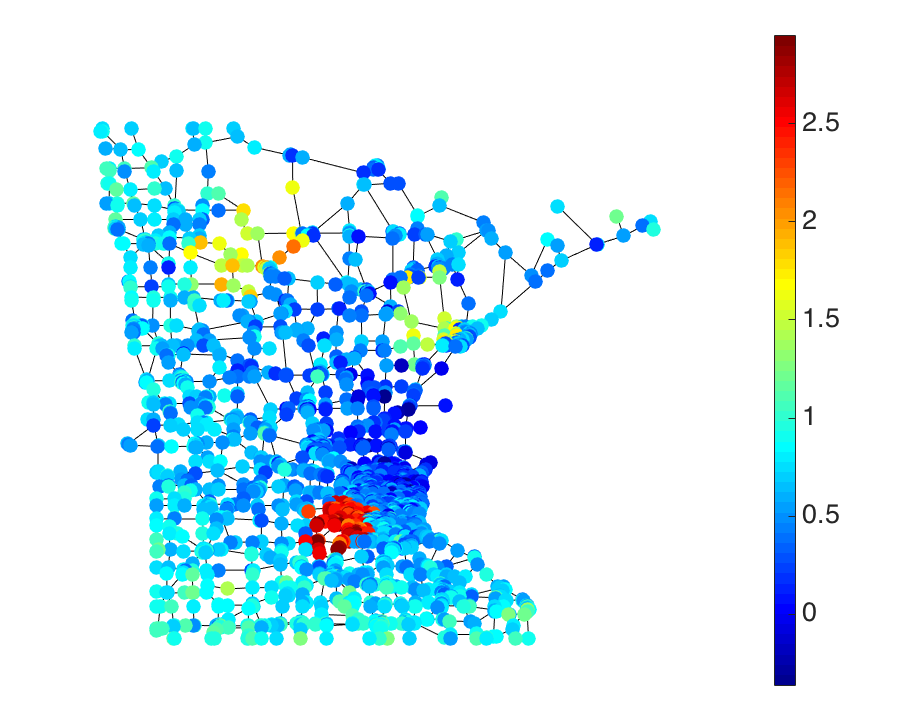}
   &
  \includegraphics[width=0.3\columnwidth]{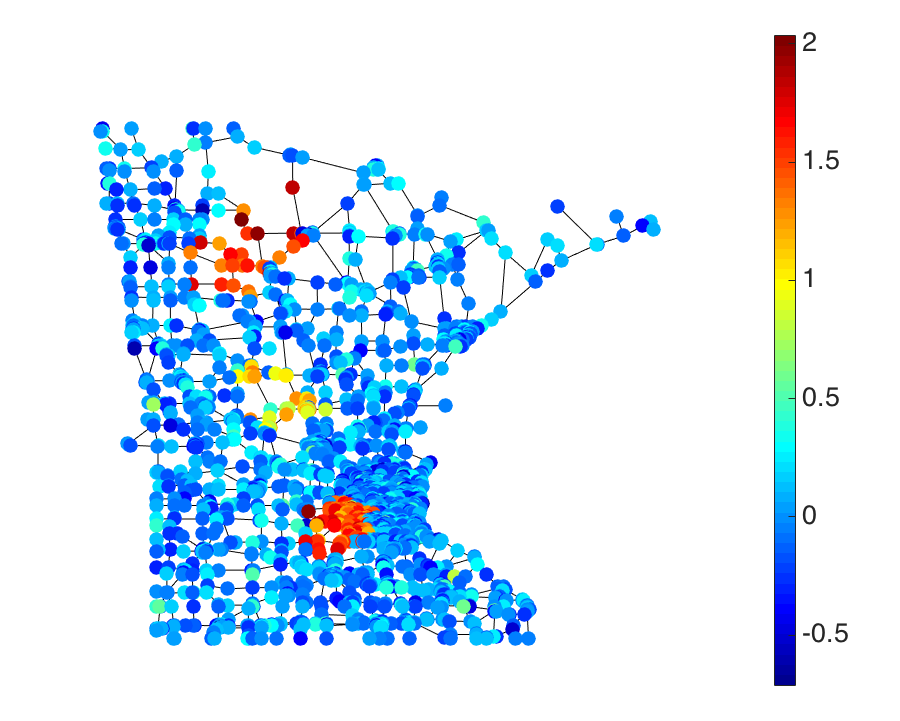}
  \\
    {\small (a) Signal 1.} &  {\small (b) Signal 2.}  &  {\small (c) Signal 3.}
    \\
   \includegraphics[width=0.3\columnwidth]{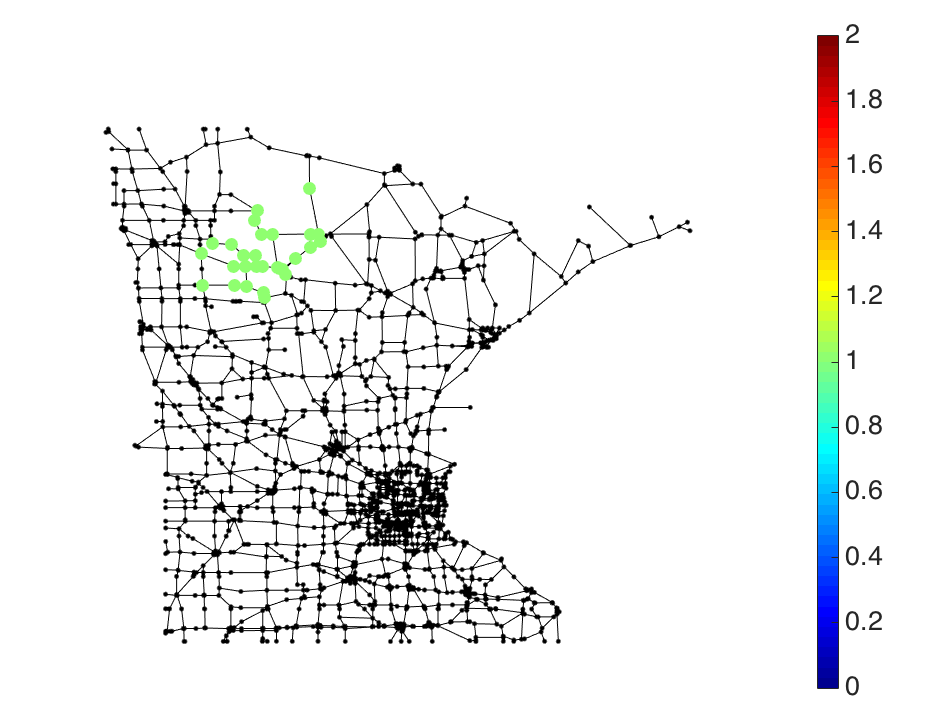}
 &
  \includegraphics[width=0.3\columnwidth]{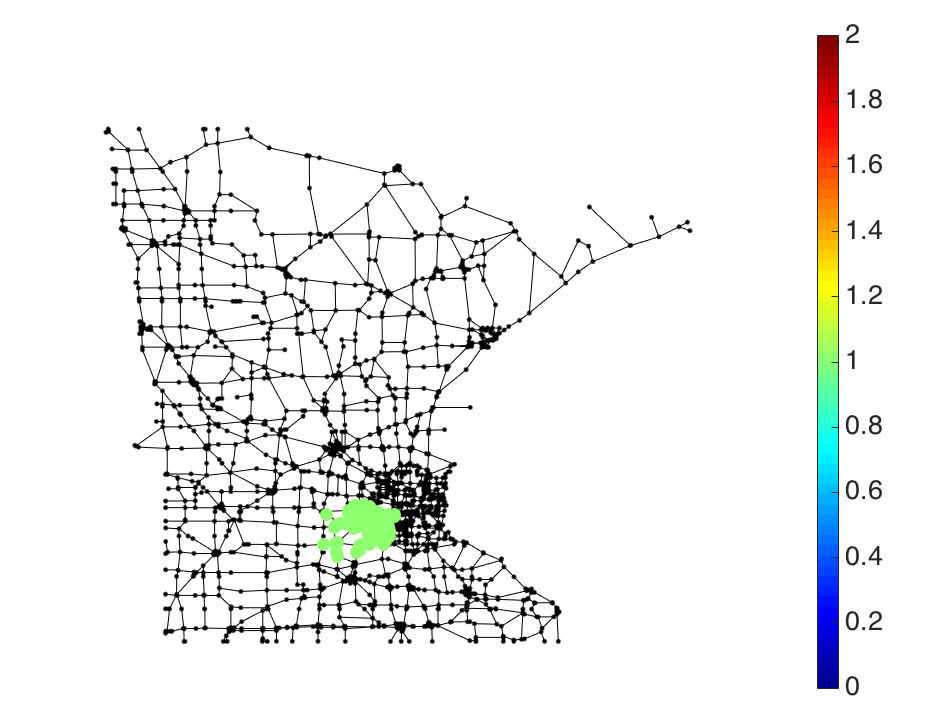}
   &
  \\
    {\small (d) Piece 1.} &  {\small (e) Piece 2.}  &
\end{tabular}
  \end{center}
  \caption{\label{fig:toy_dictionary} Dictionary learning on
    graphs. Given signals (a)--(c), dictionary learning aims to find shared
    activated pieces (d), (e). }
\end{figure}

\mypar{Dictionary learning} This task is the counterpart to sparse
dictionary learning in classical signal processing with the aim of
learning a graph dictionary of one-piece atoms from multiple graph
signals.  Instead of focusing on approximation, the proposed graph dictionary
focuses on finding shared localized patterns from a large set of graph signals.  For
example, given the signals in
Figures~\ref{fig:toy_dictionary}(a)---(c), we look for the underlying
shared activated pieces as in Figures~\ref{fig:toy_dictionary}(d) and
(e). This is relevant in many real-world applications that try to
localize common patterns in a large data volume. For example, on
weekday evenings, the area near Penn Station in Manhattan is
notoriously crowded. We can model the number of passing vehicles at
all the intersections by a graph signal on the Manhattan street
network and use dictionary learning techniques to analyze traffic
activities at various moments and automatically localize the patterns.

\mypar{Related Work} We now briefly review related work on
localization, decomposition and dictionary learning.

\paragraph{Localization} 
Signal localization has been considered from many aspects. From a
signal processing perspective, people studied signal/noise
discrimination~\cite{North:63}, matched filtering~\cite{Turin:60}, and support recovery of sparse
signals~\cite{HauptCN:09}; from an image processing
perspective, people studied object detection in natural and biomedical
images~\cite{Gonzalez:02}, and foreground detection in a video
sequence~\cite{KimCHD:05}; and from a data mining perspective, people
studied anomaly detection~\cite{Faloutsos:14}.  Within
the context of graphs, much of the literature considers detecting
smooth or piecewise-constant graph signals from signals corrupted by
Gaussian noise~\cite{CastroCD:11, HuCSFLL:13,
  SharpnackKS:13, SharpnackKS:13a}. A recent work~\cite{ChenYZSK:16}
considers detecting localized graph signals corrupted by Bernoulli
noise. In this paper, we localize activated pieces on graphs.

\paragraph{Decomposition} Signal decomposition/representation is at
the heart of most signal processing tasks~\cite{VetterliKG:12} as
signals are represented or approximated as a linear combination of
basic building blocks. Fourier analysis can be seen as spectral
decomposition of signals into sinusoidal components and is widely used
in communication systems. Wavelet and multiresolution techniques
decompose signals into components at different scales and are widely
used in image, video compression, and
elsewhere~\cite{KovacevicGV:12}. Other approaches include independent
component analysis~\cite{Bishop:06}, which aims to adaptively
decompose a signal into independent components. Within the context of
graphs, counterparts to classical transforms include the graph Fourier
transform~\cite{SandryhailaM:13, ShumanNFOV:13}, the windowed graph
Fourier transform~\cite{ShumanRV:15}, and the wavelet
transform~\cite{HammondVG:11}. In a number of works, representations
are considered based on the graph Fourier
domain~\cite{SandryhailaM:13, CoifmanM:06, NarangO:12}, designing
graph filters in the graph Fourier domain, and achieving
representations that are localized in the graph Fourier domain. Yet
others consider representations based on the vertex
domain~\cite{GavishNC:10, CrovellaK:03,SharpnackKS:13, ChenJVSK:16},
decomposing a graph into multiple sub-graphs of various sizes and
achieving representations that are particularly localized in the graph
vertex domain.  In this paper, we look at representations based on the
graph vertex domain in a data-adaptive fashion in contrast to some of
the previous work where representations are designed based on the
graph structure only.

\paragraph{Dictionary learning} Dictionary learning finds a sparse
representation of the input data as a linear combination of basic
building blocks. This task is widely studied in image processing,
computer vision, and machine learning. Popular algorithms include the
method of optimal directions~\cite{EnganAH:99},
K-SVD~\cite{AharonEB:06} and online dictionary
learning~\cite{MairalBPS:09}. Within the context of graphs,
\cite{ThanouSF:14} considers learning a concatenation of
subdictionaries, with each subdictionary being a polynomial of the
graph Laplacian matrix. This method captures the localized patterns
that are limited to the exact $K$-hop neighbors, which is not flexible
enough to capture arbitrarily shaped localized patterns. Furthermore,
to update the graph dictionary, quadratic programming with a few hard
constraints is needed, which is not easily scalable.  We here aim to
learn a graph dictionary that captures localized patterns with
arbitrary shapes (not restricted to $K$-hop neighbors). Moreover, the
learning process is scalable.

\mypar{Contributions} We start with a basic localization problem whose
goal is to localize an activated piece with unit magnitude in a noisy
one-piece graph signal. We propose an efficient and effective solver
and extend it to arbitrary magnitude. This solver is able to
efficiently search for a localized pattern with arbitrary shape and
directly finds supports of localized patterns without tuning a
threshold, which is a notoriously challenging issue in many
localization problems. Building on localization, we next consider
signal decomposition and dictionary learning on graphs. We conduct
extensive experiments to validate the proposed methods.  The results
show that cut-based localization is good at localizing the ball-shaped
class and path-based localization is good at localizing the elongated
class. We further apply our algorithms to analyze urban data, where
our methods automatically detect events and explore mobility patterns
across the city from a large volume of spatio-temporal data.

The main contributions of the paper are:
\begin{itemize}
\item A novel and efficient solver for graph signal localization
  without tuning parameters.
\item A novel and efficient solver for graph signal decomposition.
\item A novel and efficient solver for graph dictionary learning.
\item Extensive validation of the proposed solvers on both simulated
  as well as real data (Minnesota road and Manhattan street networks).
\end{itemize}

\mypar{Notation} Consider an undirected graph $G = (\V, \E )$, where
$\V = \{v_1,\ldots, v_{N}\}$ is the set of nodes and $\E =
\{e_1,\ldots, e_{M}\}$ is the set of edges. A~\emph{graph signal} $\x$
maps the graph nodes $v_n$ to the signal coefficients $x_n \in \R$; in
vector form, $\x \ = \ \begin{bmatrix} x_1 & x_2 & \ldots & x_{N}
\end{bmatrix}^T \in \R^N$. Let $C \subseteq \V $ be a subset of nodes;
we represent it by the indicator vector, 
\begin{equation}
\label{eq:indicator}
{\bf 1}_{C}  =
  \left\{
    \begin{array}{rl}
        1, & v_i \in C;\\
      0, & \mbox{otherwise}.
  \end{array} \right.
\end{equation}
that is, ${\bf 1}_{C}$ is a graph signal with $1$s in $C$ and $0$s in
the complement node set $\overline{C} = \V \setminus C$. We say that
the node set $C$ is~\emph{activated}.  When the node set $C$ forms a
connected subgraph, we call $C\in \Cc$ a~\emph{piece}, $\Cc$ is the
set of all pieces and ${\bf 1}_{C}$ a~\emph{one-piece graph signal}.

Define a \emph{piecewise-constant graph signal} as
\begin{eqnarray*}
 \x \ = \  \sum_{i = 1}^K \mu_i  {\bf 1}_{C_i},
\end{eqnarray*}
with $C_i$ a piece, $\mu_i$ a constant and $K$ the number of pieces.

\mypar{Outline of the paper} We present the three problems, graph
signal localization, decomposition, and dictionary learning in
Sections~\ref{sec:localization}, \ref{sec:decomposition}, and
\ref{sec:dictionary}, respectively. Section~\ref{sec:conclusions}
concludes the paper and provides pointers to future research
directions.

\section{Signal Localization on Graphs}
\label{sec:localization}
In this section, we propose a solver for the localization problem,
validating it by simulations.

Consider localizing an activated piece $C\in \Cc$ in a noisy,
piecewise-constant graph signal \vspace{-2mm}
\begin{eqnarray}
    \vspace{-2mm}
  \label{eq:gs_onepiece}
	\x \ = \  \mu \one_C + \epsilon, % \ \in \ \R^N,
\end{eqnarray}
where $\one_C$ is the indicator function~\eqref{eq:indicator}, $\mu$
is the signal strength and $\epsilon \sim \N(0, \sigma^2 \Id_N)$ is
Gaussian noise.  We consider two cases: $\mu$ arbitrary and $\mu = 1$.

Previously, this has been formulated as a detection problem via a scan
statistic searching for a most probable anomaly set. While this is
similar to localizing an activated piece, it is either computationally
inefficient or hindered by strong assumptions. For example,
in~\cite{CastroCD:11}, the authors analyze the theoretical performance
of detecting paths, blobs and spatial temporal sets by exhaustive
search, resulting in a costly algorithm, while
in~\cite{SharpnackKS:13a, ChenYZSK:16}, the authors
aim to detect a node set with a specific cut number, resulting in a
computationally efficient algorithm, but limited by strong
assumptions.

We aim to efficiently localize an activated piece with arbitrary shape.
The maximum likelihood estimator of the one-piece-localization problem
under Gaussian noise is
\begin{eqnarray}
    \vspace{-4mm}
  \label{eq:opfu}
  \min_{\mu, C}  \left\| \x -  \mu \one_C \right\|_2^2,
  \quad \text{subject to } C \in \Cc.
      \vspace{-4mm}
\end{eqnarray}

\vspace{-4mm}
\subsection{Methodology}

\subsubsection{Localization with unit magnitude}
Consider first \eqref{eq:gs_onepiece} with a unit-magnitude signal,
that is, $\mu=1$.
% \begin{eqnarray*}
%   \x \ = \   \one_C + \epsilon \ \in \ \R^N,
% \end{eqnarray*}
% where $C$ is a piece and $\epsilon \sim \N(0, \sigma^2 \Id_N)$. The
% goal is to localize $C$.  We require $C$ to be connected.
Then \eqref{eq:opfu} reduces to
\begin{eqnarray}
  \label{eq:opfp}
  \min_{C}  \left\| \x -  \one_C \right\|_2^2,
  \quad \text{subject to } C \in \Cc.
\end{eqnarray}
This optimization problem is at the core of this paper. In what
follows, we first look at hard thresholding as a baseline and then
propose two efficient solvers for two typical and complementary
classes: a ball-shaped class and an elongated class.  A ball-shaped class consists of activated pieces, forming subgraphs with relatively small  diameters compared to its cardinality and dense internal connections; as an opposite to a ball-shaped class, an elongated class consists of activated pieces, forming subgraphs with relatively large  diameters compared to its cardinality and sparse internal connections. An extreme case of an elongated piece is a path, which is a tree with two nodes degree 1 and the other nodes degree 2.   We then combine both solvers to obtain a general solver for~\eqref{eq:opfp}.

\mypar{Hard thresholding based localization} The difficulty in
solving~\eqref{eq:opfp} comes from the constraint, which forces the
activated nodes to form a connected subgraph (piece). Since we do not
restrict the size and shape of a piece, searching over all pieces to
find the global optimum is an NP-complete problem. We consider a
simple algorithm that solves~\eqref{eq:opfp} in two steps: we first
optimize the objective function and then project the solution onto the
feasible set in~\eqref{eq:opfp}. While this simple algorithm
guarantees a solution, it is not necessarily optimal. The objective
function can be formulated as
\begin{eqnarray*}
  \min_C  \left\| \x - \one_C \right\|_2^2
  =     \min_{\t \in \{ 0, 1 \}^N }  \left\| \x - \t \right\|_2^2
  =   \min_{ t_i \in \{ 0, 1 \} }  \sum_{i=1}^N  ( x_i - t_i )^2,
\end{eqnarray*}
where $\t \in R^N$ is a vector representation of a node set $C$. Since
$t_i$ are independent, we can optimize each individual element to
obtain $t^*_i = 1$, when $\x_i > 1/2$; and $0$, otherwise. In other
words, the optimum of the objective function is
\begin{equation*}
  \{ v_i \in \V  \mid x_i > \frac{1}{2}  \} 
  \ = \ \min_C  \left\| \x - \one_C \right\|_2^2.
\end{equation*}
This step is nothing but global hard thresholding.  We then project
this solution onto the feasible set; that is, we solve
\begin{eqnarray}
  \label{eq:hard_thr}
  C^*_{\rm thr} 
  & = &   \Pj_{\rm \Cc} ( \arg \min_C  \left\| \x - \one_C \right\|_2^2  ) \\
    \nonumber
    & = &   \Pj_{\rm \Cc} (   \{ v_i \in \V  \mid x_i > \frac{1}{2}  \}  ),
\end{eqnarray}
where the projection operator $\Pj_{\rm \Cc} ( C )$ extracts the
largest connected component (piece) in a node set $C$.  Thus, this
solver simply performs hard thresholding and then finds a connected
component among the nodes with nonzero elements on it.

\mypar{Cut-based localization for ball-shaped class}  The key idea
in~\eqref{eq:hard_thr} is to partition nodes into activated and
inactivated nodes. This is similar to graph cuts: cutting a series of
edges with minimum cost and obtaining two isolated components. A small cut cost indicates that internal connection is stronger than external connection. The difference between graph cuts and our task is that graph cuts optimize edge weights,
while~\eqref{eq:opfp} optimizes graph signal coefficients. We can adaptively update the edge weights by using graph signal coefficients and solve~\eqref{eq:opfp} by using efficient graph-cut solvers. Inspired
by~\cite{SharpnackKS:13a, ChenYZSK:16}, we add the number of edges
connecting activated and inactivated nodes to the objective function
to induce a connected component. Let $\Delta \in \R^{M \times N}$ be
the~\emph{graph incidence matrix} of $G$~\cite{Newman:10}. When the
edge $e_i$ connects the $j$th to the $k$th node ($j < k$), the the
$i$th row of $\Delta$ is
\begin{equation}
\label{eq:Delta}
 \Delta_{i, \ell} =
  \left\{
    \begin{array}{rl}
        1, & \ell = j;\\
       - 1, & \ell = k;\\
      0, & \mbox{otherwise}.
  \end{array} \right.
\end{equation}

The signal $\Delta \one_C \in \R^M$ records the first-order
differences of $\one_C$. The number of edges connecting $C$ and
$\overline{C}$ is
\begin{displaymath}
  \left\|  \Delta \one_C \right\|_0  =   \left\|  \Adj \one_C  \right\|_1 - \one_C^T \Adj \one_C,
\end{displaymath}
where $\Adj$ is the adjacency matrix, $ \left\| \Adj \one_C
\right\|_1$ is the sum of the degrees of the nodes in $C$ and
$\one_C^T \Adj \one_C$ describes the internal connections within
$C$. When there are no edges connecting  activated nodes, $ \left\|
  \Delta \one_C \right\|_0  = \left\| \Adj \one_C \right\|_1$. When
all activated nodes are well connected and form a ball-shaped piece, internal connections $\one_C^T
\Adj \one_C$ is large and $ \left\| \Delta \one_C \right\|_0 
\ll  \left\| \Adj \one_C
\right\|_1$. Thus, minimizing $\left\| \Delta \one_C \right\|_0$ induces $C$
to be a piece. We thus solve
\begin{subequations}
  \begin{eqnarray}
    \label{eq:cut_1}
    C_\lambda & = &   \Pj_{\rm \Cc} ( \arg \min_C  \left\| \x - \one_C \right\|_2^2 + \lambda \left\|  \Delta \one_C \right\|_0  ),
    \\
    \label{eq:cut_2}
    C^*_{\rm cut} & = & \arg \min_{C_\lambda}  \left\| \x - \one_{C_\lambda} \right\|_2^2.
  \end{eqnarray}
\end{subequations}
Given $\lambda$,~\eqref{eq:cut_1} solves the regularized optimization
problem and extracts the largest connected component, while
\eqref{eq:cut_2} optimizes over $\lambda$ to find a solution. The hard
thresholding~\eqref{eq:hard_thr} is a subcase
of~\eqref{eq:cut_2} with $\lambda = 0$.  We solve \eqref{eq:cut_1}
using the Boykov-Kolmogorov graph cuts
algorithm~\cite{BoykovVZ:01,KolmogorovZ:04}; we call this
\emph{cut-based localization}. 

To understand the regularized problem \eqref{eq:cut_1} better, we have
\begin{eqnarray*}
  &&  \min_C  \left\| \x - \one_C \right\|_2^2 + \lambda \left\|  \Delta \one_C \right\|_0
  \\
  & = &   \min_{\t \in \{0, 1\}^N}  \left\| \x - \t \right\|_2^2 + \lambda \left\|  \Delta \t \right\|_0
  \\
  & = &   \min_{t_i \in \{0, 1\} } \sum_i (  ( x_i - t_i )^2 + \lambda \sum_{j \in {\rm \Nei(i)}} |t_i -t_j|),
\end{eqnarray*}
where $\Nei(i)$ is the neighborhood of the $i$th node. That is, we have $t^*_i = 1$ when
\begin{eqnarray*}
  ( x_i - 1 )^2 + \lambda \sum_{j \in \Nei(i)} | x_j - 1|   <  x_i^2 + \lambda \sum_{j \in \Nei(i)} | x_j |.
\end{eqnarray*}
The solution is
\begin{equation}
\label{eq:cut_sol}
 \left( \t^*_{\rm cut} \right)_i  =
  \left\{
    \begin{array}{rl}
        1, & x_i  >  \frac{1}{2} +  \frac{\lambda}{2} \left( \alpha_0 - \alpha_1 \right);\\
      	0, & \mbox{otherwise},
  \end{array} \right.
\end{equation}
with $\alpha_1$, $\alpha_0$ the number of neighbors of $x_i$ with
value $1$ and $0$, respectively. Thus, \eqref{eq:cut_sol} is nothing
but adaptive local thresholding, where the value of each element
depends on the values of its neighbors. This explains why cut-based
localization outperforms hard thresholding.

\mypar{Path-based localization for elongated class} 
Cut-based
localization~\eqref{eq:cut_1} is
better suited to capturing ball-shaped pieces than elongated
pieces.  For example, to solve~\eqref{eq:opfp}, we
introduced $\left\| \Delta \one_C \right\|_0 $ in~\eqref{eq:cut_1} to
promote fewer connections between activated nodes and inactivated
ones. This is because when $C$ is a ball-shaped piece, $\one_C^T \Adj \one_C$ is
large (see Figures~\ref{fig:localization_ball}(a)--(b)). An extreme case is that  $C$ is fully connected, $C$ forms a ball-shaped
piece  and $\one_C^T
\Adj \one_C = |C|(|C| - 1)/2$, which is large enough for graph cuts to capture.; however, when nodes in $C$ are weakly
connected, for example, when $C$ forms a path,
$\one_C^T \Adj \one_C = |C| - 1$, which is too small for graph cuts to capture.

To address elongated pieces, we consider two methods. The first method promotes elongated shape based on convex optimization. We start
by restricting $\left\| \Adj \one_C \right\|_{\infty}$, the maximum
degree of the nodes in the activated piece to be at most $2$,
promoting an elongated shape, and then solve the following
optimization problem:
\begin{eqnarray*}
  \min_C  \left\| \x - \one_C \right\|_2^2,
  \quad \text{subject to }  \quad
  \left\| \Adj \one_C \right\|_{\infty} \leq 2.
\end{eqnarray*}
This constraint requires that each activated node connect to at most
two activated nodes, which promotes elongated shape. To solve this
efficiently, we relax the problem to a convex one,
\begin{eqnarray*}
  && \t_{p_1}^* \ = \ \arg \min_{\t \in \R^N}  \left\| \x - \t \right\|_2^2,
  \\
  && \text{subject to } \left\| \Adj \t \right\|_{\infty} \leq 2
  \quad \text{and} \quad
  0 \leq \t \leq 1.
\end{eqnarray*}
We then set various thresholds for $\t_{p_1}^*$, extract the
largest connected component and optimize over the threshold to get
\begin{eqnarray*}
  \nonumber
  C_\lambda & = &   \Pj_{\rm \Cc} ( \one_{\t_{p_1}^* \geq \lambda} ),
  \\
\label{eq:path_2}
C^*_{p_1} & = & \arg \min_{C_\lambda}  \left\| \x - \one_{C_\lambda} \right\|_2^2.
\end{eqnarray*}
The solution promotes an elongated shape, but does not constrain it to
an exact path.

The second method searches for an exact path based on the
shortest-path algorithm, leading to 
\begin{equation}
  \label{eq:path_1}
  C^*_{p_2} \ = \ \arg \min_{C {\rm~is~a~path}}  
  \left\| \x - \one_C \right\|_2^2 + \lambda | C |,
\end{equation}
where the regularization parameter $\lambda = 2 x_{\max} - 1 $ with
$x_{\max}$ the maximum element in $\x$, guaranteeing that
shortest-path algorithm will work later.  The optimization
problem~\eqref{eq:path_1} can be implemented in two steps as follows:
\begin{equation*}
  \min_{ {\rm~all~pairs~of}~s, t} 
  (\min_{C {\rm~connects}~s, t}  \left\| \x - \one_C \right\|_2^2 + \lambda | C |).
\end{equation*}
We first fix two end points $s$ and $t$ and find the optimal path $C$
connecting $s$ and $t$, and then enumerate all pairs of end points and
find the globally optimal path. We will show that the first step can
be solved exactly by the shortest-path algorithm.

The objective function of~\eqref{eq:path_1} can be rewritten as  
\begin{eqnarray*}
  \left\| \x - \one_C \right\|_2^2 + \lambda | C |
  & \ = \ &  \x^T \x - 2  \one_C^T \x + (1 + \lambda)  \one_C^T \one
  \\
  & \ = \ & \x^T \x + 2 \one_C^T \y,
\end{eqnarray*}
where $\y = \frac{1+\lambda}{2} \one - \x$. Since $C$ is the only
variable, minimizing~\eqref{eq:path_1} is equivalent to minimizing
$\one_C^T \y$.

We consider a graph with edge weight
\begin{equation*}
  \W_{i,j}  \ = \
  \left\{
    \begin{array}{rl}
      \frac{ y_i + y_j }{2}, & (i,j) \in \E;\\
      \infty, & \mbox{otherwise}.
    \end{array} \right.
\end{equation*}

Let $C$ be a path connecting two end points $s$ and $t$. The path
weight of $C$ is
\begin{eqnarray*}
  \sum_{i,j \in C} \W_{i,j} \ = \  \sum_{k \in C} y_k - \frac{y_{s} + y_{t}}{2}
    \ = \  \one_C^T \y - \frac{y_{s} + y_{t}}{2}.
\end{eqnarray*}
This indicates that given two end points $s$ and $t$, minimizing
$\one_C^T \y$ is equivalent to minimizing the path weight. Since
$\lambda = 2 x_{\max} - 1$, all edge weights are nonnegative. We can
then use the shortest-path algorithm to efficiently compute the short
paths between all node pairs~\cite{CormenLRS:01} and graph coarsening
techniques to reduce the number of nodes and boost the
speed~\cite{ChevalierS:09, LiuWG:14}.

Finally, the elongated piece is obtained by choosing the one with a
smallest objective value in~\eqref{eq:opfp}.
\begin{equation}
  \label{eq:path_opt}
  C_{\rm path}^*  \ = \	\arg \min_{C \in \{C^*_{p_1}, C^*_{p_2} \}  }  \left\| \x - \one_{C} \right\|_2^2.
\end{equation}
We call this \emph{path-based localization}.

 \mypar{Combined solvers} We  combine the cut-based localization $C_{\rm cut}^*$ and the path-based localization $C_{\rm path}^*$ by choosing the one with a smaller objective value. The final solution to localize an arbitrarily shaped piece is
\begin{equation}
\label{eq:localization_opt}
	C^* \ = \ {\rm Loc_1 (\x) }  \ = \	\arg \min_{C \in \{C^*_{\rm cut}, C^*_{\rm path} \}  }  \left\| \x - \one_{C} \right\|_2^2,
\end{equation}
where Loc$_1(\cdot)$ is the operator that finds an activated piece
with unit magnitude. This solution is not the global optimum
of~\eqref{eq:opfp}; it considers two typical and complementary
 subsets in the feasible set, and combines a graph-cut solver, a convex programming solver and a shortest-path solver, all of which are efficient. Note that detection or localization techniques usually need to set thresholds; however, this localization solver~\eqref{eq:localization_opt} is parameter-free and directly outputs the supports of a localized pattern.  In
Section~\ref{sec:localization_exp}, we validate it empirically.

\subsubsection{Localization with unknown magnitude}
\label{sec:localization_unknown}
We next consider a more general case where the magnitude of the
activated piece is unknown, \eqref{eq:gs_onepiece}, with the
associated optimization problem \eqref{eq:opfu}.
% the graph signal model is
% \begin{eqnarray*}
% 	\x \ = \   \mu \one_C + \epsilon \ \in \ \R^N,
% \end{eqnarray*}
% where the node set $C$ is connected, the signal strength $\mu$ is
% unknown and the noise $\epsilon \sim \N(0, \sigma^2 \Id_N)$.  The
% optimization problem is then
% \begin{eqnarray}
% \label{eq:opfu}
% \mu^*, C^*
%  & = &  \arg\min_{\mu, C}  \left\| \x - \mu \one_C \right\|_2^2,
% 	\\
% 	\nonumber
% 	&& \text{subject to }  C \in \Cc.
% \end{eqnarray}

We iteratively solve $C$ and $\mu$ until convergence; that is, given
$C$, we optimize over $\mu$ and then given $\mu$, we optimize over
$C$. In the $k$th iteration,
\begin{eqnarray*}
  \mu^{(k)}
  & = &  \min_{\mu}  \left\| \x - \mu \one_{C^{(k)}} \right\|_2^2
  \ =  \  \frac{\x^T \one_{C^{(k)}}}{\one_{C^{(k)}}^T \one_{C^{(k)}} },
  \\
  C^{(k)}
  & = & {\rm Loc_1 (\frac{1}{\mu} \x) }.
\end{eqnarray*}
We obtain a pair of local optima by alternately minimizing these two
variables. We denote this solver by 
\begin{equation}
\label{eq:loc_unknown}
	\mu^*, C^* \ = \ 	 {\rm Loc}_{\rm unknown} ( \x ).
\end{equation}

\vspace{-4mm}
\subsection{Experimental Validation}
\label{sec:localization_exp}
We test our localization solver on two graphs: the Minnesota road
network and the Manhattan street network. On each graph, we consider
localizing two classes of simulated graph signals with different
activated sizes under various noise levels.

\subsubsection{Minnesota road network}
We model this network as a graph with its 2,642 intersections as nodes
and 3,342 road segments between intersections as undirected edges.  We
simulate ball-shaped and elongated classes of one-piece graph signals.
 
\begin{figure}[t]
  \begin{center}
    \begin{tabular}{cc}
      {\small Radius $5$. }  &  {\small Radius $10$. }\\
      \includegraphics[width=0.45\columnwidth]{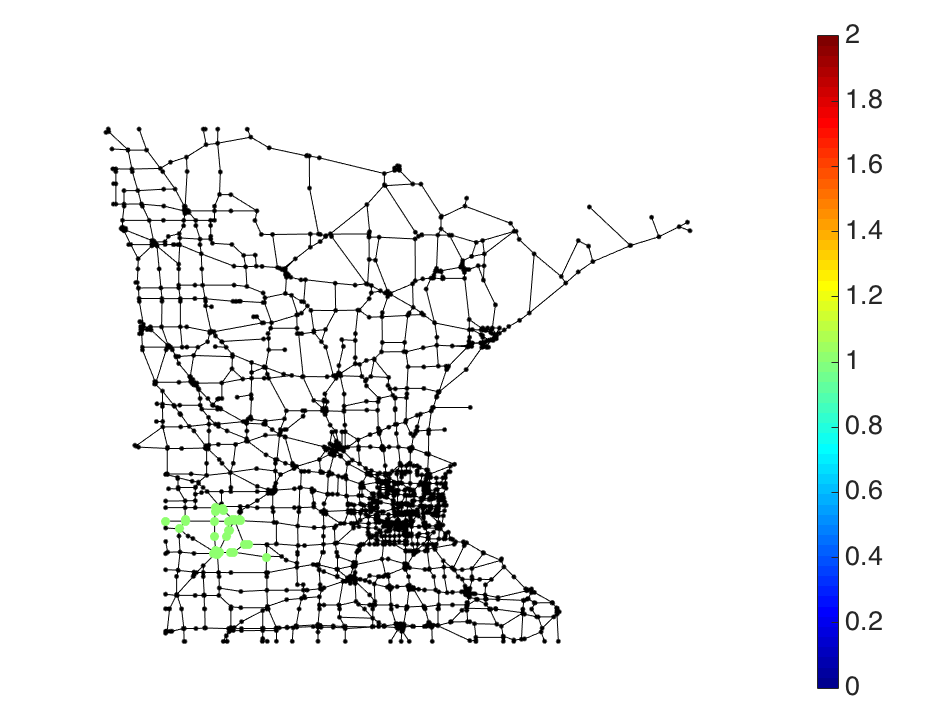}
      &
      \includegraphics[width=0.45\columnwidth]{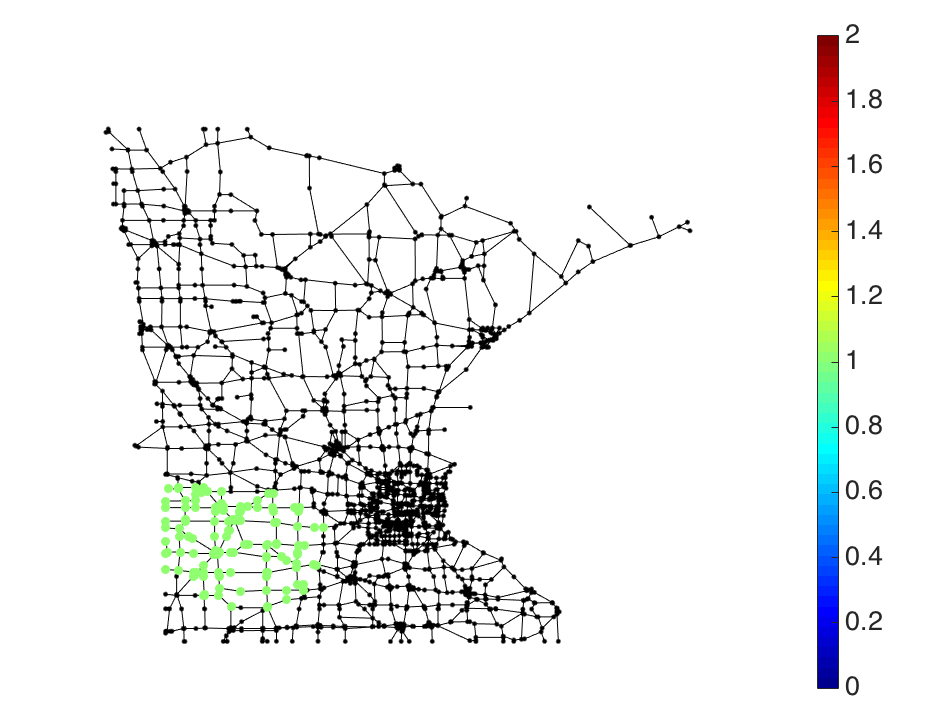}\\
      \multicolumn{2}{c}{\small (a)--(b) Ball-shaped piece.}\\
      \includegraphics[width=0.4\columnwidth]{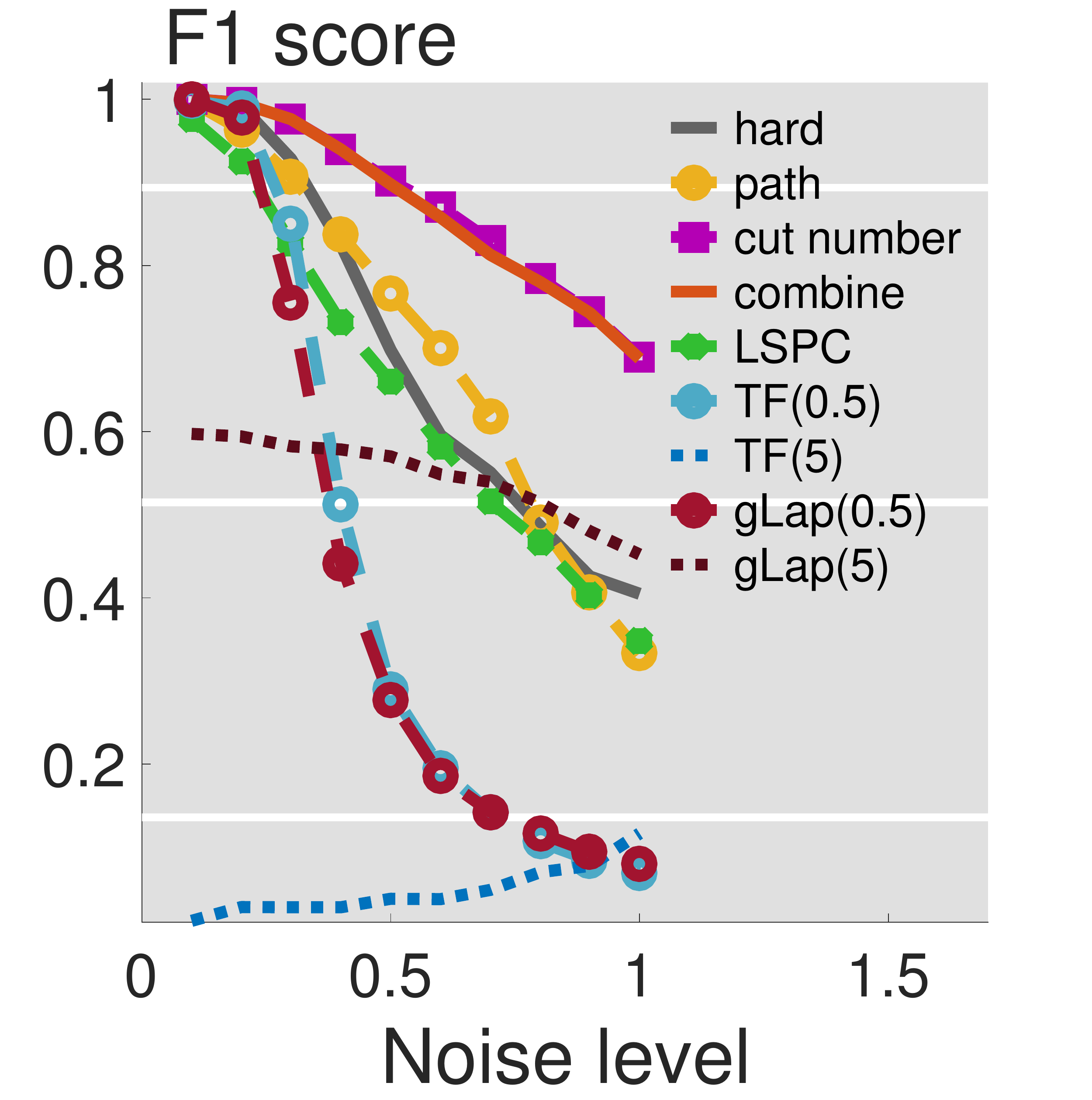}
      &
      \includegraphics[width=0.4\columnwidth]{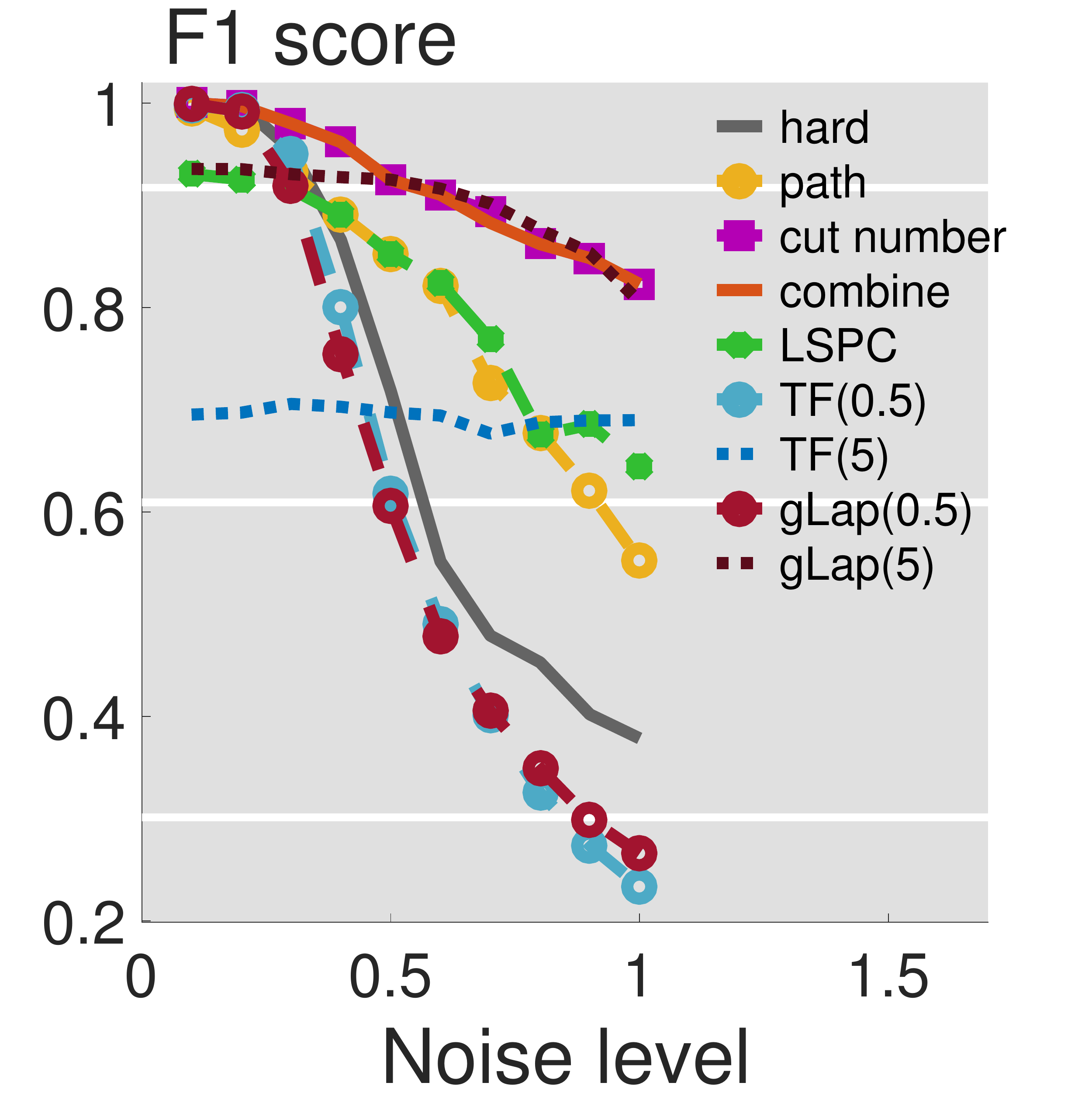}\\
      \multicolumn{2}{c}{\small (c)--(d) $F_1$ score.}\\
      \includegraphics[width=0.4\columnwidth]{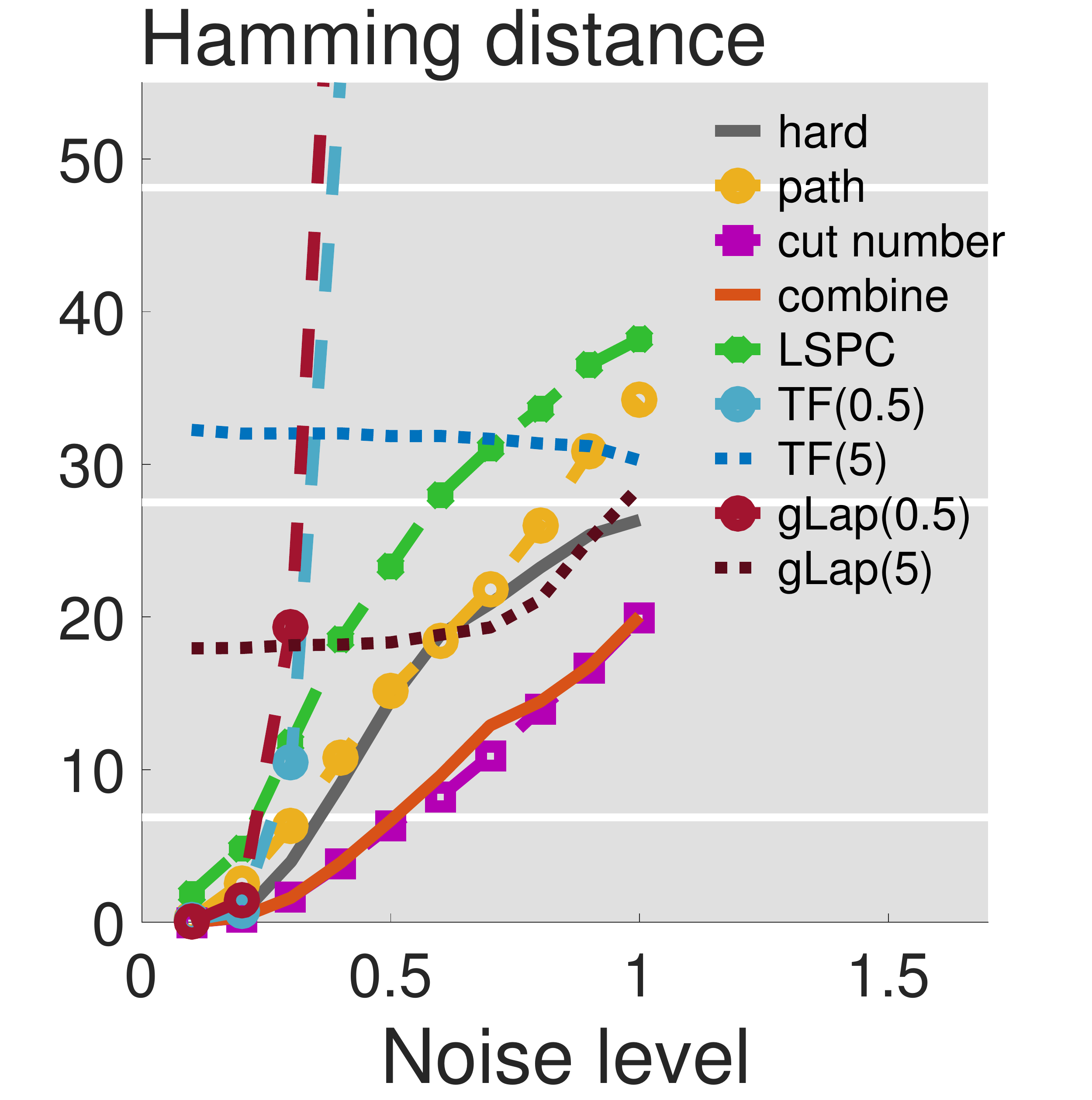}
      &
      \includegraphics[width=0.4\columnwidth]{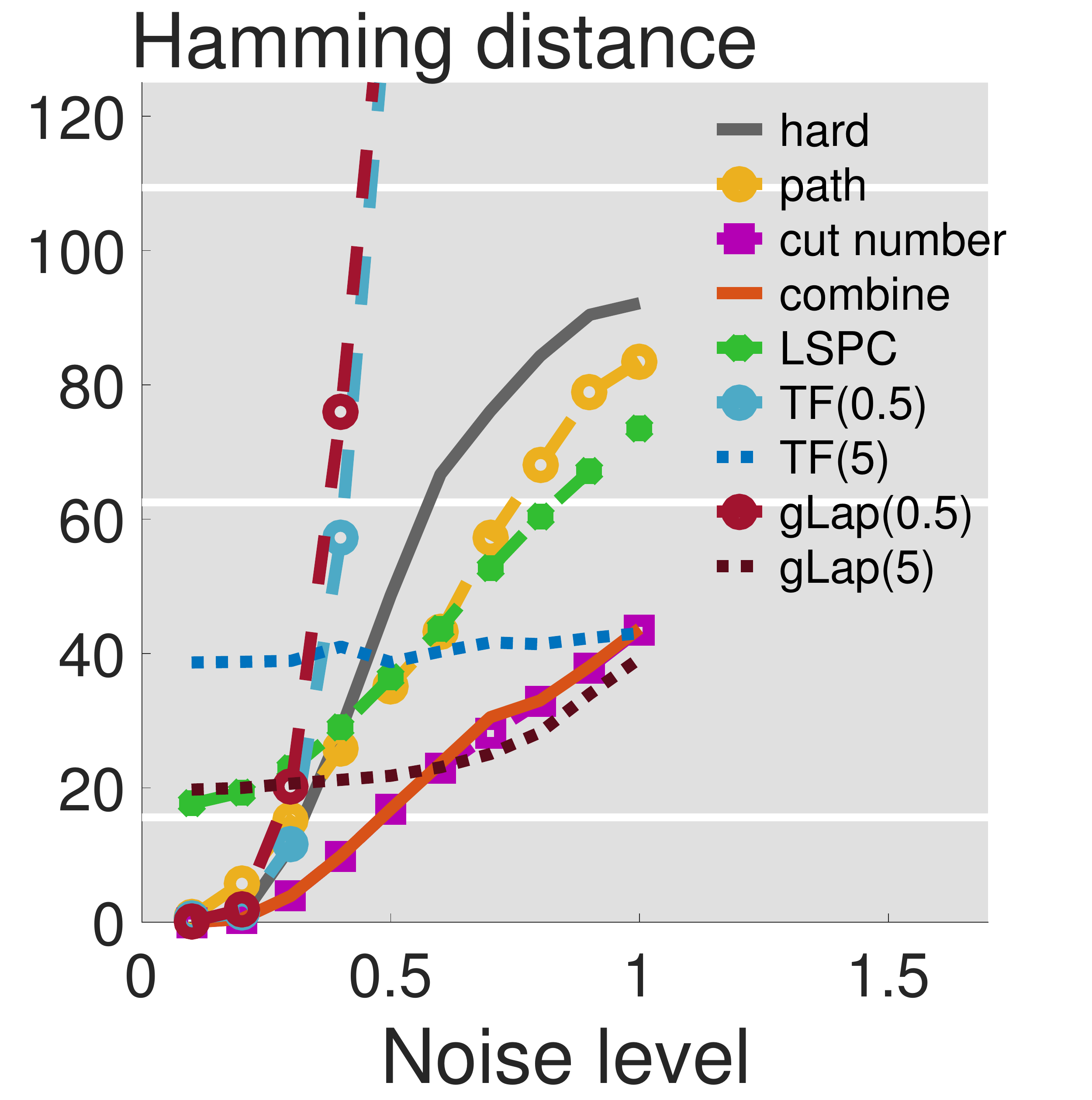}\\
      \multicolumn{2}{c}{\small (e)--(f) Hamming distance. } 
    \end{tabular}
  \end{center}
  \caption{\label{fig:localization_ball} Localizing ball-shaped pieces in the 
    Minnesota road network as a function of the noise level using hard
    thresholding (grey-solid line), path-based
    localization~\eqref{eq:path_opt} (yellow-circle line), cut-based
    localization~\eqref{eq:cut_2} (purple-square line), combined
    localization~\eqref{eq:localization_opt} (red-solid line),
    local-set-based piecewise-constant (green-square line), trending
    filtering on graphs ( $\lambda = 0.5$ light-blue-circle line,
    $\lambda = 5$ dark-blue-dashed line) and graph Laplacian denoising
    ( $\lambda = 0.5$ light-brown-circle line, $\lambda = 5$
    dark-brown-dashed line). Cut-based localization provides the most
    robust performance.  }
\end{figure}

\mypar{Ball-shaped class} We generate a ball-shaped piece by randomly
choosing one node as the center node and assigning all others within
$k=5$ or $k=10$ steps of it to an activated node set; see
Figures~\ref{fig:localization_ball}(a)--(b) for examples. Green
(lighter)/black (darker) nodes indicate activated/inactivated nodes,
respectively. We aim to localize activated pieces from noisy one-piece
graph signals with magnitude $\mu = 1$. We vary the noise variance
$\sigma^2$ from $0.1$ to $1$ with interval $0.1$ and for each randomly
generate $1,000$ noisy one-piece graph signals.

We measure the quality of the localization with two measures: the
Hamming distance and the $F_1$ score. The Hamming distance emphasizes
the difference between two sets while the $F_1$ score emphasizes their
intersection; the higher the $F_1$ score, or the lower the Hamming
distance, the better the localization quality.

The \emph{Hamming distance} is equivalent to the Manhattan distance
between two binary graph signals, that is, it counts the total number
of mismatches between the two,
\begin{eqnarray*}
  d_{\rm H}(\widehat{C}, C) \ = \ \left\| \one_{\widehat{C}} - \one_{C}  \right\|_1
  \ = \ |C \cup  \widehat{C}| - |C \cap \widehat{C}|,
\end{eqnarray*}
where $C$ is the ground truth for the activated piece and
$\widehat{C}$ is the localized piece. The lower the Hamming distance
the better the quality: it reaches its minimum value at $d_{\rm
  H}(\widehat{C}, C) = 0$.

The \emph{$F_1$ score} is the harmonic mean of the precision and
recall and measures the matching accuracy,
\begin{eqnarray*}
  F_1 (\widehat{C}, C) \ = \  
   2 \frac{   {\rm precision} (\widehat{C}, C) \cdot {\rm recall} (\widehat{C}, C) }
  {  {\rm precision} (\widehat{C}, C) +  {\rm recall} (\widehat{C}, C)},
\end{eqnarray*}
where precision is the fraction of retrieved instances that are
relevant (true positives over the sum of true positives and false
positives), while recall is the fraction of relevant instances that
are retrieved (true positive over the sum of true positives and false
negatives)~\cite{DudaHS:01}. Thus, with true positives~$= |C \cap
\widehat{C}|$, sum of true positives and false positives~$=
|\widehat{C}|$ and the sum of the true positives and false
negatives~$= |C|$, we have
\begin{eqnarray*}
  F_1 (\widehat{C}, C) 
  \ = \ 
  2 \frac{ |C \cap \widehat{C}| }{ |C| + |\widehat{C}|}.
\end{eqnarray*}
The higher the $F_1$ score the better the quality: it reaches its
maximum value at $F_1=1$ and minimum value at $F_1=0$.

\begin{figure}[t]
  \begin{center}
    \begin{tabular}{cc}
      {\small Length $15$. }  &  {\small Length $91$. }\\
      \includegraphics[width=0.45\columnwidth]{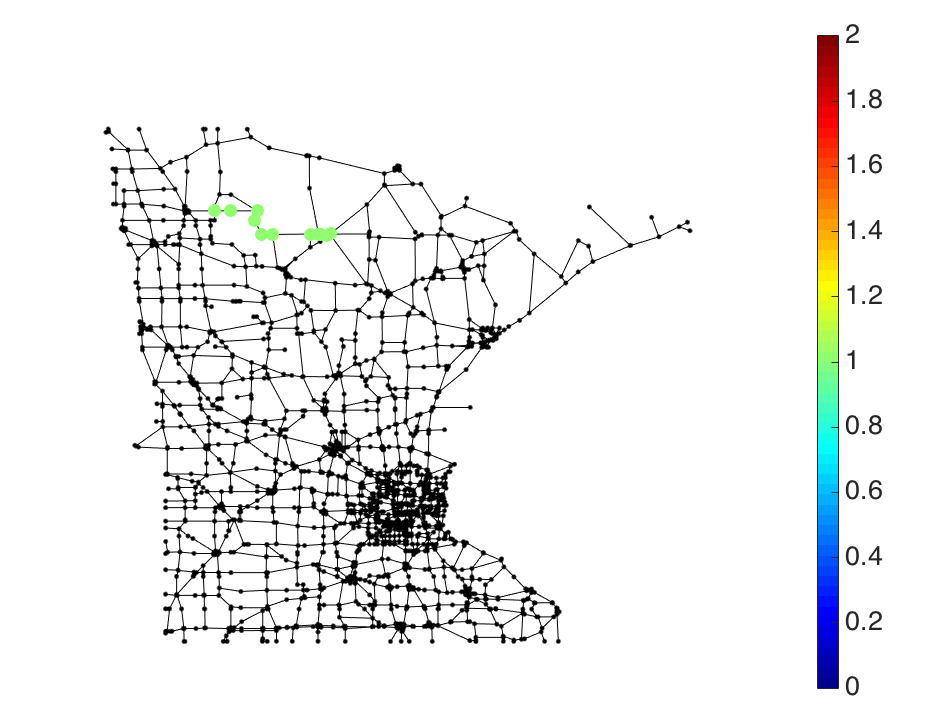} &
      \includegraphics[width=0.45\columnwidth]{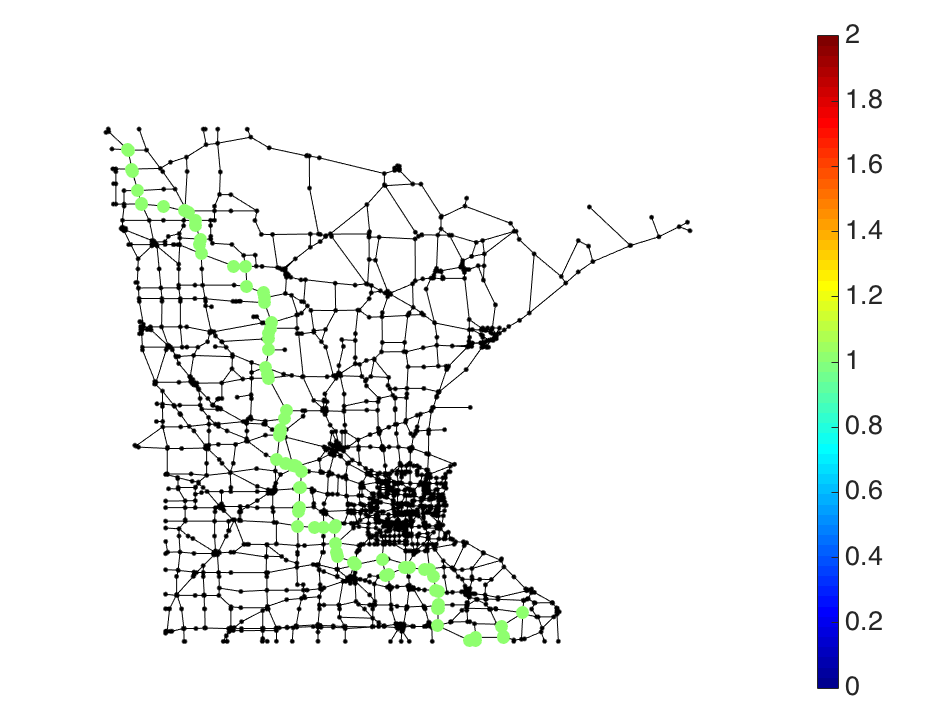}\\
      \multicolumn{2}{c}{\small (a)--(b) Path.}\\
      \includegraphics[width=0.4\columnwidth]{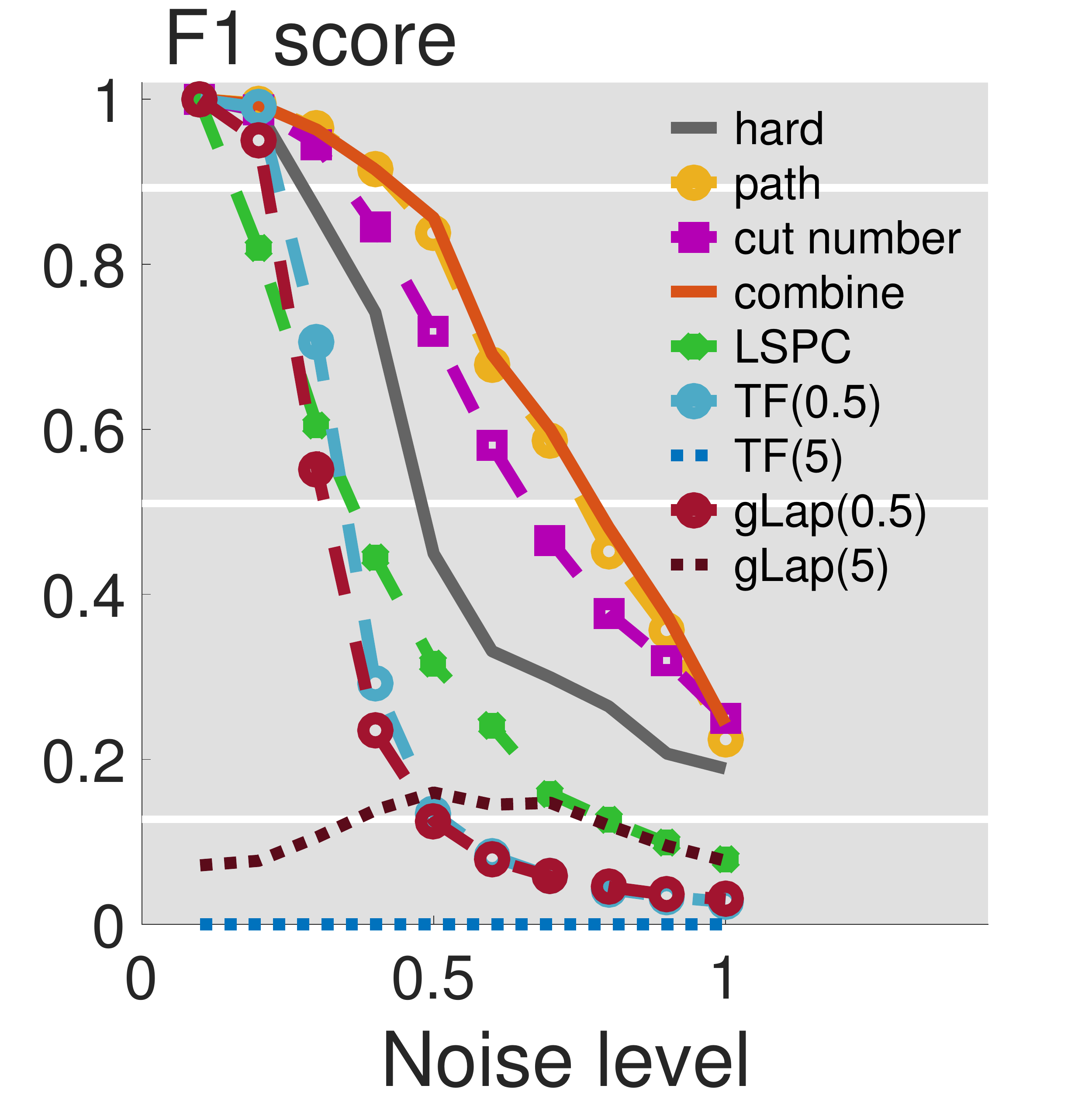} &
      \includegraphics[width=0.4\columnwidth]{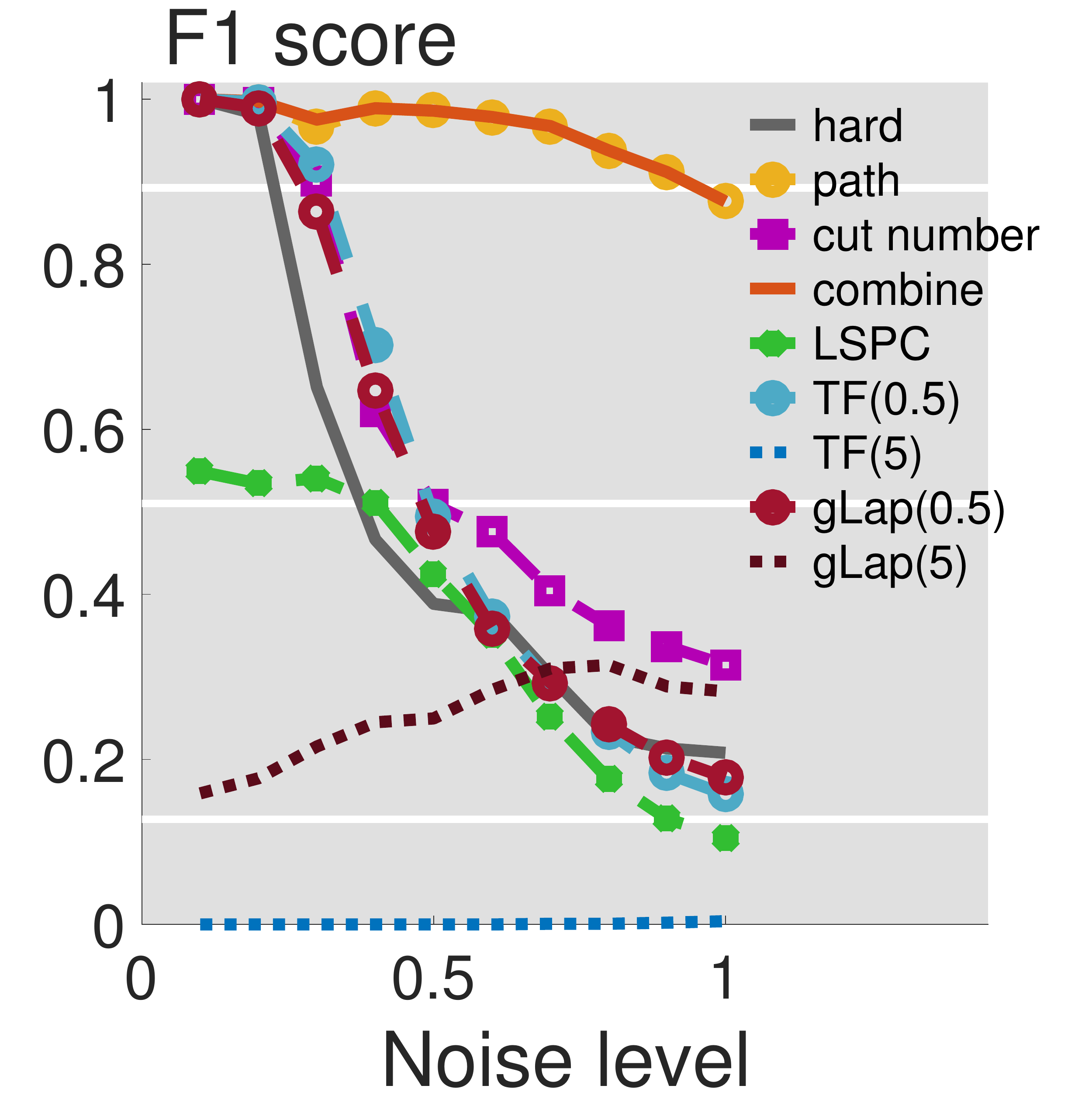} \\
      \multicolumn{2}{c}{\small (c)--(d) $F_1$ score.}\\
      \includegraphics[width=0.4\columnwidth]{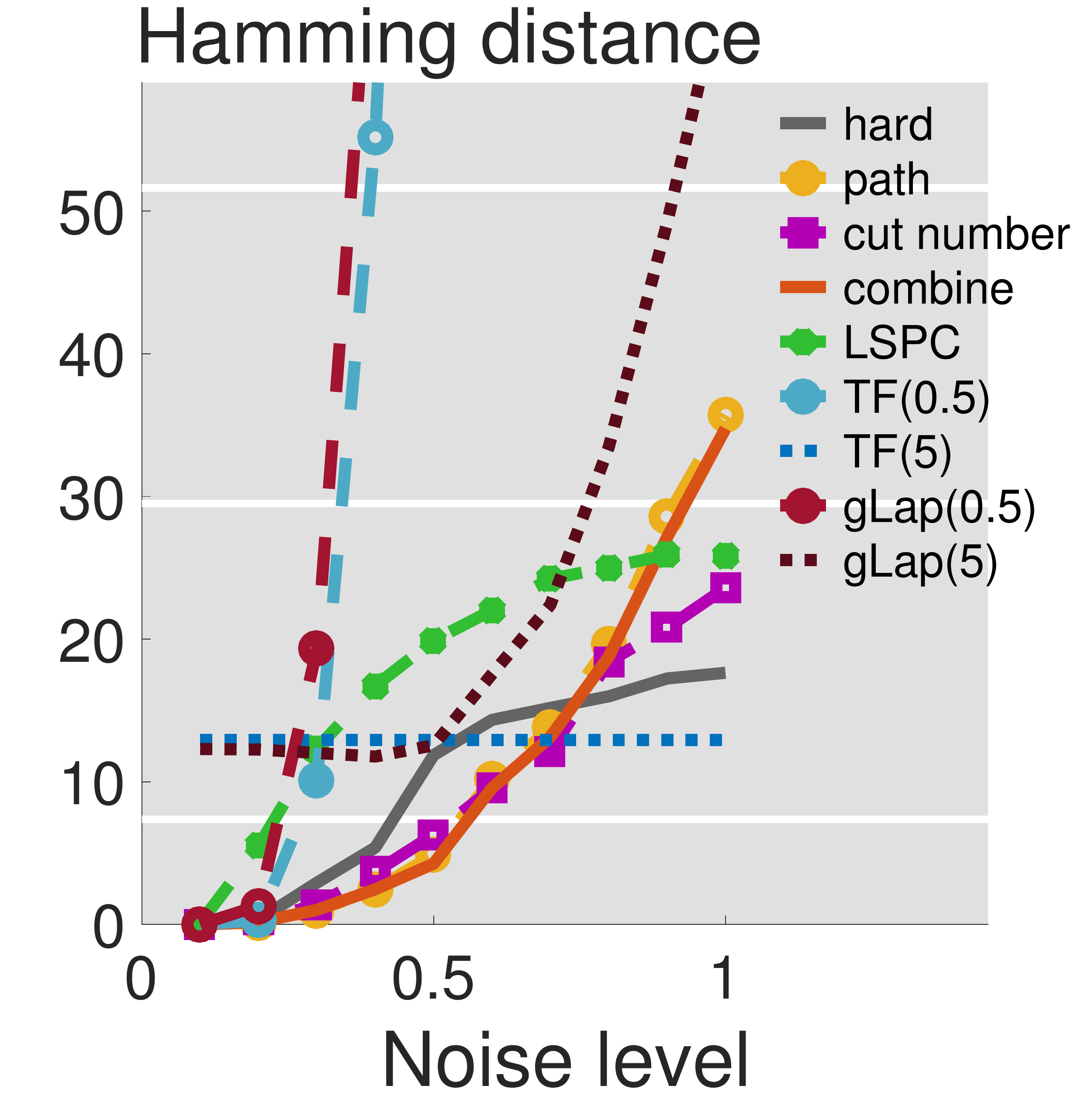} &
      \includegraphics[width=0.4\columnwidth]{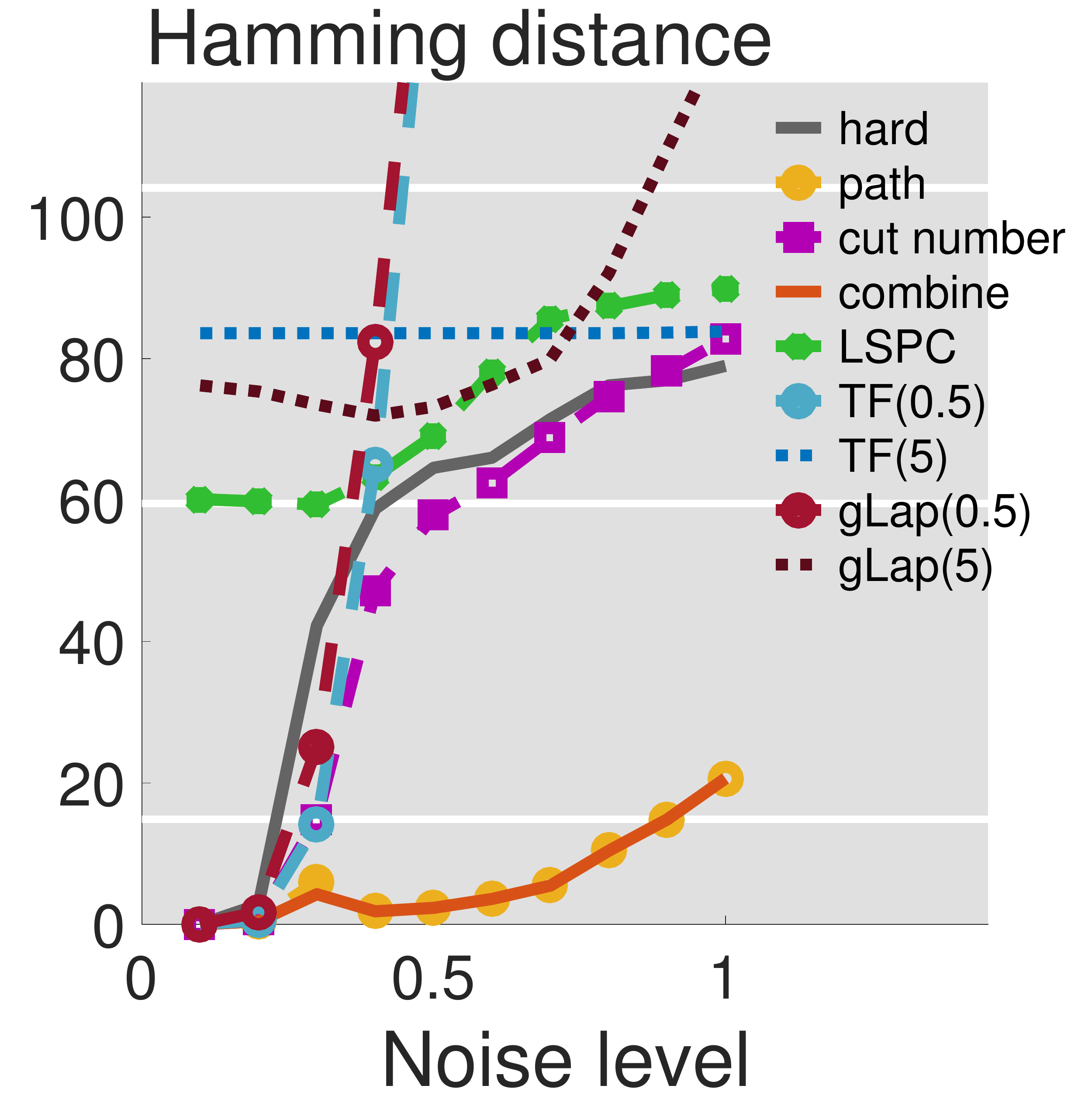} \\
      \multicolumn{2}{c}{\small (e)--(f) Hamming distance. } 
    \end{tabular}
  \end{center}
  \caption{\label{fig:localization_path} Localizing paths in the
    Minnesota road network as a function of noise level using hard
    thresholding (grey-solid line), path-based
    localization~\eqref{eq:path_opt} (yellow-circle line), cut-based
    localization~\eqref{eq:cut_2} (purple-square line), combined
    localization~\eqref{eq:localization_opt} (red-solid line),
    local-set-based piecewise-constant (green-square line), trending
    filtering on graphs ( $\lambda = 0.5$ light-blue-circle line,
    $\lambda = 5$ dark-blue-dashed line) and graph Laplacian denoising
    ( $\lambda = 0.5$ light-brown-circle line, $\lambda = 5$
    dark-brown-dashed line). Path-based localization provides the most
    robust performance.}
\end{figure}

Figures~\ref{fig:localization_ball}(c)--(d) show the $F_1$ scores and
Figures~\ref{fig:localization_ball}(e)--(f) the Hamming distances of
localizing ball-shaped pieces with radii $k=5$ and $10$ (left and
right columns, respectively) as functions of the noise variance. Each
figure compares hard thresholding~\eqref{eq:hard_thr} (hard,
grey-solid line), path-based localization~\eqref{eq:path_opt} (path,
yellow-circle line), cut-based localization~\eqref{eq:cut_2} (cut
number, purple-square line), combined
localization~\eqref{eq:localization_opt} (combine, red-solid line),
local-set-based piecewise-constant dictionary (LSPC, green-square
line)~\cite{ChenJVSK:16},  trend filtering on graphs
(TF)~\cite{WangSST:15} and graph Laplacian denoising
(gLap)~\cite{ShumanNFOV:13}.  LSPC is a predesigned graph dictionary
and we use the matching pursuit algorithm to choose one atom to
localize an activated piece. Trend filtering on graphs and graph
Laplacian denoising are denoising algorithms, which solve the
following two optimization problems, respectively,
\begin{eqnarray*} {\rm TF}(\lambda): && \min_{\t} \left\| \x - \t
  \right\|_2^2 + \lambda \left\| \Delta \t \right\|_1,
  \\
  {\rm gLap}(\lambda): && \min_{\t} \left\| \x - \t \right\|_2^2 +
  \lambda \t^T \LL \t,
\end{eqnarray*}
where $\lambda$ is a tuning parameter, $\Delta$ is the graph incidence
matrix~\eqref{eq:Delta}, and $\LL$ is the graph Laplacian matrix.
$\left\| \Delta \t \right\|_1$ promotes localized adaptivity and $\t^T
\LL \t$ promotes smoothness. When $\lambda$ is small, both solutions
are close to the noisy graph signal; when $\lambda$ is large, both
solutions are regularized to be either localized or smooth. We expect
that a small $\lambda$ works better in a noiseless case and a large
$\lambda$ works better in a noisy case. After obtaining the denoised
solution $\t^*$, we implement hard thresholding with threshold $\mu/2
= 0.5$ to localize the activated piece. To test the sensitivity to the
tuning parameter, we also vary $\lambda$ as either $0.5$ or $5$ in our
experiments.  Trend filtering on graphs and graph Laplacian denoising
involve two tuning parameters: regularization parameter $\lambda$ and
threshold; however, neither method provides a guideline by which to
choose those parameters in practice. Our proposed localization
method~\eqref{eq:localization_opt}, on the other hand, is
parameter-free.

We observe that: (1) Cut-based localization provides the most robust
performances; hard thresholding is a special case of cut-based
localization, path-based localization is designed for capturing
elongated pieces, LSPC is a data-independent dictionary and cannot
adapt the shape of its atom to the given piece, graph Laplacian
denoising with careful parameter selection works well for large
activated pieces but still fails to localize small activated pieces as
the smooth assumption cannot capture localized variation. Trending
filtering on graphs works well in denoising of a piecewise-constant
graph signal; however, it may not work well in localization as the
output is real and the threshold is hard to choose. Here the threshold
is $\mu/2 = 0.5$, which sometimes works well and sometimes rules out
the entire output. We also see that both graph Laplacian denoising and
trending filtering on graphs are sensitive to the tuning parameter
$\lambda$. The advantages of cut-based localization compared to other
methods are: it is parameter-free, scalable and robust to noise. (2)
Noise level influences localization performance; as it grows, the
$F_1$ score decreases and the Hamming distance increases.  (3) The
size of the activated piece influences localization; localizing a
piece with radius $10$ is easier than localizing a piece with radius
$5$; this was also observed in~\cite{ChenYZSK:16}.

\mypar{Elongated class} We generate an elongated path by randomly
choosing two nodes as starting and ending nodes and computing the
shortest path between the two. We look at two classes of path lengths:
between $10$ to $15$ and longer than $80$~\footnote{The path length is the geodesic distance between the two end nodes.}; see
Figures~\ref{fig:localization_path}(a)--(b) for examples.  We aim to
localize activated paths with magnitude $\mu = 1$ from noisy path
graph signals. We vary the noise variance $\sigma^2$ from $0.1$ to $1$
with interval $0.1$ and for each randomly generate $1,000$ noisy path
graph signals. We measure the quality of the localization by the
Hamming distance and $F_1$ score.

Figures~\ref{fig:localization_path}(c)--(d) and (e)--(f) show the
$F_1$ scores and the Hamming distances, respectively, of localizing
elongated pieces with lengths $15$ and $91$ as functions of the noise
variance. We compare the same methods as in
Figure~\ref{fig:localization_ball}.
% : hard
% thresholding (grey-solid line), path-based
% localization~\eqref{eq:path_opt} (yellow-circle line), cut-based
% localization~\eqref{eq:cut_2} (purple-square line), combined
% localization~\eqref{eq:localization_opt} (red-solid line),
% local-set-based piecewise-constant ( LSPC, green-square line),
% trending filtering on graphs ( TF($\lambda = 0.5$) light-blue-circle
% line, TF($\lambda = 5$) dark-blue-dashed line) and graph Laplacian
% denoising ( gLap($\lambda = 0.5$) light-brown-circle line,
% gLap($\lambda = 5$) dark-brown-dashed line).

\begin{figure}[t]
  \begin{center}
    \begin{tabular}{ccccc}
      \includegraphics[width=0.35\columnwidth]{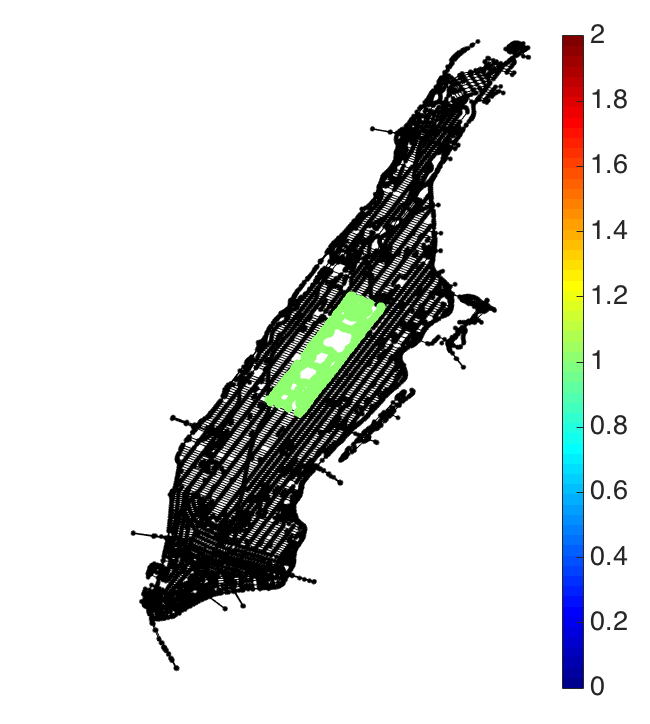}
      &
      \includegraphics[width=0.35\columnwidth]{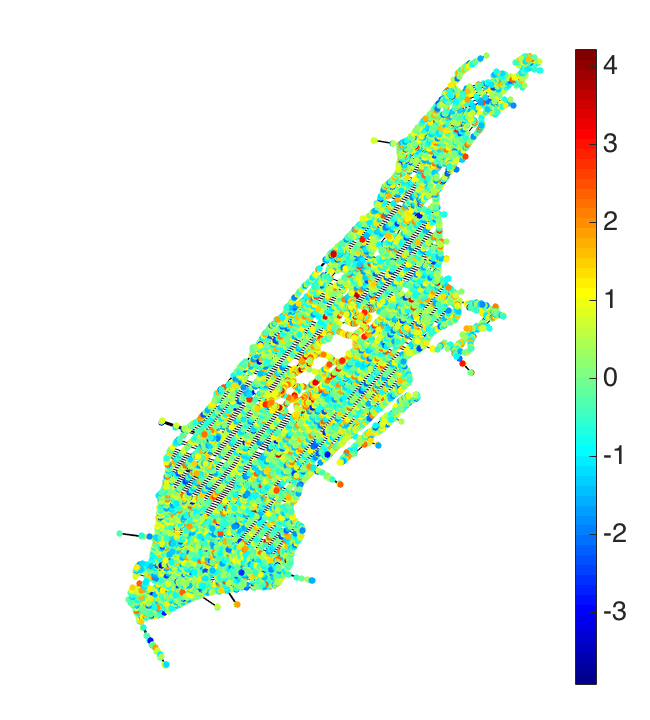}
      \\
      {\small (a) Signal.}  &  {\small (b) Noisy signal.}
      \\
      \includegraphics[width=0.35\columnwidth]{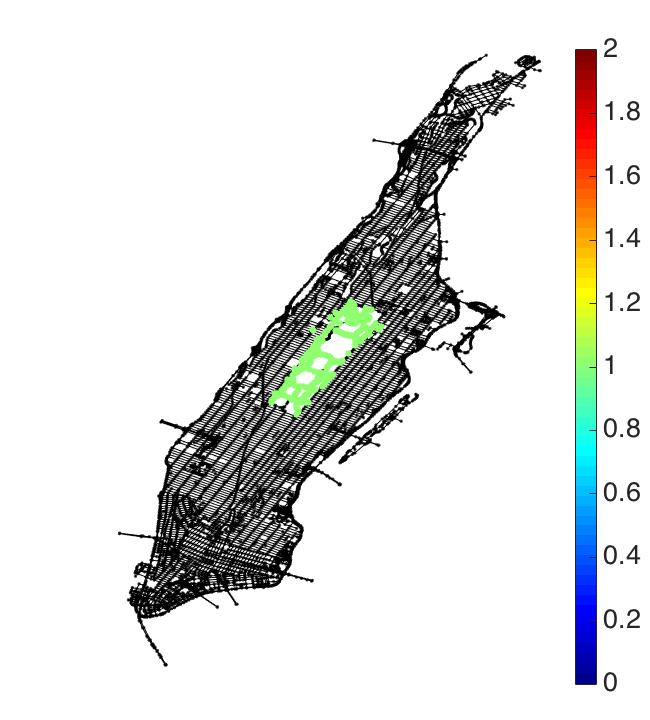}
      &
      \includegraphics[width=0.35\columnwidth]{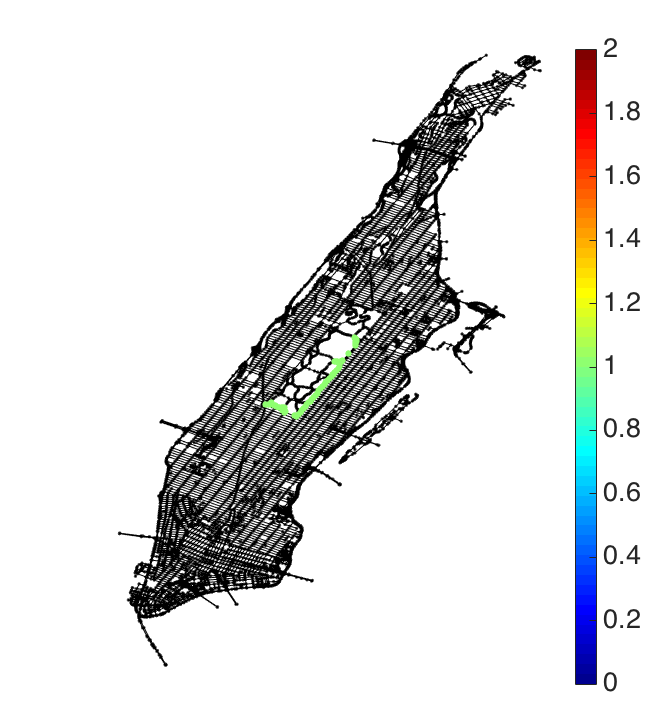}
      \\
      {\small (c) Activated piece (cut).}  &  {\small (d) Activated piece (path).}
      \\
      \includegraphics[width=0.4\columnwidth]{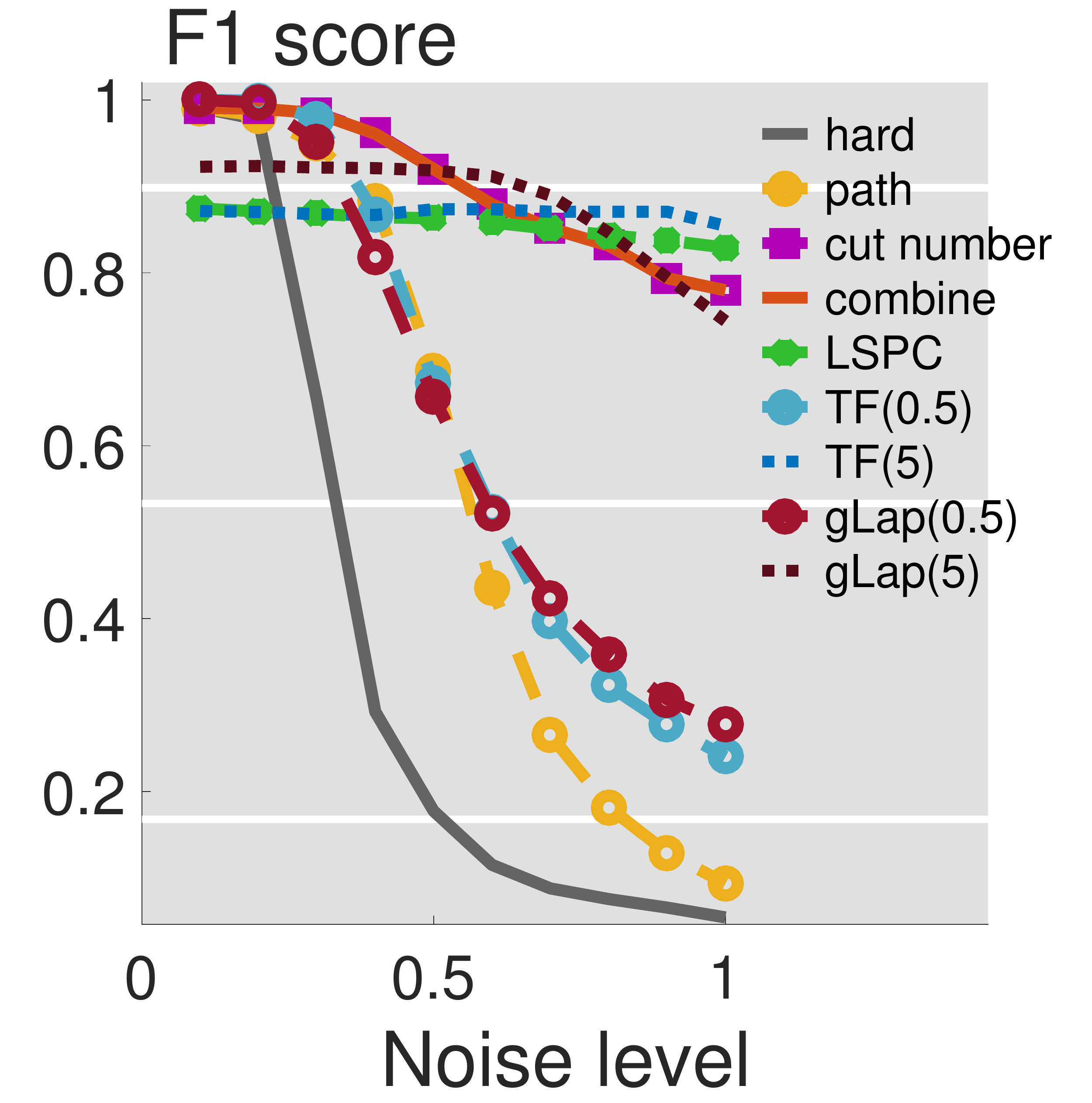}
      &
      \includegraphics[width=0.4\columnwidth]{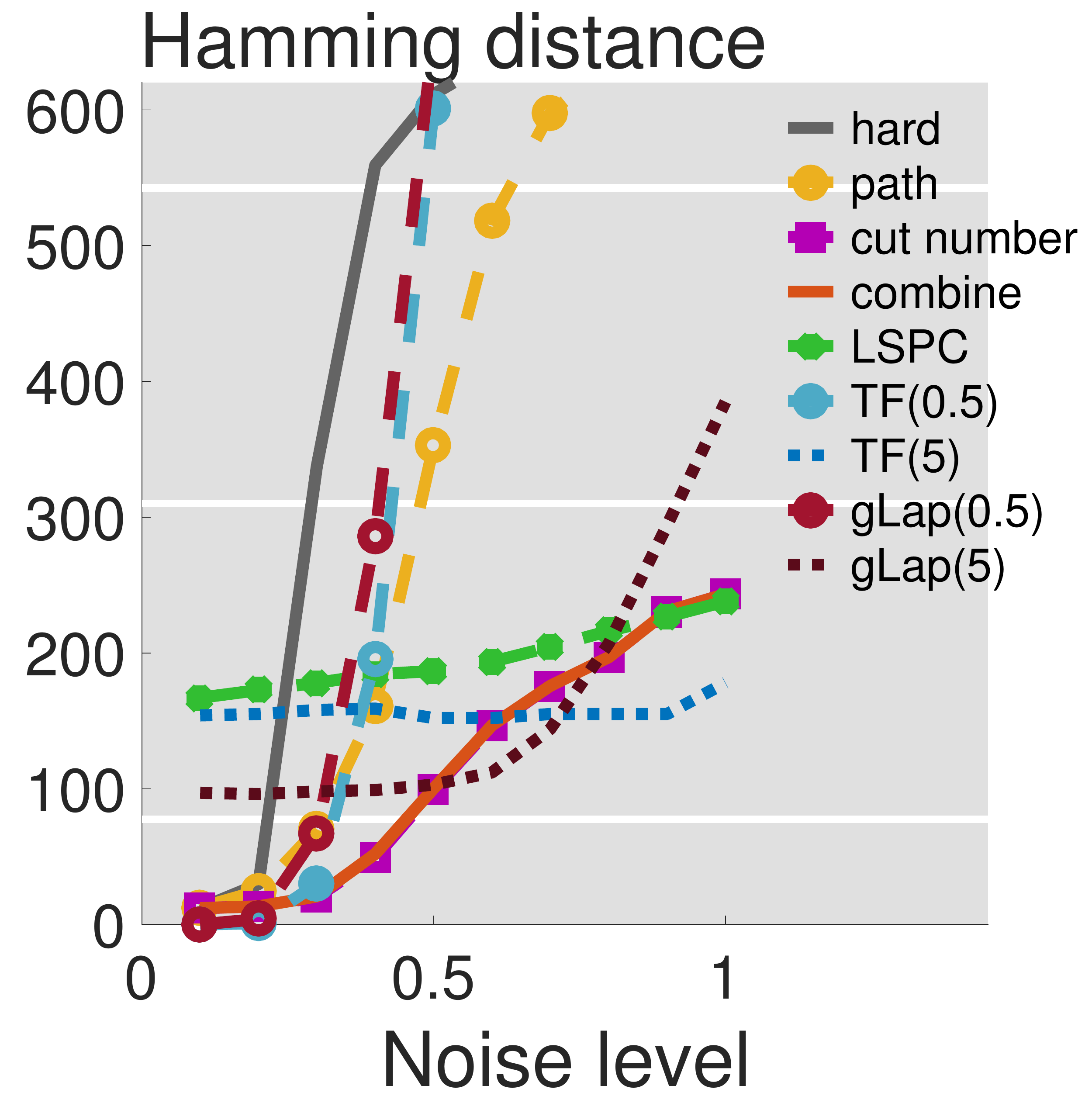}
      \\
      {\small (e) F1 score.}
      &  {\small (f) Hamming distance.}
    \end{tabular}
  \end{center}
  \caption{\label{fig:localization_manhattan_cpark} Localizing (a)
    Central Park in the Manhattan street network as a function of the
    noise level from (b) its noisy version. (c) Activated piece
    obtained by cut-based localization with $F_1=0.81$ and $d_{\rm H}
    = 214$. (d) Activated piece obtained by path-based localization
    with $F_1=0.24$ and $d_{\rm H} = 586$. Cut-based localization
    outperforms path-based localization. }
\end{figure}

\begin{figure}[t]
  \begin{center}
    \begin{tabular}{cc}
      \includegraphics[width=0.35\columnwidth]{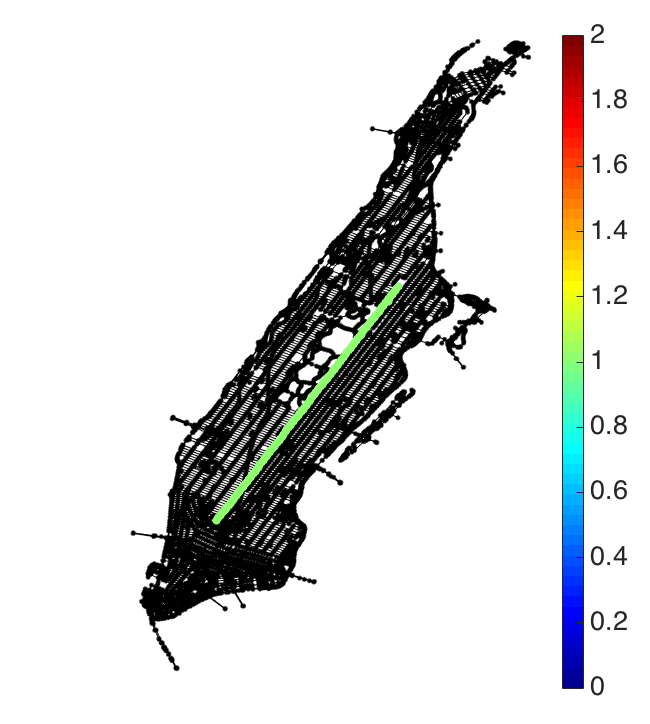}
      &
      \includegraphics[width=0.35\columnwidth]{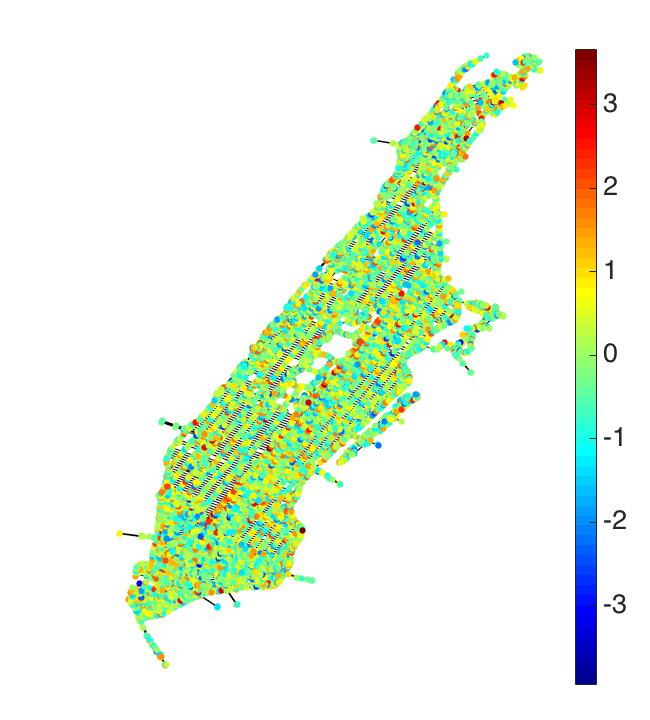}
      \\
      {\small (a) Signal.}  &  {\small (b) Noisy signal.}
      \\
      \includegraphics[width=0.35\columnwidth]{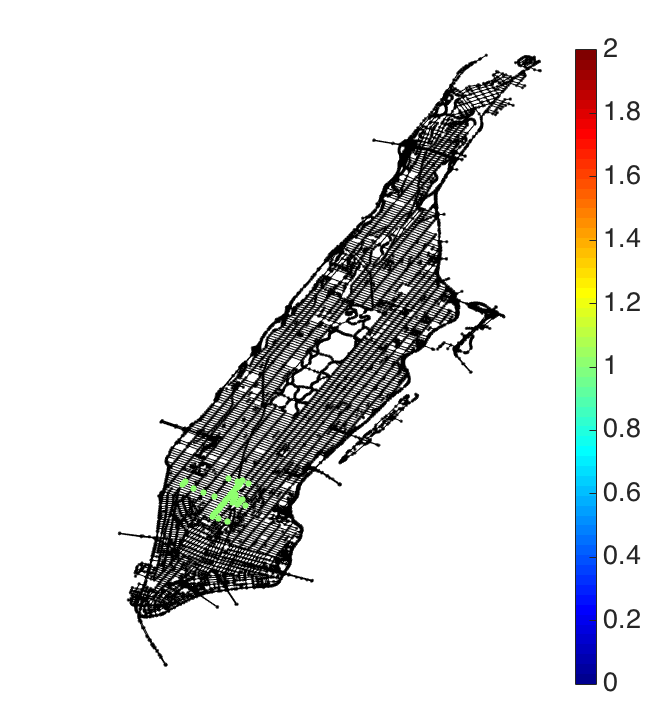}
      &
      \includegraphics[width=0.35\columnwidth]{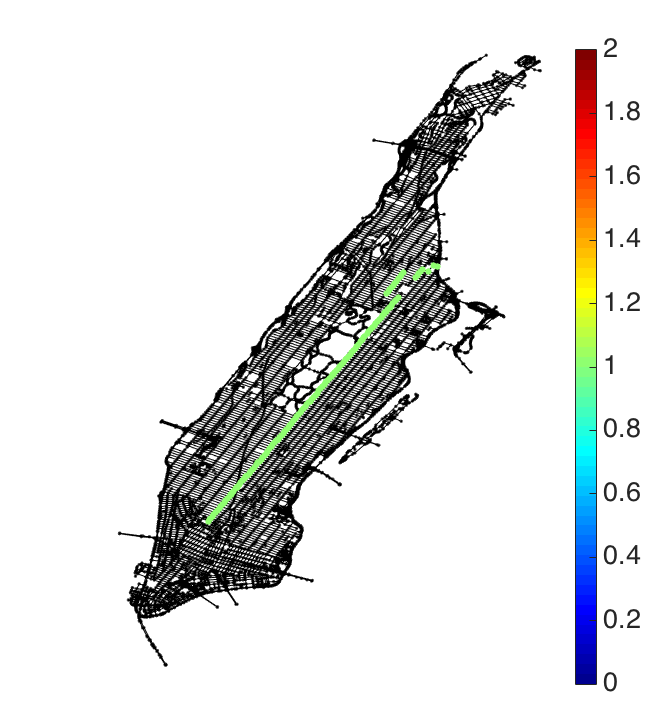}
      \\
      {\small (c) Activated piece (cut).}  &    {\small (d) Activated piece (path).}
      \\
      \includegraphics[width=0.4\columnwidth]{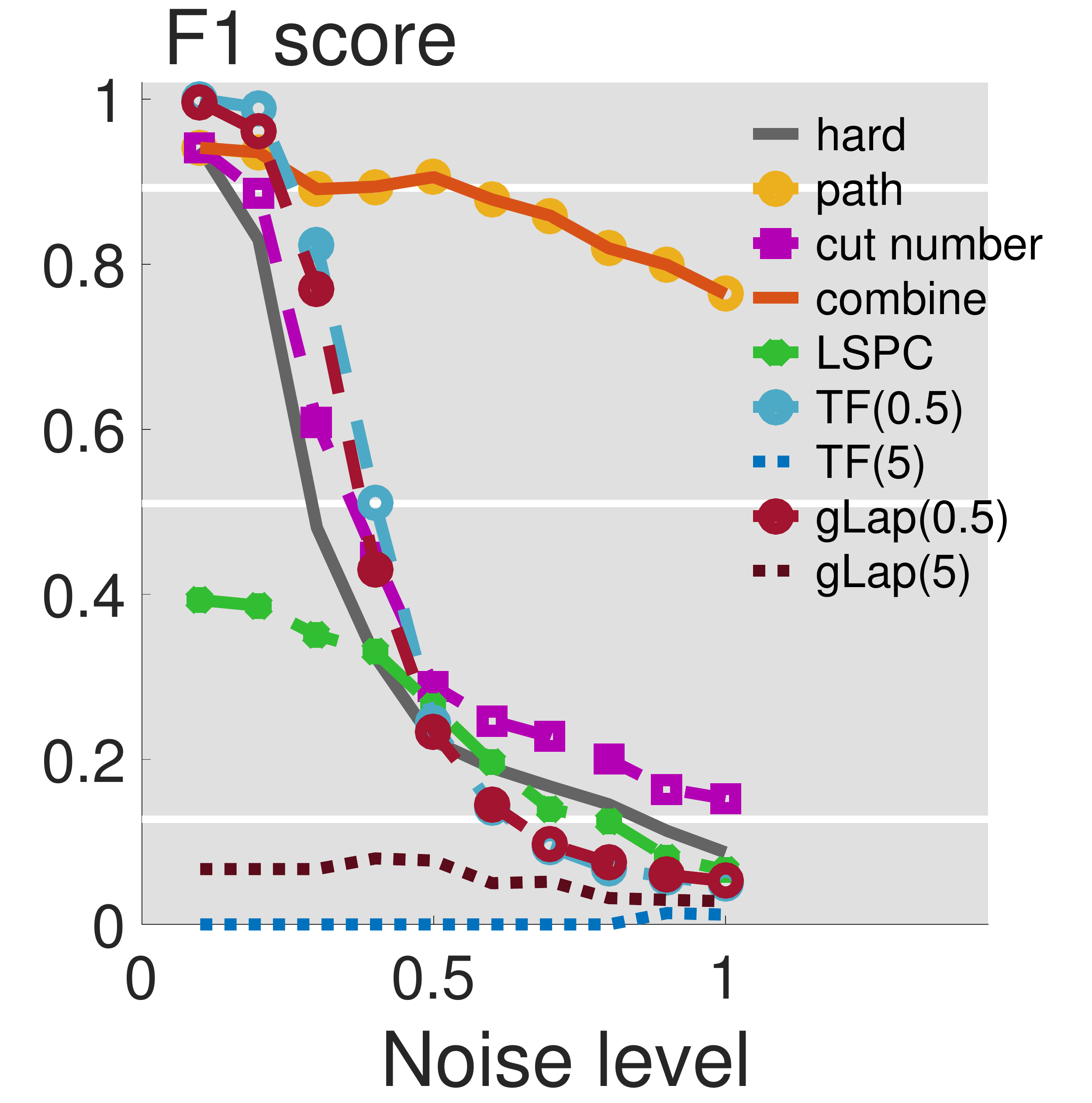}
      &
      \includegraphics[width=0.4\columnwidth]{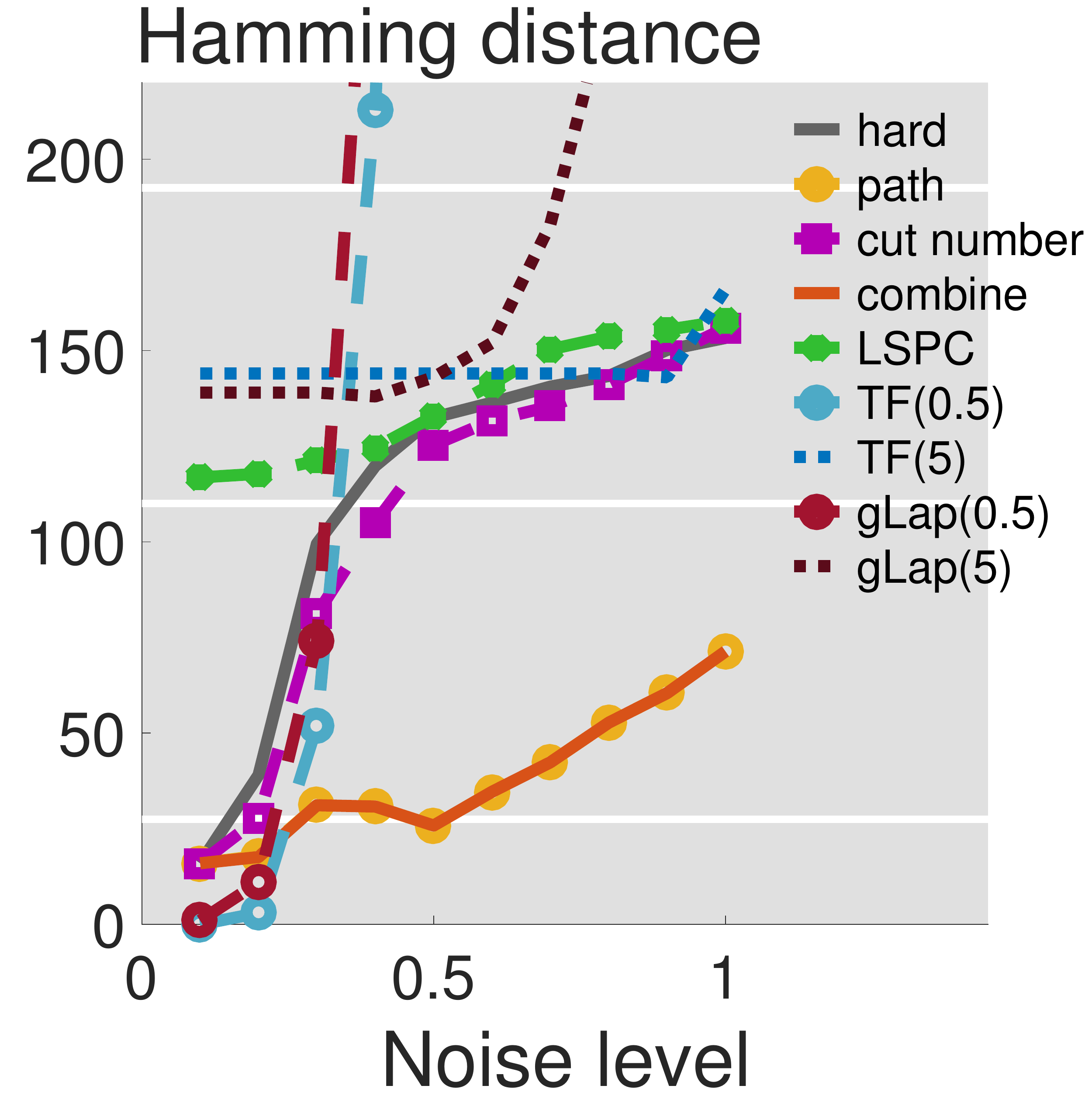}
      \\
      {\small (e) F1 score.}   &  {\small (f) Hamming distance.}
    \end{tabular}
  \end{center}
  \caption{\label{fig:localization_manhattan_5avenue} Localizing (a)
    Fifth Avenue in the Manhattan street network as a function of the
    noise level from (b) its noisy version. (c) Activated piece
    obtained by cut-based localization with $F_1=0.24$ and $d_{\rm H}
    = 144$. (d) Activated piece obtained by path-based localization
    with $F_1=0.85$ and $d_{\rm H} = 43$. Path-based localization
    outperforms cut-based localization. }
\end{figure}
We observe that: (1) Path-based localization performs the best; it is
also parameter-free, scalable and robust to noise. (2) When the path
is short, cut-based localization and path-based localization perform
similarly; when the path is long, path-based localization
significantly outperforms cut-based localization. (3) Noise level
influences localization performance; as it grows, the $F_1$ score
decreases and the Hamming distance increases. (4) The path length
influences localization performance.  Similarly to what we observed in
the ball-shaped case, it is much harder to localize a short path
because a small activation provides less information.

\subsubsection{Manhattan street network}
We model this network as a graph with its 13,679 intersections as
nodes and 17,163 city streets as undirected edges.  We generate two
classes of one-piece graph signals: a ball-shaped Central Park piece
and a Fifth Avenue path.  We compare results with those from the
Minnesota road network to validate our conclusions.

\mypar{Central Park}
Figures~\ref{fig:localization_manhattan_cpark}(a)--(d) show a piece
activating the nodes in Central Park, a noisy version with noise
variance $\sigma^2 = 1$ and the activated piece provided by cut-based
and path-based localization, respectively. Even when the graph signal
is corrupted by a high level of noise, cut-based localization still
provides accurate localization, while path-based localization fails to
localize Central
Park. Figures~\ref{fig:localization_manhattan_cpark}(e)--(f) show the
$F_1$ score and Hamming distance when localizing Central Park as a
function of the noise level, respectively. The results are averaged
over 1,000 runs.  Cut-based localization provides the most robust
performance, which is consistent with what we observed with the
Minnesota road network.  Both trend filtering and graph Laplacian
denoising work well at times, but are sensitive to the tuning
parameter $\lambda$. Since neither method provides a guideline by
which to choose the parameter, their performances may be
unstable. LSPC is less sensitive to noise and slightly outperforms the
other methods when the noise level is high. Since LSPC is
data-independent, it chooses the most relevant predesigned atom to fit
a noisy signal; on the other hand, as a data-adaptive method,
cut-based localization designs an atom from the noisy signal and
easily fits noisy data.  Overall, the localization performance of LSPC
depends highly on the shape of its atoms: when LSPC has a predesigned
atom matching the ground truth, it is robust to noise and provides
effective localization performance; when this is not the case, it
fails to localize well.

\mypar{Fifth Avenue}
Figures~\ref{fig:localization_manhattan_5avenue}(a)--(d) show a piece
activating the nodes along Fifth Avenue, a noisy version with noise
variance $\sigma^2 = 1$ and the activated piece provided by cut-based
localization and the activated piece provided by path-based
localization, respectively. When the graph signal is corrupted by a
high level of noise, both cut-based localization and path-based
localization fail to localize Fifth
Avenue. Figures~\ref{fig:localization_manhattan_5avenue}(e)--(f) show
the $F_1$ score and Hamming distance when localizing Fifth Avenue as a
function of the noise level, respectively. The results are averaged
over 1,000 runs. We see that the path-based localization significantly
outperforms all the other localization methods under various noise
levels. This is similar to what we observed with the Minnesota road
network.

%We also observe that localizing Fifth Avenue is harder than localizing Central Park because the results in especially when the noise level is high, which is again consistent with what we saw in the Minnesota road network.

\section{Signal Decomposition on Graphs}
\label{sec:decomposition}
We now extend localization discussion by considering multiple
activated pieces making it a decomposition problem.  We extend the the
localization problem solver in Section~\ref{sec:localization} to the
decomposition problem and validate it through simulations.

Consider localizing $K$ activated pieces $C_i$, $i = 1, \, 2, \,
\ldots, \, K$, in a noisy, piecewise-constant graph signal
\begin{eqnarray*}
	\x & = &   \sum_{i=1}^K \mu_i \one_{C_i} + \epsilon, % \ \in \ \R^N,
%	\\	
%	& = &
%	\begin{bmatrix}
%	\one_{C_1} &   \one_{C_2} & \cdots & \one_{C_K}
%	\end{bmatrix}  \mu + \epsilon,
\end{eqnarray*}
where $\one_C$ is the indicator function~\eqref{eq:indicator}, $K \ll
N$, $C_i$ are connected and $\epsilon \sim \N(0, \sigma^2 \Id_N)$ is
Gaussian noise.

Our goal is to develop an algorithm to efficiently decompose such a
graph signal into several pieces. The corresponding optimization
problem is
\begin{eqnarray}
  \label{eq:PCD}
  % && 
  \min_{\mu_i, C_i}  \left\| \x - \sum_{i=1}^K \mu_i \one_{C_i} \right\|_2^2,
  % \\
  % \nonumber
  \quad \text{subject to } C_i \in \Cc.
\end{eqnarray}
When the number of activated pieces $K = 1$, the decomposition
problem~\eqref{eq:PCD} becomes the localization
problem~\eqref{eq:opfp}. This problem is similar to independent
component analysis in classical signal processing, whose goal is to
decompose a signal into several independent components. Here we
consider a piece in the graph vertex domain as a component and we
allow the  pieces to overlap for flexibility.

\vspace{-4mm}
\subsection{Methodology}
% Consider a graph signal,
% \begin{eqnarray*}
% 	\x & = &   \sum_{i=1}^K \mu_i \one_{C_i} + \epsilon \ \in \ \R^N
% % 	\\	
% % 	& = &
% 	\begin{bmatrix}
% 	\one_{C_1} &   \one_{C_2} & \cdots & \one_{C_K}
% 	\end{bmatrix}  \mu + \epsilon,
% \end{eqnarray*}
% where $K \ll N$, $C$ is connected and $\epsilon \sim \N(0, \sigma^2
% \Id_N)$. In the localization problem, we are concerned with one
% piece that has anomaly behavior in a graph signal. When we extend
% this idea to multiple pieces, we are actually approximating a graph
% signal by a linear combination of several pieces.

The goal now is not only to denoise or approximate a graph signal,
which would only give $\sum_{i=1}^K \mu_i \one_{C_i}$ but to analyze
it through decomposition as well; that is, we localize each piece and
estimate its corresponding magnitude, yielding $\mu_i$ and $C_i$.
% The pieces $C_i$ are basic constituents to construct a graph signal
% and do not necessarily have anomalies.
Since a graph signal is decomposed into several one-piece components,
we call this~\emph{piecewise-constant decomposition}.

% To solve the optimization problem
% \begin{eqnarray}
% \label{eq:PCD}
% 	&& \min_{\mu, C_i}  \left\| \x - \sum_{i=1}^K \mu_i \one_{C_i} \right\|_2^2,
% 	\\
% 	\nonumber
% 	&& \text{subject to }  C_i \in \Cc.
% \end{eqnarray}
To solve the optimization problem~\eqref{eq:PCD}, we update one piece
at a time by coordinate descent~\cite{BoydV:04}. When we freeze other
variables, optimizing over $\mu_i$ and $C_i$ is equivalent to
localization with unknown magnitude in
Section~\ref{sec:localization_unknown}, because
\begin{displaymath}
 \x - \sum_{j \neq i} \mu_j \one_{C_j} \ = \ \mu_i \one_{C_i} + \epsilon
\end{displaymath}
is a one-piece graph signal. Thus, for each piece, we solve the
localization problem~\eqref{eq:opfu} by using the localization solver~\eqref{eq:loc_unknown},
\begin{eqnarray*}
  \mu_i^*, C_i^* \ = \ 	 {\rm Loc}_{\rm unknown} ( \x - \sum_{j \neq i} \mu_j \one_{C_j} ).
\end{eqnarray*}
We iteratively update each piece until convergence to obtain a
series of solutions providing a local optimum~\cite{BoydV:04}.

\vspace{-4mm}
\subsection{Experimental Validation}
\label{sec:decomposition_exp}
We validate the piecewise-constant decomposition solver on the
Manhattan street network through two tasks: (1) We simulate a series
of piecewise-constant graph signals by a linear combination of
one-piece signals, each with a particular ZIP code; our goal is to
recover ZIP codes from signal observation; (2) We decompose
and analyze two real graph signals in Manhattan, restaurant
distribution and taxi-pickup activity.  For restaurant distribution,  we extract data from the restaurant inspection results of New York City provided by Department of Health and Mental Hygiene\footnote{Data from~\url{https://data.cityofnewyork.us/Health/DOHMH-New-York-City-Restaurant-Inspection-Results/xx67-kt59}};
for taxi-pickup activity,  we use a 2015 public dataset of
taxi pickups \footnote{Data
  from~\url{http://www.nyc.gov/html/tlc/html/about/trip_record_data.shtml}.}.

\subsubsection{Signal  Decomposition on Graphs: Localizing ZIP codes}
\label{sec:loc_zipcode}
\begin{figure}[t]
  \begin{center}
    \begin{tabular}{ccc}
     \includegraphics[width=0.35\columnwidth]{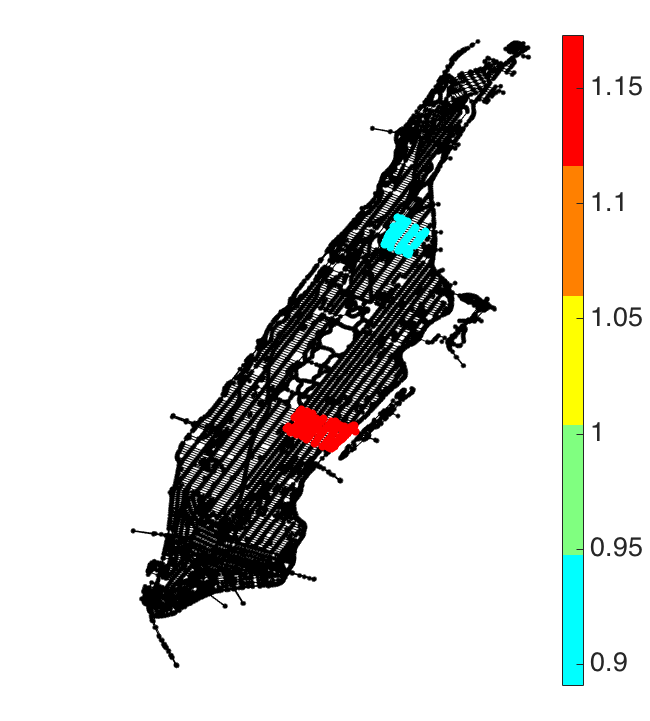}
 &
    \includegraphics[width=0.35\columnwidth]{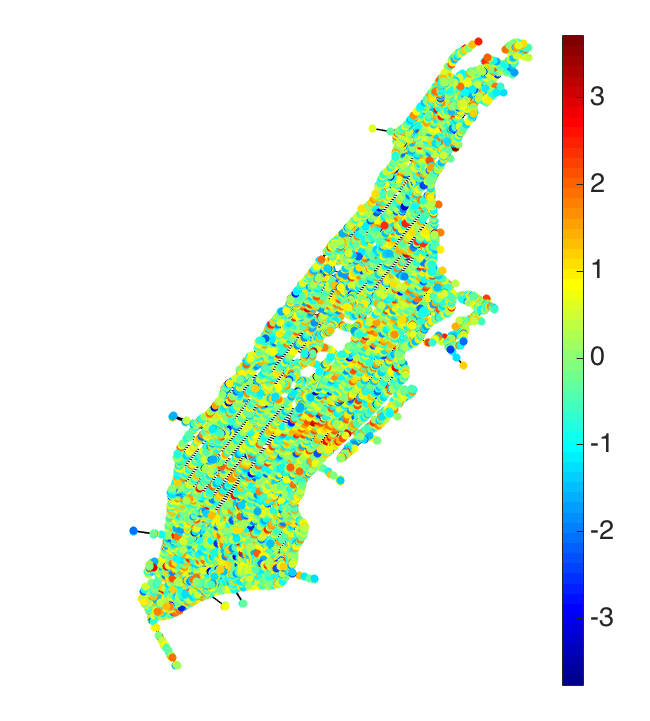}
      \\
    {\small (a) Signal. }  &  {\small (b) Noisy signal. }
    \\
         \includegraphics[width=0.35\columnwidth]{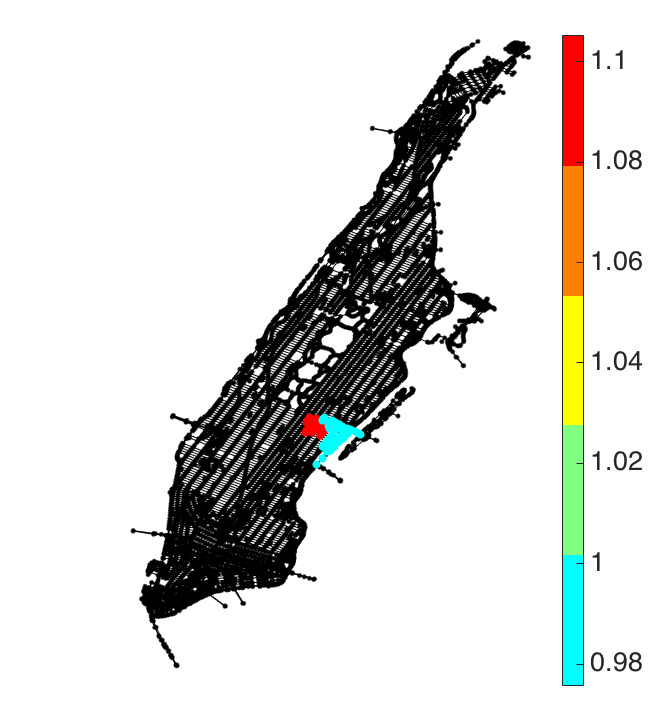}
 &
    \includegraphics[width=0.35\columnwidth]{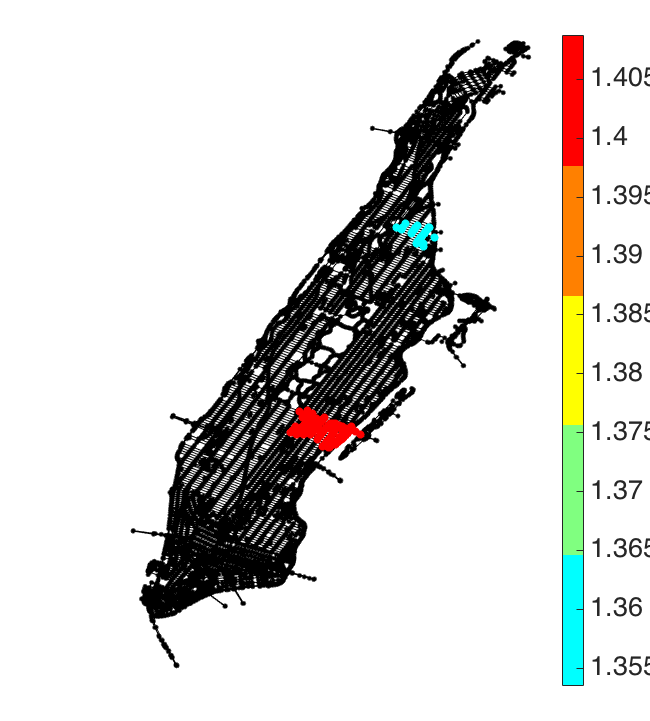}
    \\
    {\small (c) Decomposition (LSPC). }  &  {\small (d) Decomposition (PCD). }
    \\
\end{tabular}
  \end{center}
    \vspace{-3mm}
  \caption{\label{fig:localization_zipcode_example} Localizing a
    piecewise-constant graph signal associated with ZIP codes 10128
    and 10030. LSPC fails to localize ZIP code 10030, but PCD
    successfully localizes both ZIP codes. The $F_1$ scores of LSPC
    and PCD are $0.41$ and $0.57$, respectively, and the Hamming
    distances of LSPC and PCD are $162$ and $141$, respectively.}
      \vspace{-3mm}
\end{figure}

There are 43 ZIP codes in Manhattan. In each study, we randomly
select two of those and generate a piecewise-constant graph signal,
\begin{equation*}
  \x \ = \   \mu_1 \one_{{\rm ZIP}_1} + \mu_2 \one_{{\rm ZIP}_2} + \epsilon,
\end{equation*}
where the signal strength is uniformly distributed $\mu_i \sim \U(0.5,
1.5)$; node set ZIP$_i$ indicates the nodes belonging to the selected ZIP code; and the noise $\epsilon \sim \N(0, \sigma^2 \Id)$
with $\sigma^2$ varying from $0.1$ to $1$ with interval $0.1$. At each
noise level, we generate $1,000$ piecewise-constant graph
signals. Figure~\ref{fig:localization_zipcode_example}(a) shows a
piecewise-constant signal generated by ZIP codes 10022 and 10030 and
(b) shows a noisy graph signal with noise variance $\sigma^2 = 1$. We
use the local-set-based piecewise-constant
dictionary (LSPC)~\cite{ChenJVSK:16} and piecewise-constant
decomposition~\eqref{eq:PCD} (PCD) to decompose this noisy graph
signal into two activated
pieces.  We did not compare with hard thresholding, trend filtering and graph Laplacian denoising because it is difficult to determine thresholds for multiple pieces with various activated magnitudes. Figures~\ref{fig:localization_zipcode_example}(c)--(d) show
the results provided by LSPC and PCD, respectively. We see that LSPC
fails to localize ZIP code $10030$, but PCD successfully localizes
both ZIP codes and provides acceptable localization performance.

\begin{figure}[t]
  \begin{center}
    \begin{tabular}{ccc}
     \includegraphics[width=0.4\columnwidth]{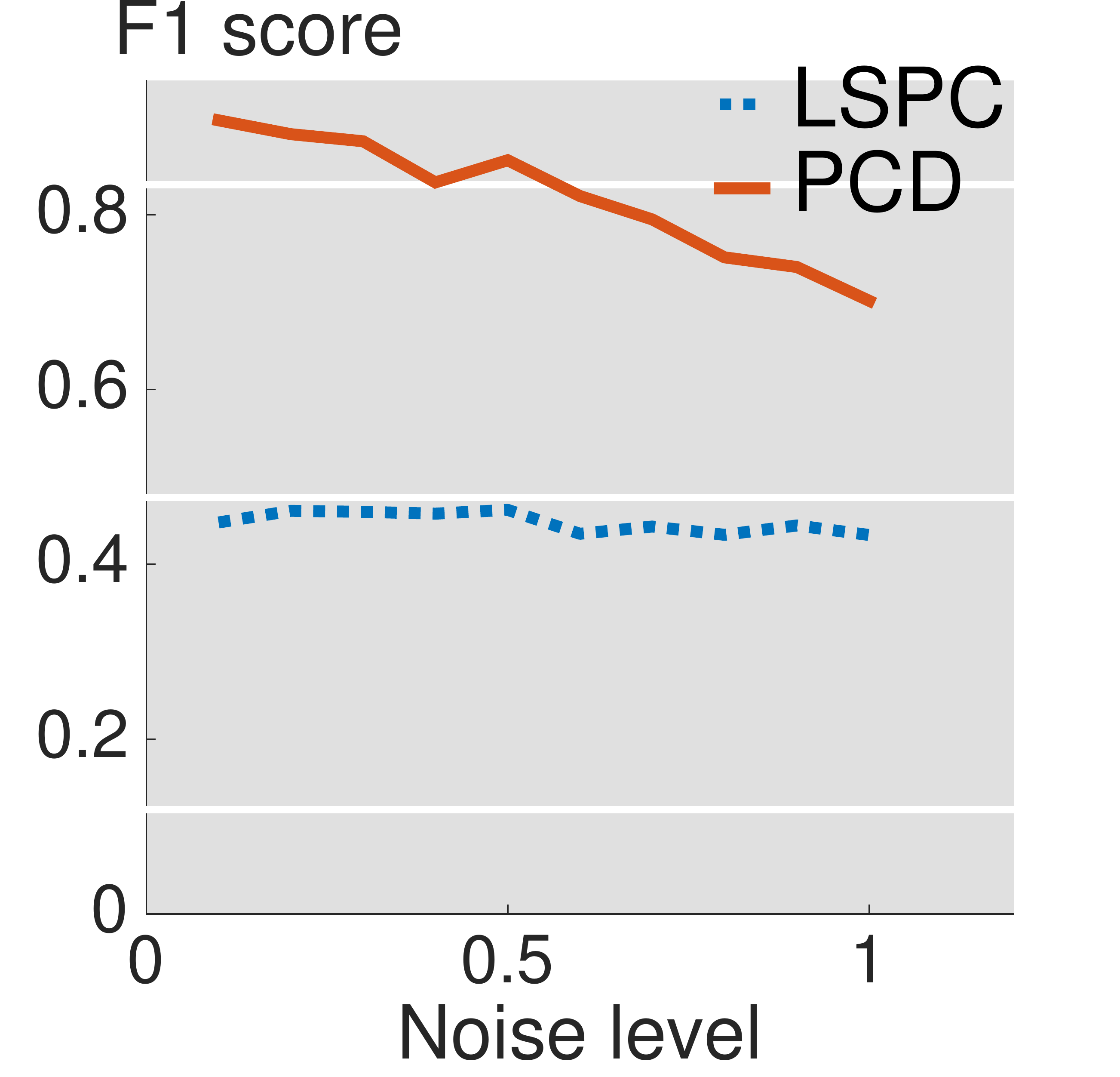}
 &
    \includegraphics[width=0.4\columnwidth]{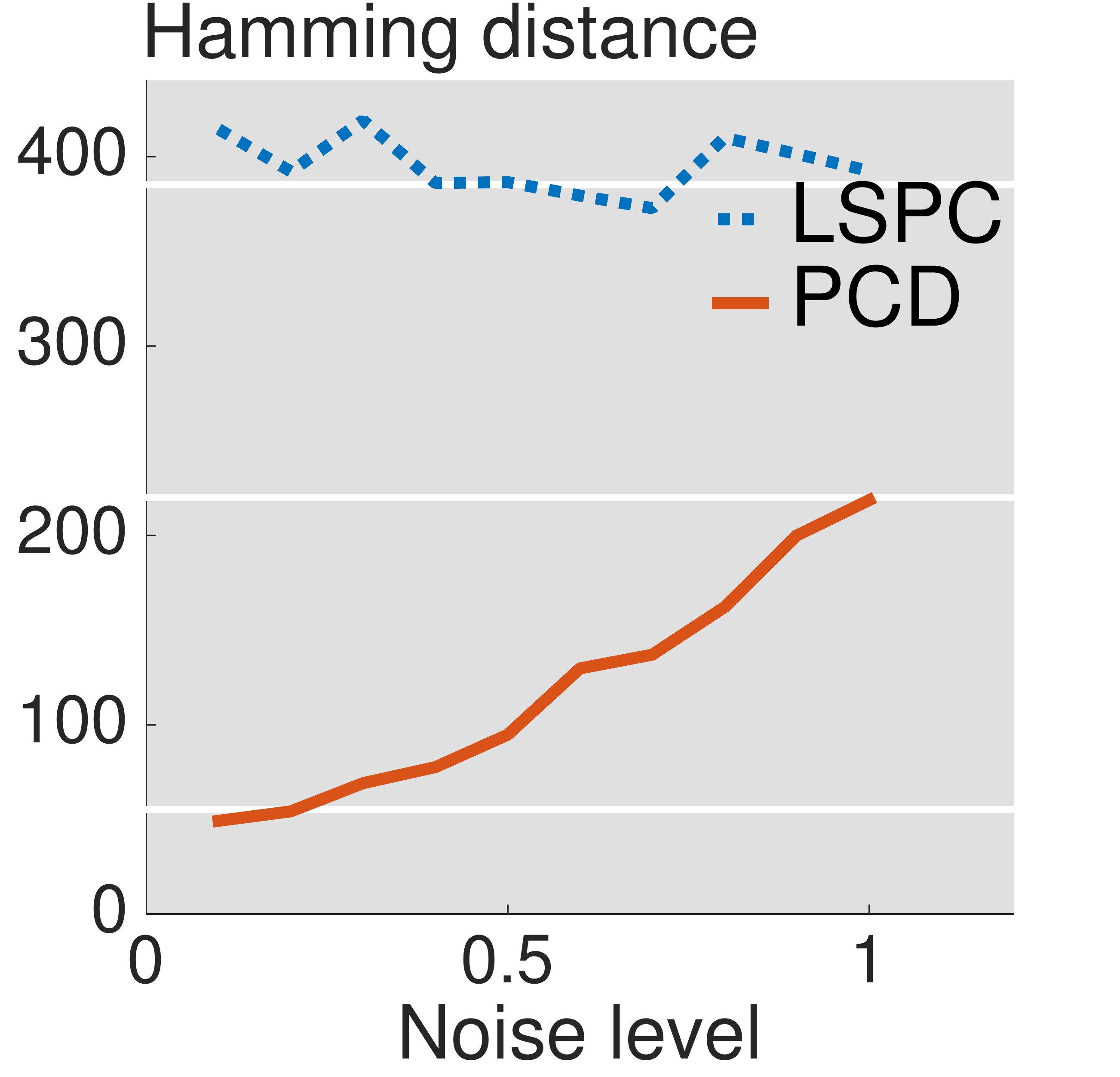}
      \\
    {\small (a) $F_1$ score. }  &  {\small (b) Hamming distance. }
    \\
\end{tabular}
  \end{center}
       \vspace{-3mm}
  \caption{\label{fig:localization_zipcode_performance} Localizing
    multiple ZIP codes as a function of noise level. The proposed PCD
    significantly outperforms LSPC. }
         \vspace{-3mm}
\end{figure}

To quantify the localization performance, we again use the $F_1$ score
and the Hamming distance.
Figures~\eqref{fig:localization_zipcode_performance}(a)--(b) show the
$F_1$ score and Hamming distance when localizing ZIP codes in
Manhattan as a function of noise level. The results at each noise
level are averaged over $1,000$ runs. Again, we see that PCD provides
significantly better performance than LSPC, but PCD is sensitive to
noise, while LSPC is robust to noise.


\begin{figure*}[t]
  \begin{center}
    \begin{tabular}{ccccc}
    \includegraphics[width=0.33\columnwidth]{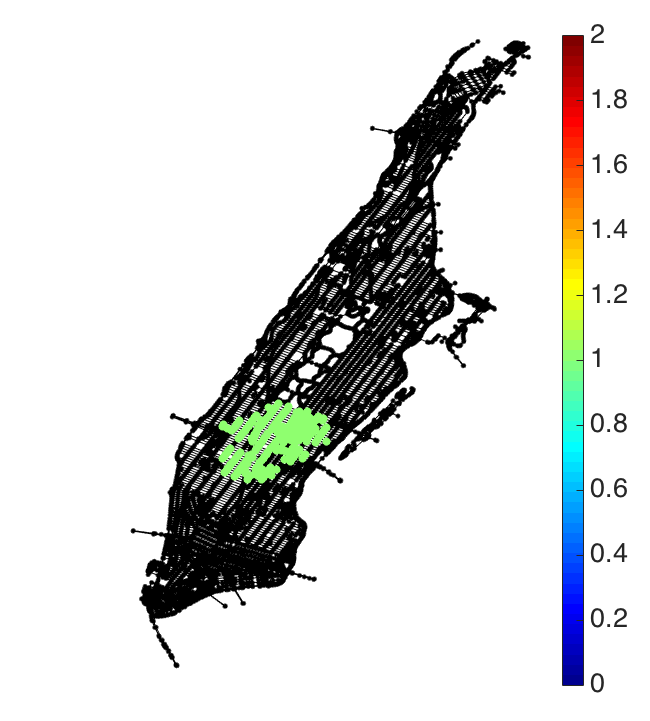}
 &
    \includegraphics[width=0.33\columnwidth]{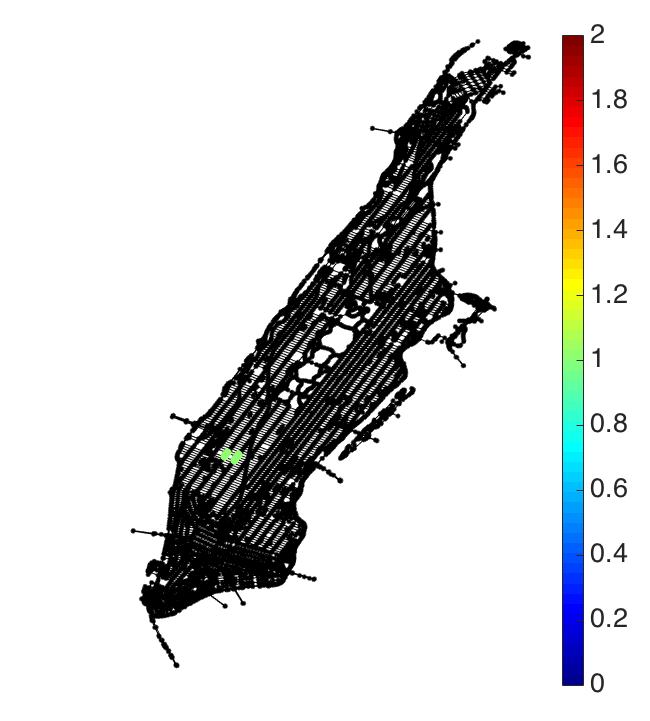}
 &
     \includegraphics[width=0.33\columnwidth]{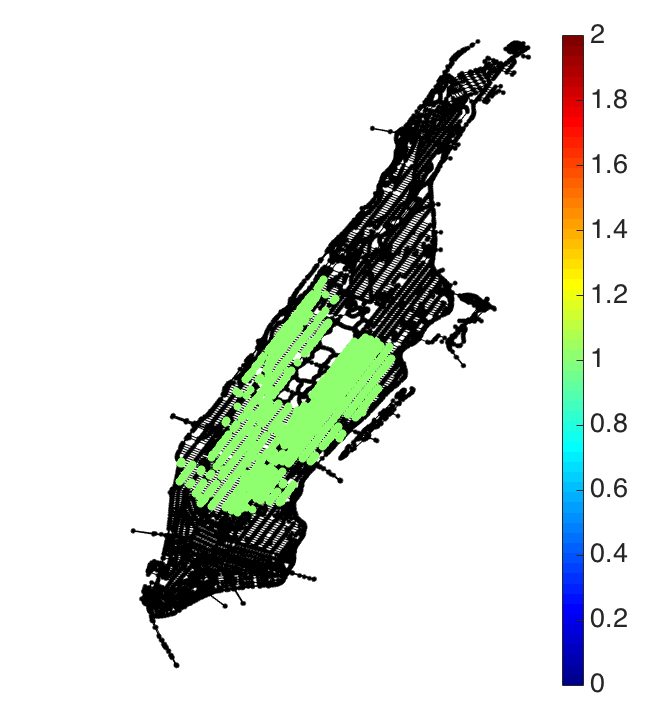}
 &
    \includegraphics[width=0.33\columnwidth]{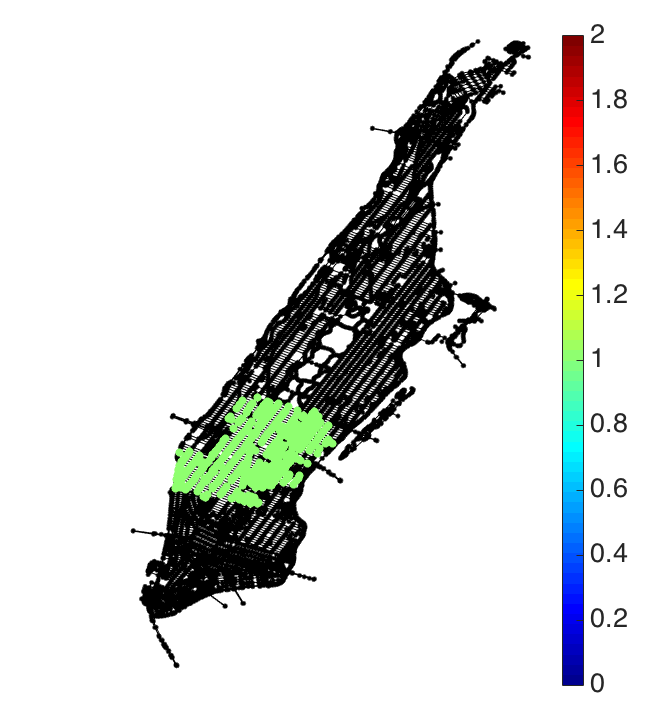}
   &
    \includegraphics[width=0.33\columnwidth]{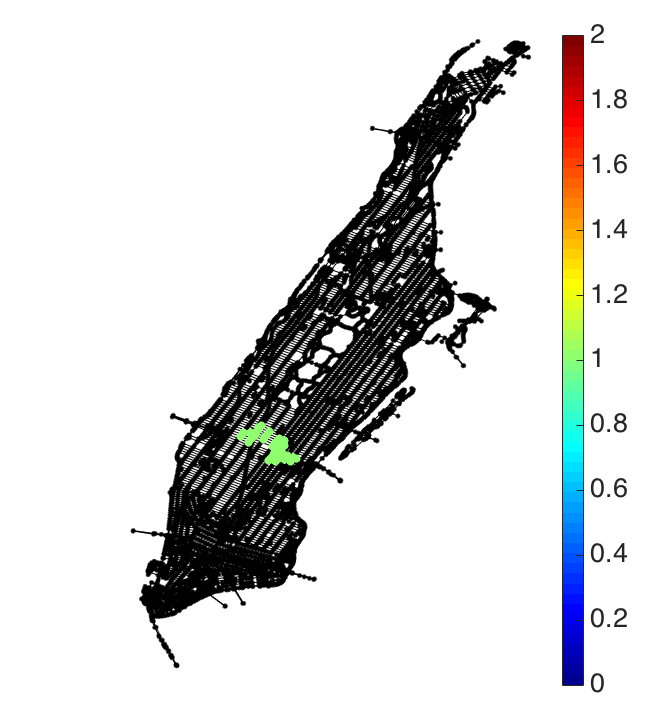}
\\
    {\small (a) Piece 1. }  &  {\small (b)  Piece 2. }  &
    {\small (c) Piece 3. }  &  {\small (d)  Piece 4. }  &
    {\small (e) Piece 5. }
\end{tabular}
  \end{center}
   \caption{\label{fig:manhattan_atoms} Top five most frequently-used pieces learned from taxi-pickup activities during rush hours in $2015$. }
\end{figure*}

\begin{figure}[t]
  \begin{center}
    \begin{tabular}{ccc}
     \includegraphics[width=0.4\columnwidth]{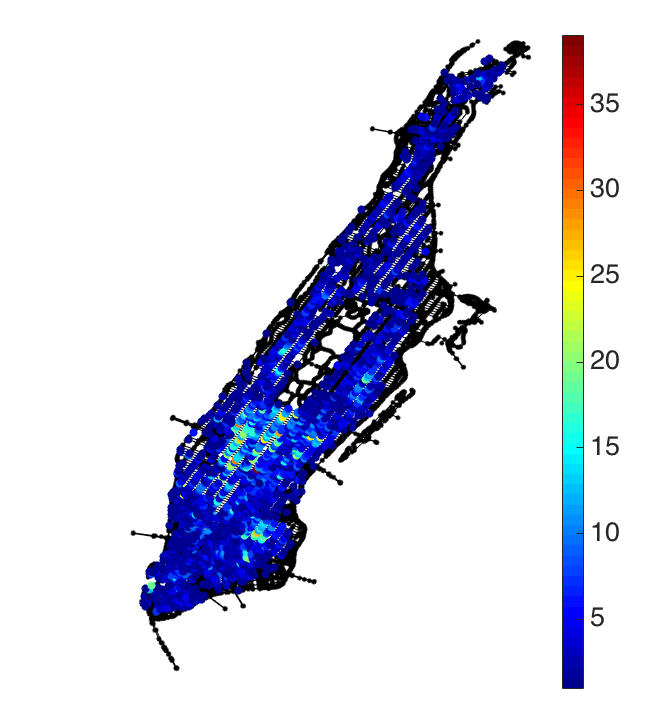}
 &
    \includegraphics[width=0.48\columnwidth]{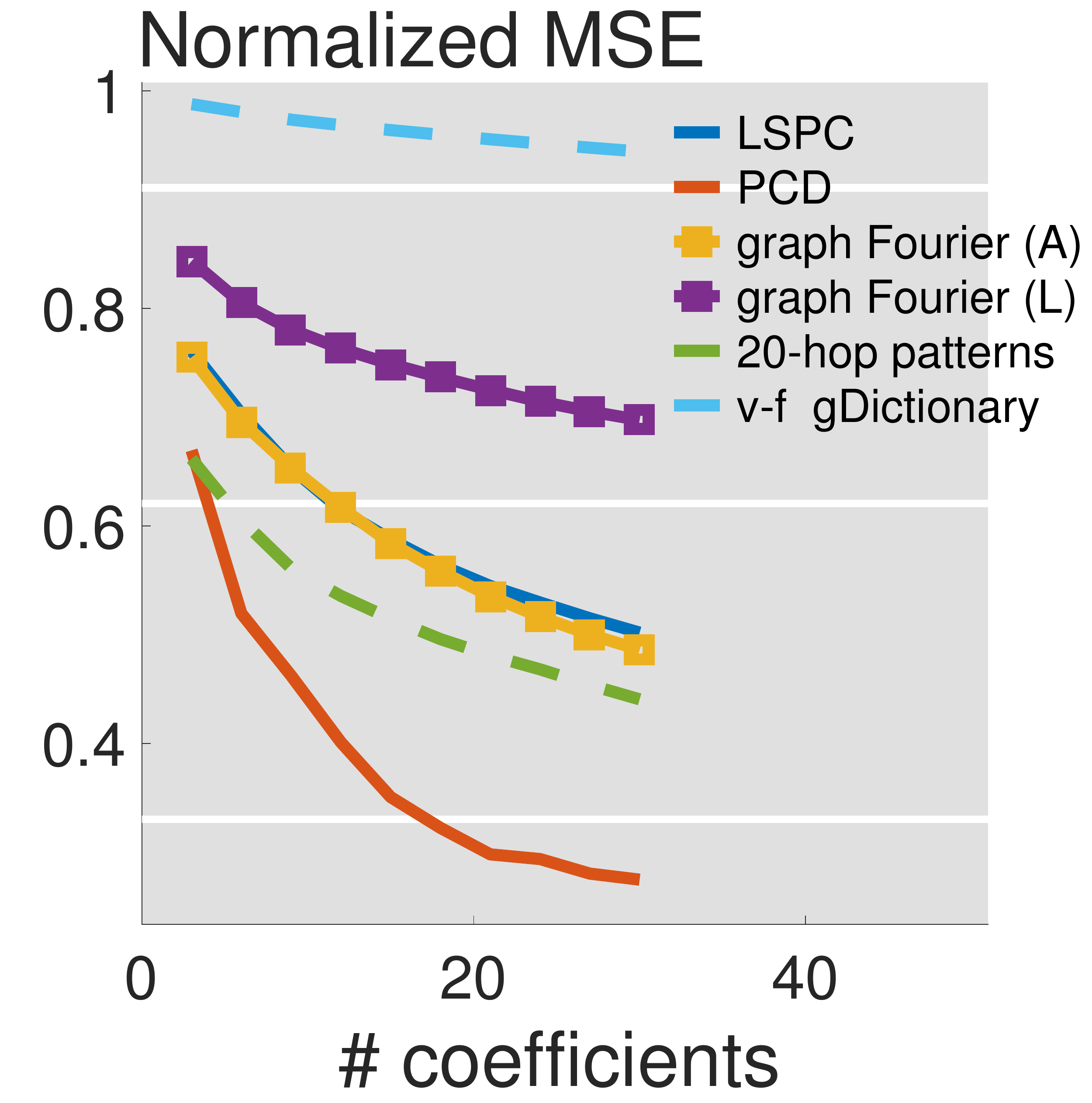}
      \\
    {\small (a) Signal. }  &  {\small (b) Approximation error. }
    \\
      \includegraphics[width=0.4\columnwidth]{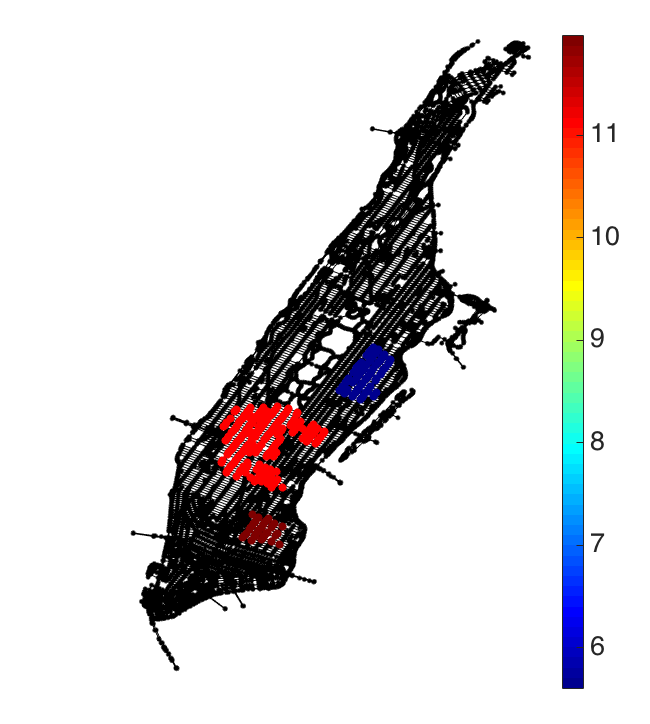}
     &
           \includegraphics[width=0.4\columnwidth]{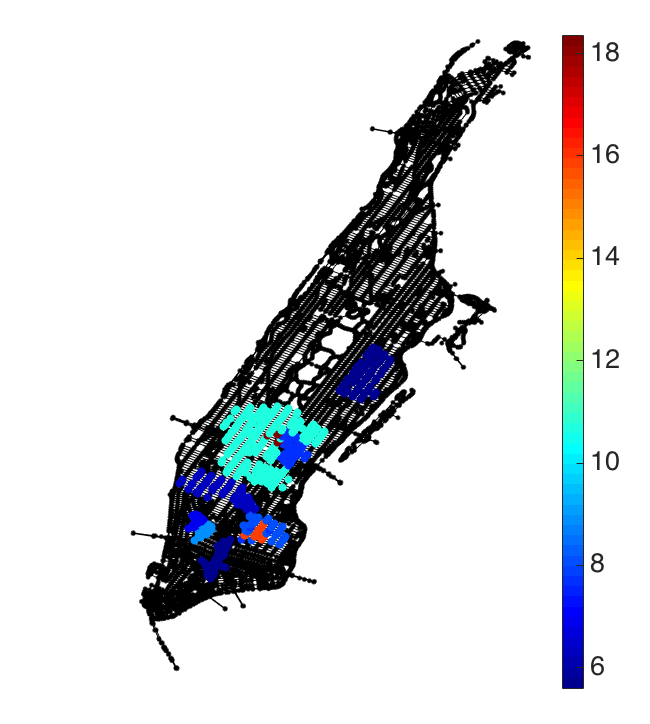}
  \\
    {\small (c) Approximation }  &  {\small (d) Approximation }
      \\
    {\small by three pieces. }  &  {\small   by nine pieces. }
    \\
\end{tabular}
  \end{center}
  \caption{\label{fig:decomp_resturant} Decomposing the restaurant
    distribution in Manhattan using a piecewise-constant decomposition. Piecewise-constant decomposition (PCD)~\eqref{eq:PCD} outperforms the other methods in the task of approximation.}
         \vspace{-3mm}
\end{figure}
\begin{figure}[t]
  \begin{center}
    \begin{tabular}{ccc}
     \includegraphics[width=0.4\columnwidth]{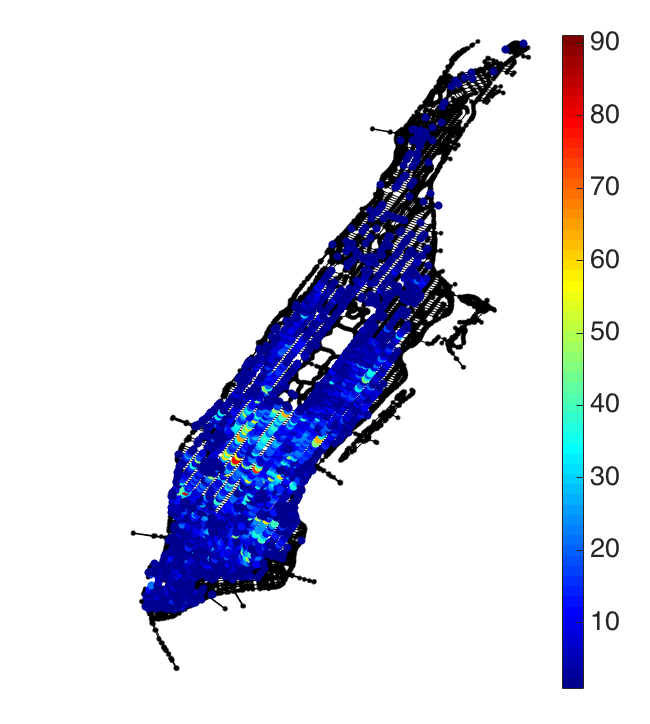}
 &
    \includegraphics[width=0.48\columnwidth]{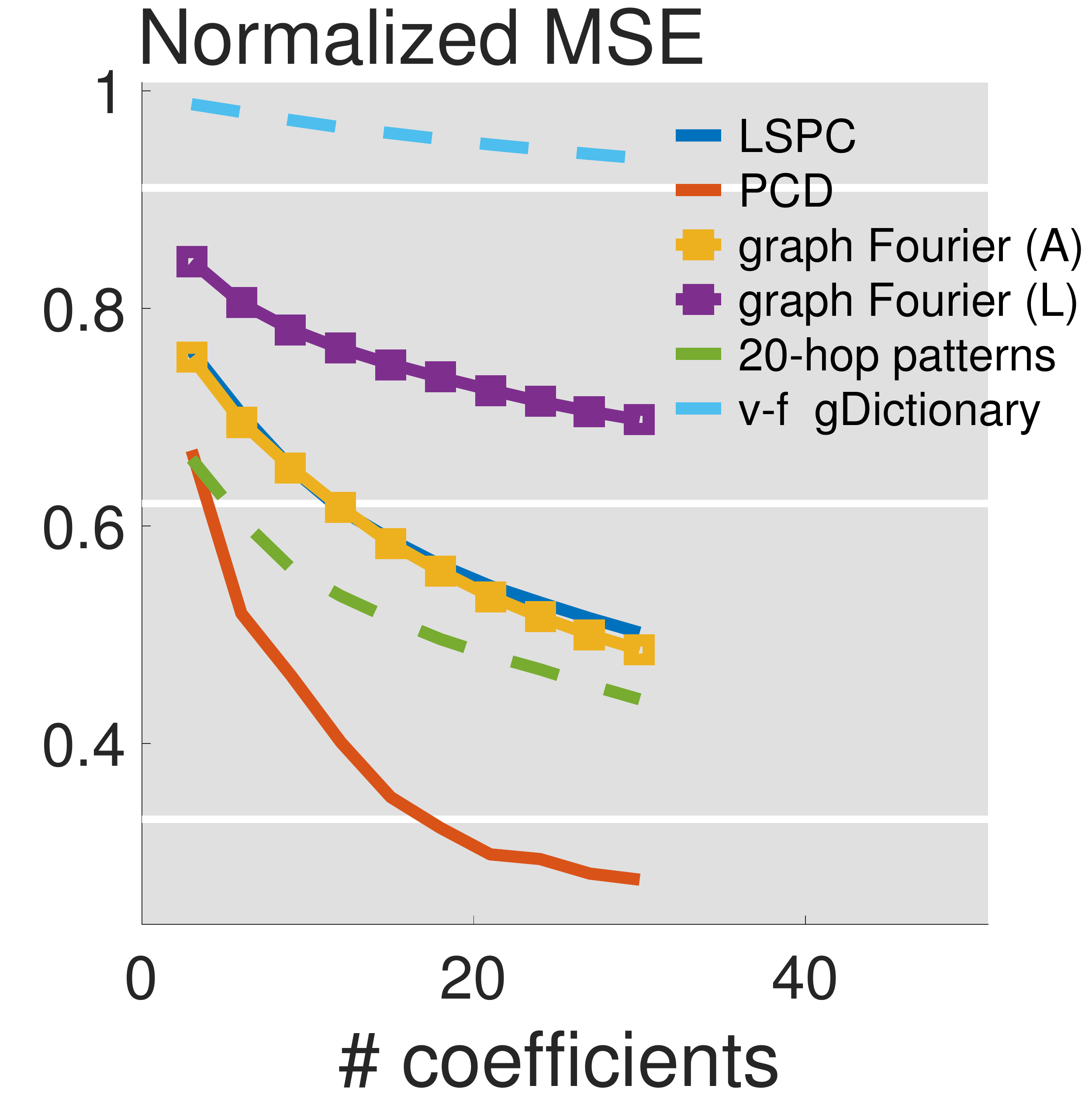}
      \\
    {\small (a) Signal. }  &  {\small (b) Approximation error. }
    \\
      \includegraphics[width=0.4\columnwidth]{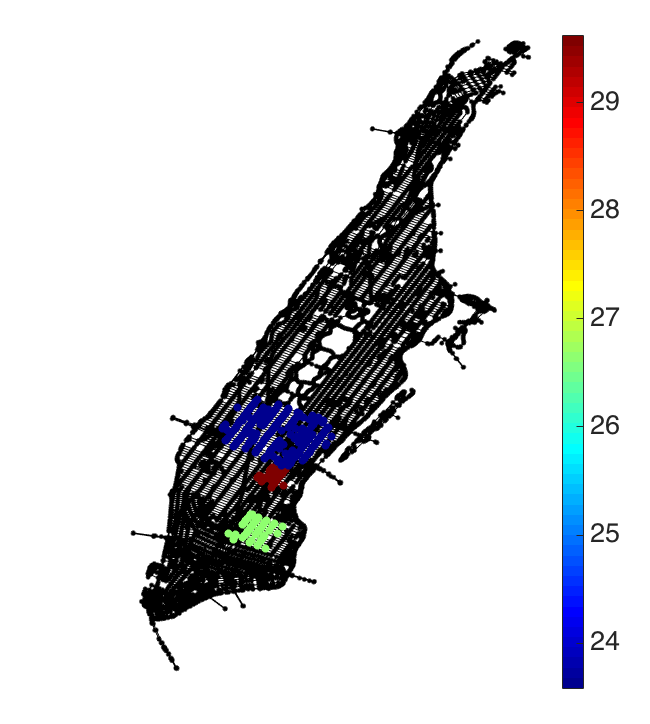}
     &
           \includegraphics[width=0.4\columnwidth]{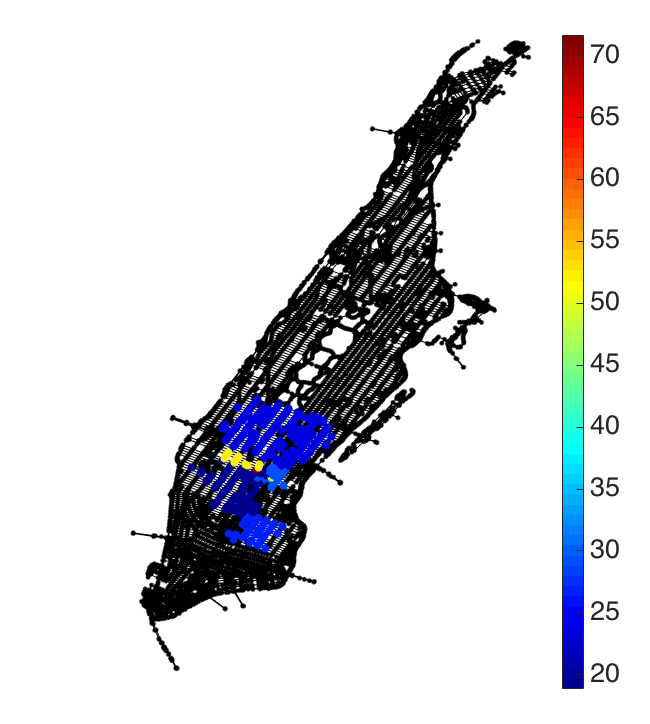}
  \\
    {\small (c) Approximation }  &  {\small (d) Approximation }
      \\
    {\small by three pieces. }  &  {\small   by nine pieces. }
    \\
\end{tabular}
  \end{center}
  \caption{\label{fig:decomp_taxipickup} Decomposing the taxi-pickup
    activity in Manhattan On Friday, June 5th, at 11 pm, using a
    piecewise-constant decomposition. Piecewise-constant decomposition (PCD)~\eqref{eq:PCD} outperforms the other methods in the task of approximation. }
         \vspace{-3mm}
\end{figure}

\begin{figure}[h]
  \begin{center}
    \begin{tabular}{ccc}
      \includegraphics[width=0.35\columnwidth]{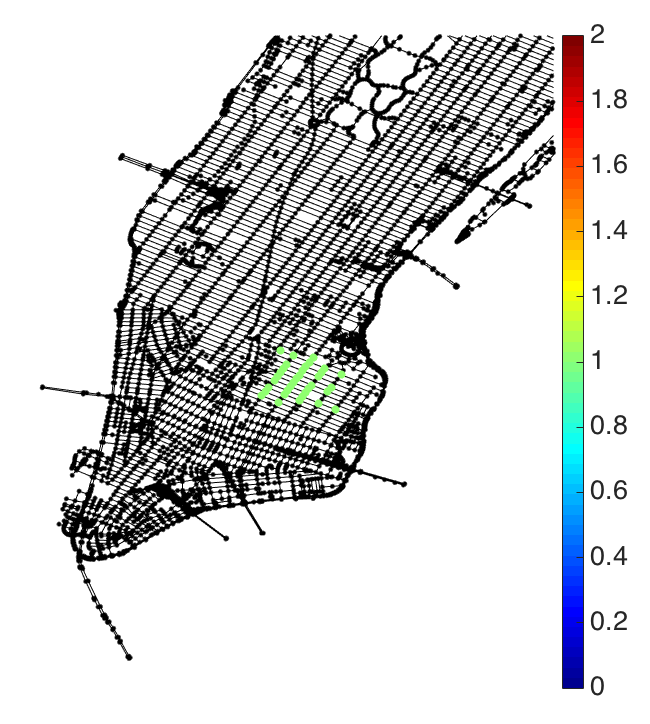}
      &
      \includegraphics[width=0.35\columnwidth]{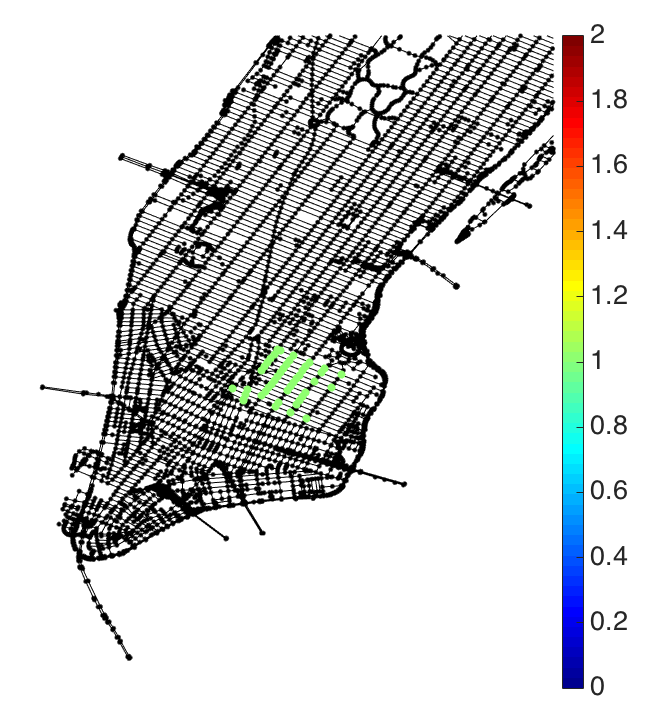}
      \\
      {\small (a) \emph{East Village} taxi pickups.}  &  {\small (b) \emph{East Village} restaurants.}
      \\
 \includegraphics[width=0.35\columnwidth]{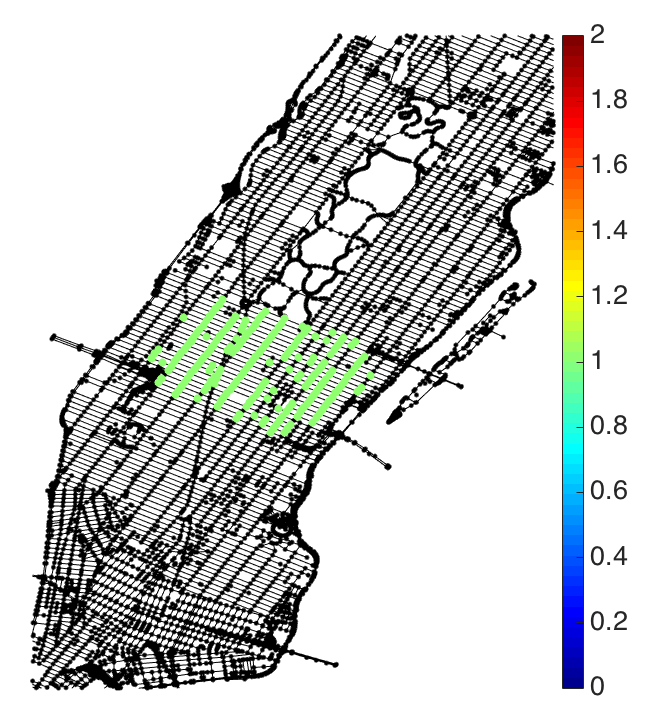}
     &
     \includegraphics[width=0.35\columnwidth]{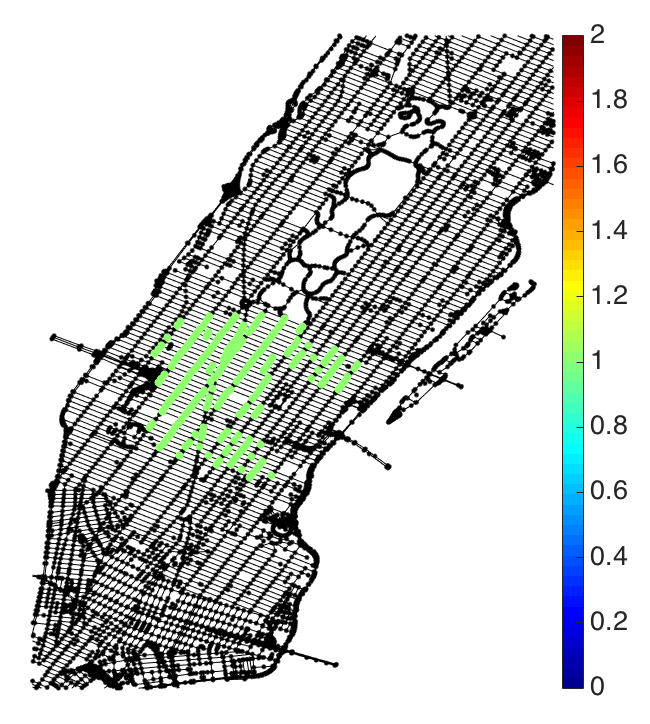}
  \\
    {\small (c) \emph{Times Square} pickups.}  &  {\small (d)  \emph{Times Square} restaurants. }
\end{tabular}
  \end{center}
  \caption{\label{fig:decomp_similar_piece} Both restaurant
    distribution and taxi-pickup activity on Friday nights activate
    the same areas around East Village and Times Square. The $F_1$ score
    between the two \emph{Times Square} pieces is $0.65$ and the $F_1$
    score between the two \emph{East Village} pieces is $0.59$,
    indicating a high overlap. }
\end{figure}

\subsubsection{Restaurant and taxi-pickups}
We now analyze restaurant density and taxi-pickup activity in
Manhattan with the aim of understanding their components through
decomposition.

Figure~\ref{fig:decomp_resturant}(a) shows the restaurant density
including the positions of 10,121 restaurants in Manhattan. We project
each restaurant to its nearest intersection and count the number of
restaurants at each intersection. We use piecewise-constant
decomposition to decompose this graph
signal. Figures~\ref{fig:decomp_resturant}(c)--(d) show the
decomposition by using three and nine pieces, respectively. Through
the decomposition, we can get a rough idea about the distribution of
restaurants. For example, areas around East Village and Times Square
have more restaurants. By using more pieces, we can obtain a finer
resolution and a better approximation.

Figure~\ref{fig:decomp_resturant}(b) shows the approximation error
where the $x$-axis is the number of expansion coefficients and the
$y$-axis is the normalized mean-squared error.  We compare nonlinear
approximation by using graph Fourier basis based on the adjacency
matrix~\cite{SandryhailaM:13} (graph Fourier (A), yellow-square line),
nonlinear approximation by using graph Fourier basis based on the
graph Laplacian matrix~\cite{ShumanNFOV:13} (graph Fourier (L),
purple-square line), sparse coding based on vertex-frequency based
graph dictionary~\cite{ShumanRV:15} (v-f gDictionary, blue-dashed
line), egonet based localized patterns suggested in~\cite{ThanouSF:14}
(20-hop patterns, green-dashed line), local-set-based
piecewise-constant graph dictionary~\cite{ChenJVSK:16} (LSPC,
blue-solid line) and piecewise-constant decomposition~\eqref{eq:PCD}
(PCD, red-solid line).  The vertex-frequency based graph dictionary, or
windowed graph Fourier transform, is a graph-frequency domain based
representation, which produces the graph spectrogram and enables
vertex-frequency analysis. In the experiments, we use four graph
filters for vertex-frequency based graph dictionary, which includes
54,716 atoms in total.  \cite{ThanouSF:14} suggests a graph
dictionary where each atom activates a $K$-hop neighbor.  In the
experiments, we put all neighbors up to 20 hops into a graph
dictionary, which includes 273,580 atoms. Note that the
local-set-based piecewise-constant graph dictionary includes 27,357
atoms and the proposed piecewise-constant decomposition adaptively
learns a small number of atoms (equal to the number of coefficients)
from the given graph signal.

Figure~\ref{fig:decomp_resturant}(d) shows that by using the same
number of expansion coefficients (pieces), piecewise-constant
decomposition significantly outperforms other methods; egonet-based
localized patterns also provide good results but are not flexible
enough to capture localized patterns with arbitrary shapes; the
vertex-frequency based graph dictionary fails because as it does not
capture localized variations as well as the vertex-domain based
representations; and graph Fourier basis based on adjacency
matrix~\cite{SandryhailaM:13} outperforms the one based on the
Laplacian matrix~\cite{ShumanNFOV:13}. Overall, piecewise-constant
decomposition provides an effective data-adaptive and
structure-related representation.

Figure~\ref{fig:decomp_taxipickup}(a) shows the Manhattan taxi-pickup
activity on Friday, June 5th, at 11 pm. We project each taxi pickup to
its nearest intersection, count the number of pickups and use
piecewise-constant decomposition to decompose this graph
signal. Figures~\ref{fig:decomp_taxipickup}(c)--(d) show the
decomposition with three and nine pieces, respectively. This
decomposition allows us to get a rough idea of the taxi pickup
distribution; for example, areas around East Village and Times Square
are busier.

Figure~\ref{fig:decomp_taxipickup}(b) shows the approximation
error. Similarly to the results for the restaurant distribution, we
see that using the same number of expansion coefficients (pieces),
piecewise-constant decomposition significantly outperforms the other
methods and graph Fourier basis based on the adjacency
matrix~\cite{SandryhailaM:13} outperforms the one based on the graph
Laplacian matrix~\cite{ShumanNFOV:13}.

Note that both the restaurant distribution and the taxi-pickup
activity on Friday nights activate the areas around East Village and
Times Square (see Figure~\ref{fig:decomp_similar_piece}). Each piece
in Figure~\ref{fig:decomp_similar_piece} comes from one of the three
pieces in Figures~\ref{fig:decomp_resturant}(c)
and~\ref{fig:decomp_taxipickup}(c); we observe that the corresponding
pieces highly overlap. The $F_1$ score between the two \emph{Times
  Square} pieces is $0.65$ and the $F_1$ score between the two
\emph{East Village} pieces is $0.59$ indicating that the taxi pickups
on Friday nights are highly correlated with the restaurant
distribution, confirming a well-understood pattern of urban lifestyle.

\vspace{-3mm}
\section{Dictionary Learning on Graphs}
\label{sec:dictionary}
We now extend signal decomposition, where we find activated pieces
from a single graph signal, to dictionary learning where the aim is to
find activated pieces  shared by multiple graph signals. In
other words, we learn activated pieces as building blocks that are
used to represent multiple graph signals.  We extend the decomposition
solver from Section~\ref{sec:decomposition} to the dictionary learning
problem and apply it to mine traffic patterns in the Manhattan data
set.

Consider a matrix of $L$ graph signals as columns,
\begin{eqnarray*}
  \X \ = \  \Dd  \Z + \Ee \ \in \ \R^{N \times L},
\end{eqnarray*}
with $\Dd = \begin{bmatrix} \one_{C_1} & \one_{C_2} & \cdots &
  \one_{C_K} \end{bmatrix}$ the graph dictionary with a predefined
number of $K$ one-piece atoms as columns, the coefficient matrix $\Z
\in \R^{K \times L}$ is sparse, $C_i$ are connected and $\Ee_{i,j}
\sim \N(0, \sigma^2)$ is Gaussian noise.  We store activated pieces in
$\Dd$ as building blocks. Thus, each graph signal (column in $\X$) is
approximated by a linear combination of several pieces from $\Dd$. We
learn the pieces $C_i$ from $\X$ by solving 
\begin{eqnarray}
  \label{eq:dictionary_learning}
  &&  \min_{\Z, C_i}  \left\| \X -  \begin{bmatrix}
      \one_{C_1} &   \one_{C_2} & \cdots & \one_{C_K}
    \end{bmatrix} \Z \right\|_2^2,
  \\
  \nonumber && \text{subject to } C_i \in \Cc \quad \text{and} \quad
  \left\| \Z \right\|_{0, \infty} \leq S,
\end{eqnarray}
where the sparsity level $S$ helps avoid overfitting and $\left\| \Z
\right\|_{0, \infty} = \max_{i=1,\cdots,N} \left\| \z_i \right\|_0$
with $\z_i$ the $i$th column in $\Z$, which is the maximum number of
nonzero elements in each column. When the number of graph signals $L =
1$, the dictionary learning problem becomes the decomposition
problem~\eqref{eq:PCD}.

Note that this graph dictionary is not designed to learn from
arbitrary graph signals. Due to the strong connectivity constraints,
\eqref{eq:dictionary_learning} is not flexible enough to approximate a
graph signal as well as using the classical K-SVD; the advantage,
however, is that it can find shared localized patterns from a large
set of graph signals, which K-SVD cannot do as it would require
careful threshold tuning for each selected atom.

\vspace{-4mm}
\subsection{Methodology}
% Consider a matrix of graph signals,
% \begin{eqnarray*}
% 	\X \ = \  \begin{bmatrix}
% 	\one_{C_1} &   \one_{C_2} & \cdots & \one_{C_K}
% 	\end{bmatrix}  \Z + \Ee
% 	 \ \in \ \R^{N \times L},
% \end{eqnarray*}
% where the coefficient matrix $\Z \in \R^{K \times L}$ is sparse,
% each piece $C_i$ is connected and the noise matrix $\Ee_{i,j} \sim
% \N(0, \sigma^2)$. Each column of $\X$ forms a piecewise-constant
% graph signal. Each piecewise-constant graph signal is a linear
% combination of several pieces from a graph dictionary.  We force the
% coefficient matrix $\Z$ to be sparse to avoid overfitting.

Our goal is to learn shared pieces and the corresponding coefficients
from a matrix of graph signals. While it would also be possible to use
the decomposition techniques from the previous section to localize
pieces from each graph signal and store all activated pieces into the
dictionary, that approach would not reveal correlations among graph
signals. We use the learning approach to find shared pieces among
several graph signals as it allows the same piece to be repeatedly
used as a basic building block thus revealing correlations.
% Consider the following
% optimization problem
% \begin{eqnarray*}
%   &&  \min_{\Z, C_i}  \left\| \X -  \begin{bmatrix}
%       \one_{C_1} &   \one_{C_2} & \cdots & \one_{C_K}
%     \end{bmatrix} \Z \right\|_F^2,
%   \\
%   && \text{subhect to}~ C_i \in \Cc, \quad \text{and} \quad \left\| \Z
%   \right\|_{0, \infty} \leq S,
% \end{eqnarray*}
% where the second constraint requires each column of $\Z$ to be sparse;
% that is, each graph signal is approximated by a linear combination of
% only a few pieces. The main idea is to obtain graph-dictionary-based
% sparse representations for a given matrix of graph signals.

To solve \eqref{eq:dictionary_learning}, we update the dictionary and
the coefficient matrix successively; that is, given the dictionary, we
optimize over the coefficient matrix and then, given the coefficient
matrix, we optimize over the dictionary.  When updating the
coefficient matrix, we fix $C_i$ and consider
\begin{eqnarray}
\label{eq:learn_gdictionary1}
	&&  \min_{\Z}  \left\| \X -  \begin{bmatrix}
	\one_{C_1} &   \one_{C_2} & \cdots & \one_{C_K}
	\end{bmatrix}  \Z \right\|_F^2,
	\\
	\nonumber
	&& \text{subject to } \left\| \Z \right\|_{0, \infty} \leq S.
\end{eqnarray}
where $\left\| \cdot \right\|_F$ denotes the Frobenius norm. This is a
common sparse coding problem, which can be efficiently solved by
orthogonal matching pursuit~\cite{PatiRK:93}.  When updating the graph
dictionary, we fix $\Z$ and consider
\begin{eqnarray}
  \label{eq:learn_gdictionary2}
  &&  \min_{C_i}  \left\| \X -  \begin{bmatrix}
      \one_{C_1} &   \one_{C_2} & \cdots & \one_{C_K}
    \end{bmatrix} \Z \right\|_F^2,
  \\
  \nonumber && \text{subject to } C_i \in \Cc.
\end{eqnarray}

Similarly to the piecewise-constant decomposition, we iteratively
update each piece. Let $\z_i$ be the $i$th row of $\Z$ and $\Rr_j = \X
- \sum_{i \neq j} \one_{C_i} \z_i^T \in \R^{N \times L}$. Then,
\begin{eqnarray*}
  && \left\| \X -  \begin{bmatrix}
      \one_{C_1} &   \one_{C_2} & \cdots & \one_{C_K}
    \end{bmatrix} \Z \right\|_F^2 \stackrel{(a)}{=} \left\| \Rr_j -
    \one_{C_j} \z_j^T \right\|_F^2
  \\
  &\stackrel{(b)}{=} & {\rm Tr} ( \Rr_j^T \Rr_j - \Rr_j^T
    \one_{C_j} \z_j^T - \z_j \one_{C_j}^T \Rr_j + \z_j \one_{C_j}^T
    \one_{C_j} \z_j^T )
  \\
  & \stackrel{(c)}{=}& {\rm Tr} ( \Rr_j^T \Rr_j ) + (
    \z_j^T \z_j \one - 2 \Rr_j \z_j )^T \one_{C_j},
\end{eqnarray*}
where (a) follows from the denotation of $\Rr_j$, (b) from the
definition of Frobenius norm and the trace operator, and (c) from
Tr$(\X \Y) =$ Tr$(\Y \X)$. Since the first term is a constant, to
update each piece, we freeze all the other pieces and solve
\begin{eqnarray*}
  &&  \min_{C_i}   (\z_j^T \z_j \one - 2 \Rr_j \z_j)^T \one_{C_j},~~ \text{subject to } C_i \in \Cc,
\end{eqnarray*}
where the objective function is an inner product of two vectors. This
is equivalent to solving the localization problem~\eqref{eq:opfp} by
setting $\x = \Rr_j \z_j /\z_j^T \z_j$; that is,
\begin{eqnarray*}
  && \left\| \frac{\Rr_j \z_j}{\z_j^T \z_j} - \one_{C_i} \right\|_2^2
  \ =  \  \frac{1}{ \left( \z_j^T \z_j \right)^2} \left\| \Rr_j \z_j - \z_j^T \z_j \one_{C_i} \right\|_2^2 
  \\
%   & =  &  \frac{1}{ \left( \z_j^T \z_j \right)^2} 
%   \bigg( 
%   (\Rr_j \z_j)^T(\Rr_j \z_j) - 2 (\Rr_j \z_j)^T  \z_j^T \z_j \one_{C_i} 
%   \\
%   && 
%   + (\z_j^T \z_j)^2 \one_{C_i}^T \one_{C_i}
%   \bigg) 
%   \\
  & =  &
  \frac{\z_j^T \Rr_j^T \Rr_j \z_j }{ \left( \z_j^T \z_j \right)^2} 
  + \frac{1}{  \z_j^T \z_j }  ( \z_j^T \z_j \one_{C_i}^T \one_{C_i} - 2 (\Rr_j \z_j)^T \one_{C_i} ),
\end{eqnarray*}
where $\z_j^T \z_j $ is a scalar and $\one_{C_i}^T \one_{C_i} = \one^T
\one_{C_i}$ because $\one_{C_i}$ is a binary vector. In other words,
we update each piece in the graph dictionary by using the localization
solver~\eqref{eq:localization_opt}.  We iteratively update each piece
until convergence and we obtain a series of solutions providing a
local optimum~\cite{BoydV:04}.

\vspace{-4mm}
\subsection{Experimental Validation}
We use the dictionary learning
techniques~\eqref{eq:learn_gdictionary1}--\eqref{eq:learn_gdictionary2}
to mine taxi-pickup patterns in Manhattan, as taxis are valuable sensors of city life~\cite{FerreiraPVFS:13,
  DoraiswamyFDFS:14,DeriM:15}, providing insight into economic activity, human behavior and mobility patterns.
We consider the 2015 public dataset of
taxi pickups (used in Section~\ref{sec:decomposition_exp}) during rush hour (6-8pm). We add
taxi pickups within each hour to obtain 1,095 ($365 \times
3$) graph signals; thus, each graph signal is a measure of taxi pickups at an
intersection in Manhattan at a specific hour. We mine pickup
patterns through two tasks: detecting events in Manhattan and checking
whether weekdays and weekends exhibit differences in traffic patterns.

\subsubsection{Dictionary Learning on Graphs: Event detection}
\label{sec:even_detection}
We consider two types of events: common events and special
events.  A common event usually leads to a place that is frequently or periodically crowded, , which shows a natural traffic
behavior, and a special event usually leads to a place that is rarely crowded, which shows an anomaly in traffic behavior.  Traffic accidents
or holiday celebrations usually lead to special events. We are going
to use the learned pieces and their corresponding coefficients to
analyze the taxi-pickup activities during rush hours. We set the size
of the dictionary to $K = 500$ and its sparsity to $S = 30$.  This
  means that $K = 500$ activated pieces are learned from $L = 1,095$
  graph signals of dimension $N = 13,679$ ($13,679$ intersections in
  Manhattan) and each graph signal is approximated by at most $30$
  pieces. When a piece is used to represent a graph signal, it means
that the corresponding area is particularly crowded at a specific time
and we need this piece to capture this traffic information. On the
other hand, when a piece is not used, it just means that the
corresponding area is not particularly crowded compared to other
areas; it does not necessarily mean that there are no taxi pickups in
the corresponding area because those taxi pickups may activate other
related pieces.

\begin{figure}[h]
  \begin{center}
    \begin{tabular}{cc}
    \includegraphics[width=0.33\columnwidth]{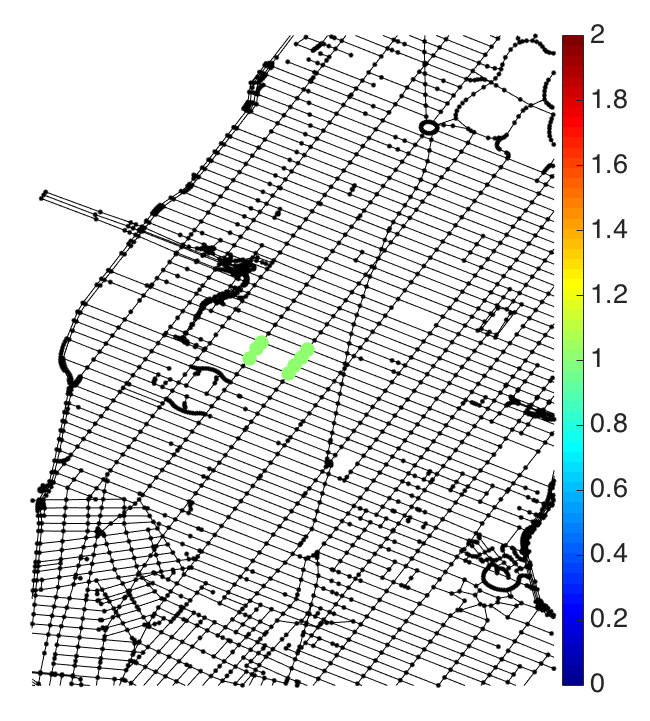}
 &
    \includegraphics[width=0.62\columnwidth]{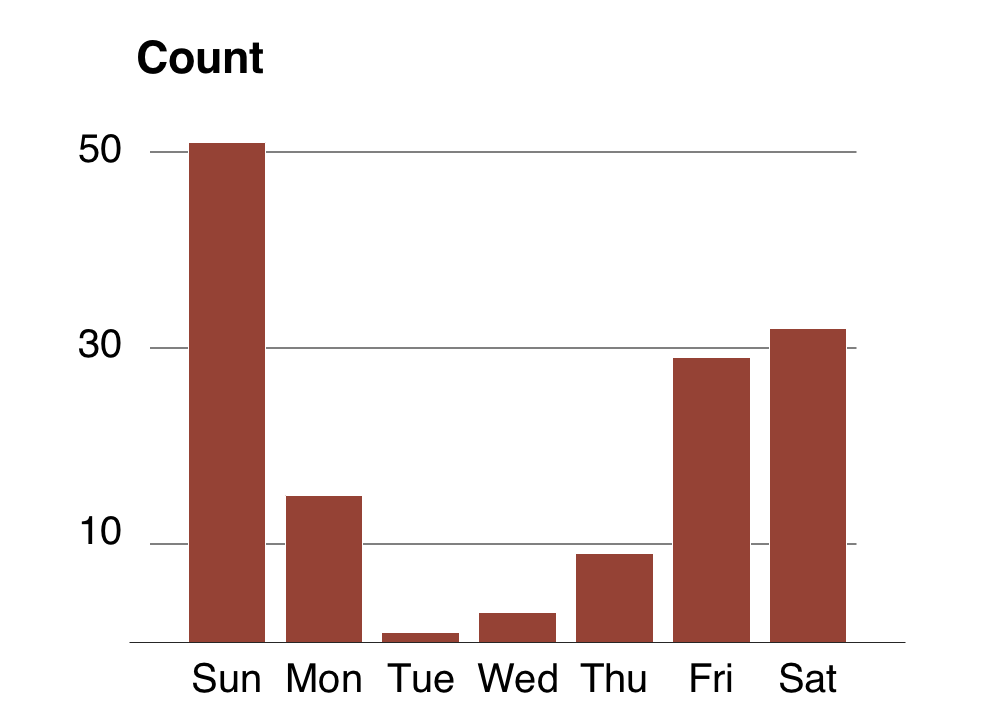}
  \\
    {\small (a) Piece 2 \emph{Penn Station}. }  &  {\small (b)  Histogram. }
    \\
\end{tabular}
  \end{center}
  \caption{\label{fig:Penn_station_crowd} When is Penn Station
    particularly crowded compared to other places? (a) Precise
    locations of Piece 2. (b) Histogram indicating that Penn Station
    is significantly more crowded than other places during weekends. }
\end{figure}

\mypar{Common events} Common events show natural traffic behaviors,
which are detected by frequently-used pieces. When a row in the
coefficient matrix has many nonzero entries, the corresponding piece
is frequently used to represent graph signals. The top five most
frequently-used pieces learned from taxi-pickup activities during rush
hours are shown in Figure~\ref{fig:manhattan_atoms}. For example,
Figure~\ref{fig:manhattan_atoms}(b) shows that a small area around
Penn Station is frequently crowded and we use Piece~2 to capture this
information. We then check which day uses this piece.
Figure~\ref{fig:Penn_station_crowd}(b) shows a histogram of the usage,
where we see that Piece 2 is more frequently used on Sundays (as well
as Fridays and Saturdays) during the entire $2015$. This indicates
that, compared to other places in Manhattan, Penn Station is
particularly crowded during weekends. This is intuitive due to a large
number of commuters going through Penn Station and visiting the nearby
Madison Square Garden on weekends.

\begin{figure}[t]
  \begin{center}
    \begin{tabular}{cc}
     \includegraphics[width=0.4\columnwidth]{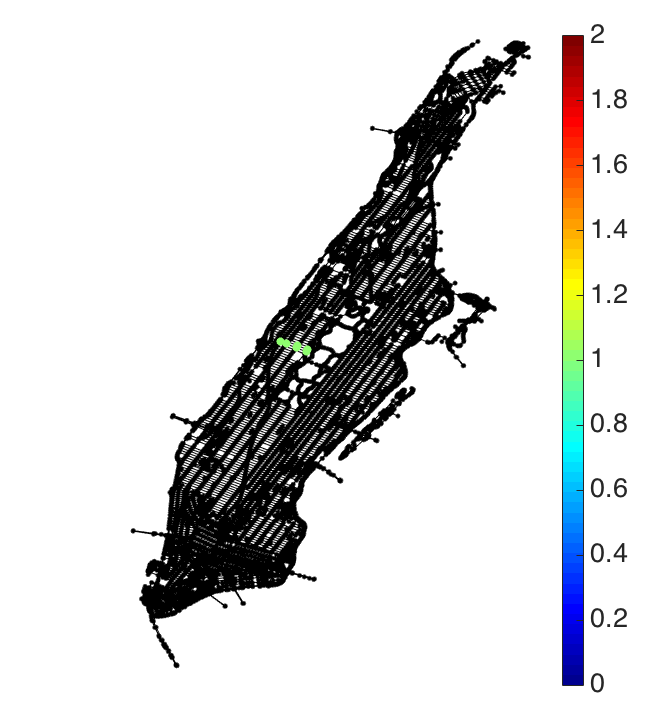}
 &
      \includegraphics[width=0.4\columnwidth]{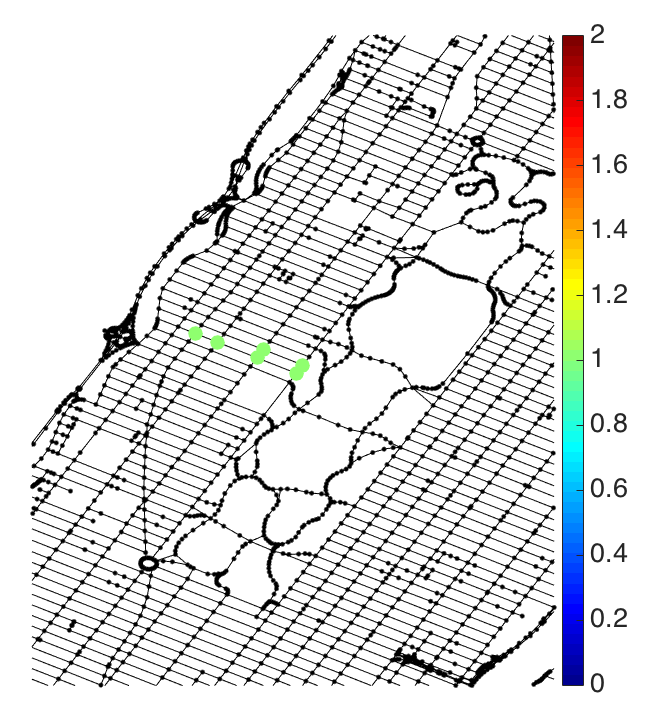}
       \\
    {\small (a) Piece \emph{W 82nd Street}. }  &  {\small (b)  Zoom-in plot. }
 \\
    \includegraphics[width=0.4\columnwidth]{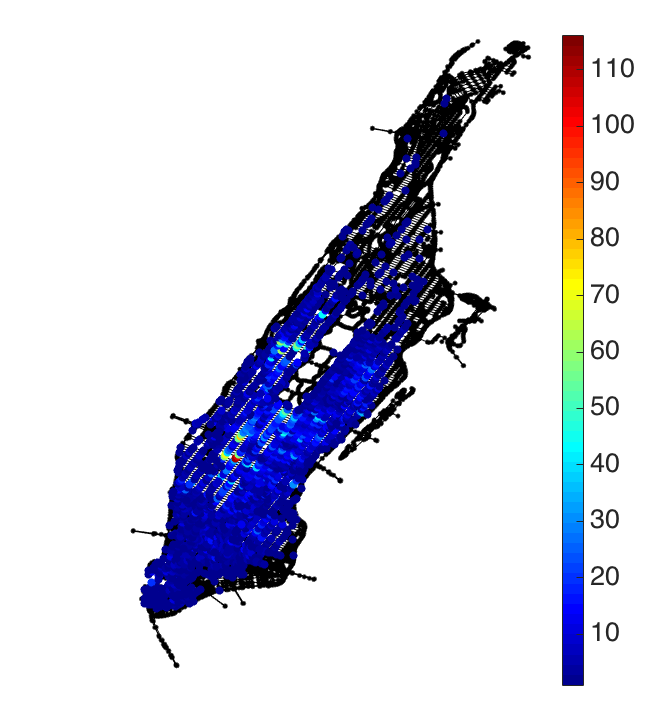}
&
    \includegraphics[width=0.4\columnwidth]{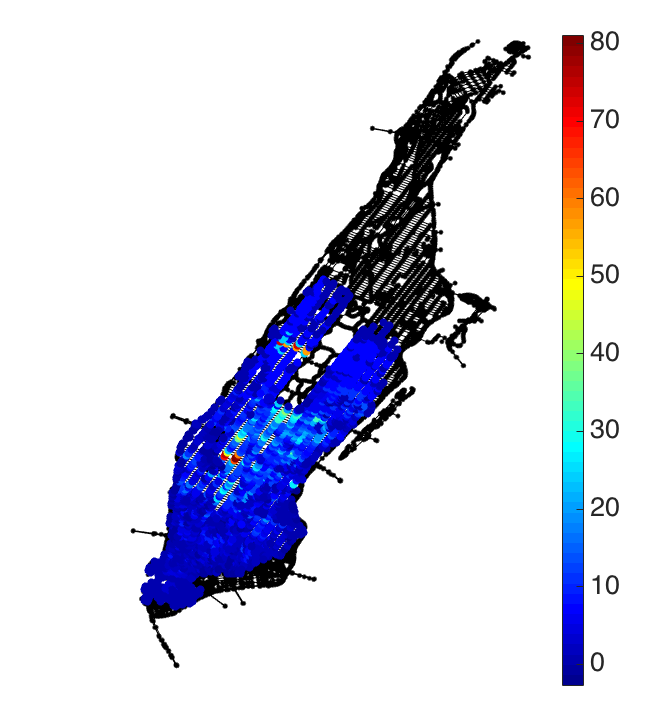}
  \\
    {\small (c) Taxi-pickup activity }  &  {\small (d)  Approximation }
    \\
        {\small  at 9 pm, Nov. 25th. }  &  {\small by 30 pieces from the dictionary. }
    \\
\end{tabular}
  \end{center}
  \caption{\label{fig:thanksgiving_piece} Inflation of the balloons
    the day before Macy's Thanksgiving Parade is detected.  6-8 pm on
    Nov 25th activates the $466$th piece, indicating locations around
    W 82nd Street are particularly crowded.  }
      \vspace{-3mm}
\end{figure}

\mypar{Special events} Special events show
anomalous traffic behaviors, which are detected by the rarely-used
pieces. When a row in the coefficient matrix has a few nonzero
entries, the corresponding piece is rarely used to represent graph
signals. Figure~\ref{fig:thanksgiving_piece}(a) shows the $466$th
piece in the graph dictionary activating the area around West 82nd street and (b) shows a zoom-in plot. This piece is only used
three times during the entire year of $2015$, between 6-8 pm on
November 25th, the night before Thanksgiving Day, indicating that the
area around West 82nd street was much more crowded compared to other
areas on that night only. Figures~\ref{fig:thanksgiving_piece}(c)--(d)
show the taxi-pickup activities at 9 pm on November 25th and its
approximation by using the graph dictionary, respectively. This event actually corresponds to the inflation of the balloons for the Macy's
Thanksgiving parade that happens on 77th and 81st streets between
Columbus and Central Park West, purely from the taxi-pickup activity.

\begin{figure}[t]
  \begin{center}
    \begin{tabular}{cc}
     \includegraphics[width=0.4\columnwidth]{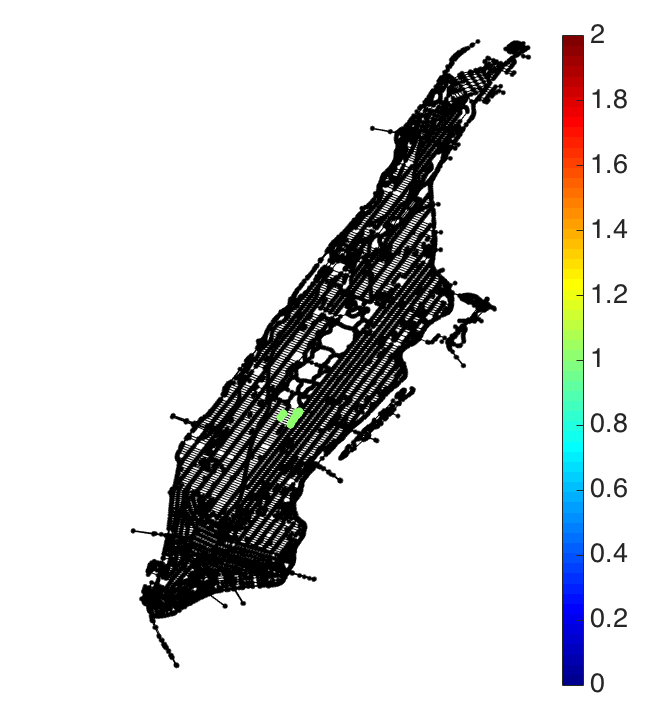}
 &
      \includegraphics[width=0.4\columnwidth]{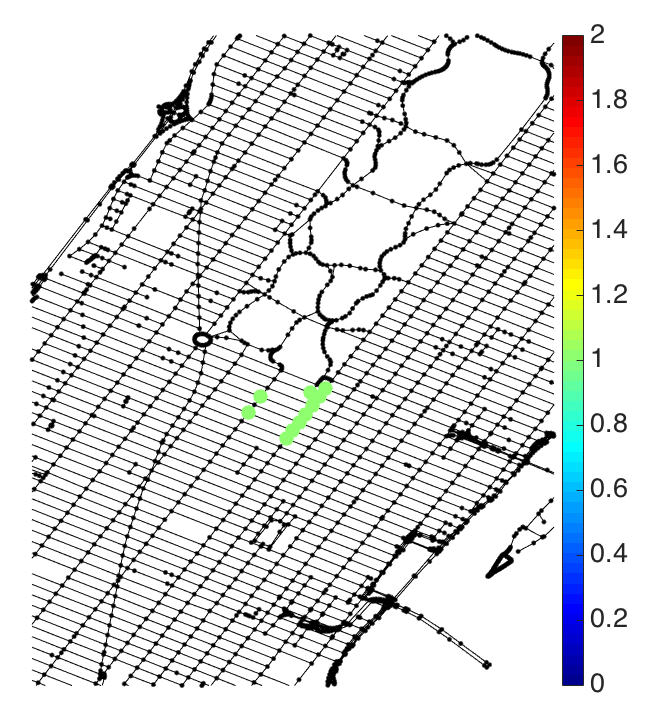}
 \\
     {\small (a) Piece \emph{Apple Store, 5th Avenue}.}  &  {\small (b)  Zoom-in plot. }
 \\
%          {\small  \emph{Apple Store, 5th Avenue}. }  &
%  \\
    \includegraphics[width=0.4\columnwidth]{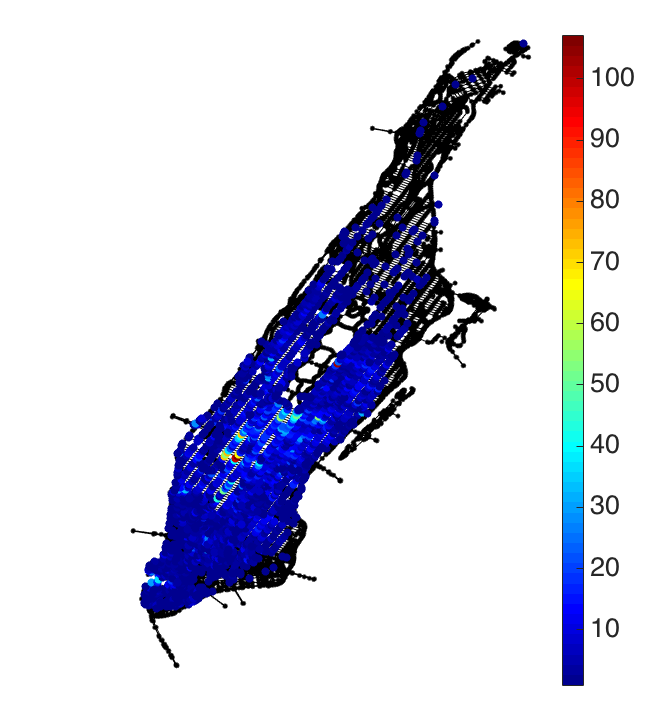}
&
    \includegraphics[width=0.4\columnwidth]{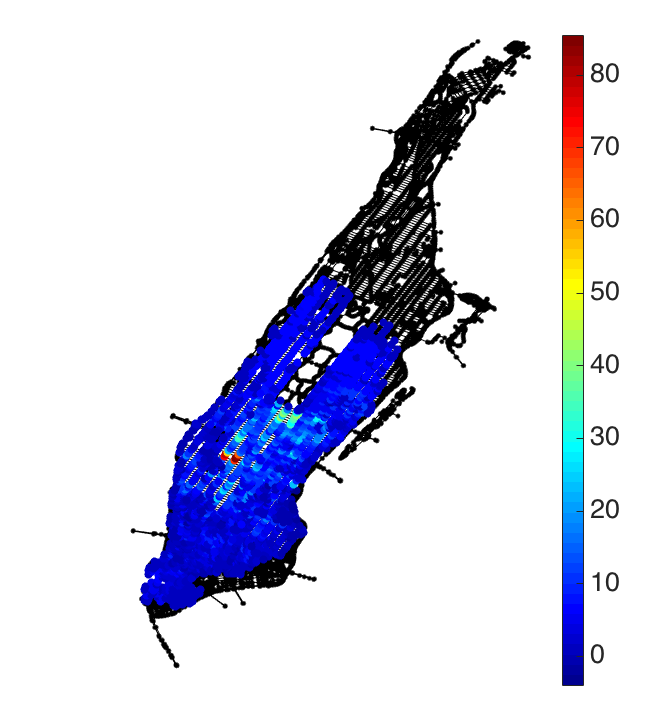}
  \\
      {\small (c) Taxi-pickup activity }  &  {\small (d)  Approximation }
    \\
        {\small  at 8 pm, July 3rd. }  &  {\small by 30 pieces from the dictionary. }
    \\
\end{tabular}
  \end{center}
  \caption{\label{fig:national_day_piece} 8 pm on July 3rd activates
    the $377$th piece, indicating that the 5th Ave. Apple store area
    is particularly crowded.  }
      \vspace{-3mm}
\end{figure}

Figure~\ref{fig:national_day_piece}(a) shows the $377$th piece in the
graph dictionary activating the area around Apple Store on 5th Avenue
and (b) shows a zoom-in plot. This piece is only used one time during
the entire year of $2015$, that is, 8 pm on July 3rd, the night before
Independence Day. This indicates that the area around Apple Store on
5th Avenue was much more crowded compared to other areas on that night
only. Figures~\ref{fig:national_day_piece}(c)--(d) show the
taxi-pickup activities at 8 pm on July 3rd and its approximation by
using the graph dictionary, respectively.

\subsubsection{What day is today in Manhattan?}
Are traffic patterns on weekends different from traffic patterns on
weekdays? We use the learned graph dictionary to answer this question.

Graph dictionary learning not only provides traffic-correlated pieces,
but also extracts traffic-correlated features through approximation.
Similarly to principal component analysis, graph-dictionary-based
sparse representation corresponds to unsupervised learning; it reduces
the dimension of the representation, thereby extracting key
information. Unlike principal component analysis, which is unaware of
the graph structure, graph-dictionary-based sparse representation
extracts traffic-correlated features based on the graph
structure. Since there are 500 pieces in the graph dictionary, we
reduce the dimension of each graph signal from 13,679 to 500 and the
corresponding 500 expansion coefficients are traffic-correlated
features.

% \vspace{-3mm}
\begin{figure}[h]
  \begin{center}
     \includegraphics[width=0.4\columnwidth]{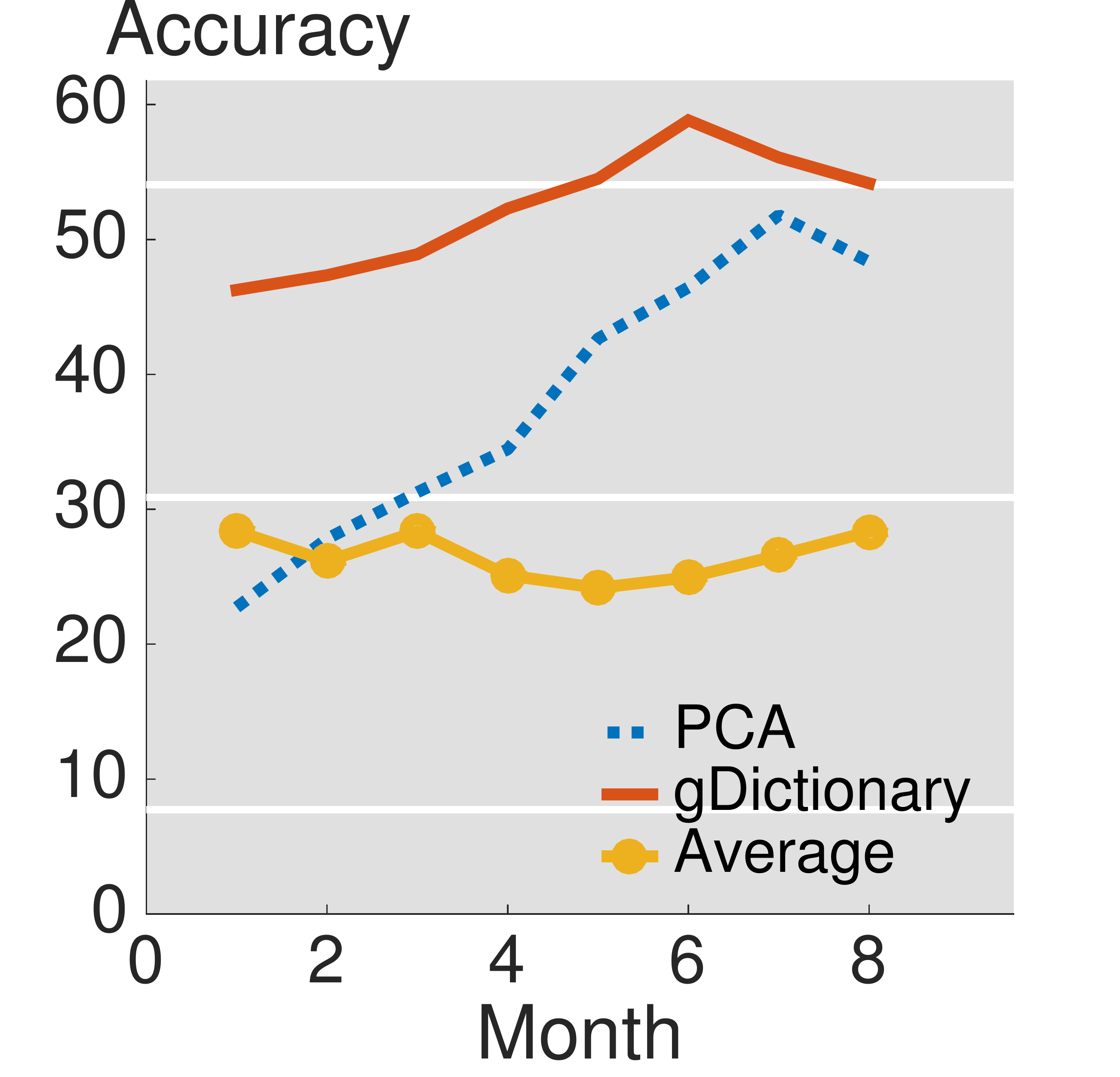}
  \end{center}
  \vspace{-3mm}
  \caption{\label{fig:day_classification_month} Classifying to which
    day a graph signal belongs. Graph-dictionary-based sparse
    representation significantly outperforms principal component
    analysis.}
      \vspace{-1mm}
\end{figure}
We now consider a seven-class classification task: given a graph
signal, we aim to identify to which day of the week it belongs by
using traffic-correlated
features. Figure~\ref{fig:day_classification_month} shows the
classification accuracy as a function of training data, where the
$x$-axis is the number of months used in training and the $y$-axis is
the classification accuracy. For example, when the month is $x = 1$,
we use 93 ($31 \times 3$) graph signals in January as training data
and use the remaining 1002 graph signals as testing data. Since this
task involves seven classes, classification accuracy of a random guess
is $14.29\%$. We compare three methods: principal component analysis
with top 500 components (PCA, blue dotted line), average value of each
graph signal (Average, yellow-circle line) and the
graph-dictionary-based sparse representation (gDictionary, red solid
line).  We see that graph-dictionary-based sparse representation
significantly outperforms principal component analysis and the average
value. The classification accuracy of graph-dictionary-based sparse
representation increases as the number of training data grows before
July after which it drops slightly, showing that summer traffic
patterns are distinctive.

\begin{figure}[htb]
  \vspace{-3mm}
  \begin{center}
    \begin{tabular}{cc}
      \includegraphics[width=0.4\columnwidth]{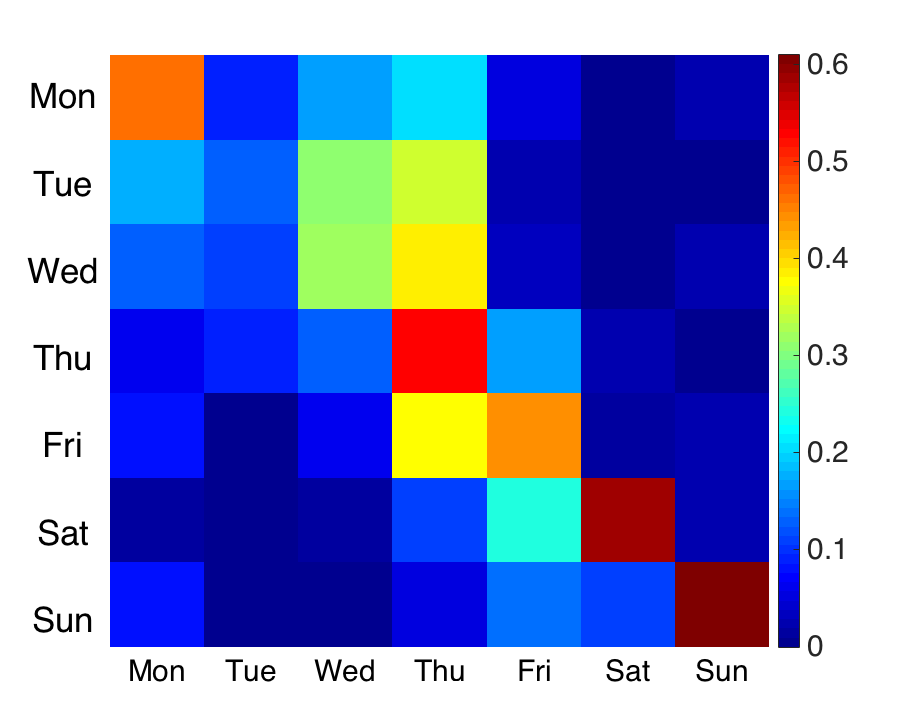} &
      \includegraphics[width=0.4\columnwidth]{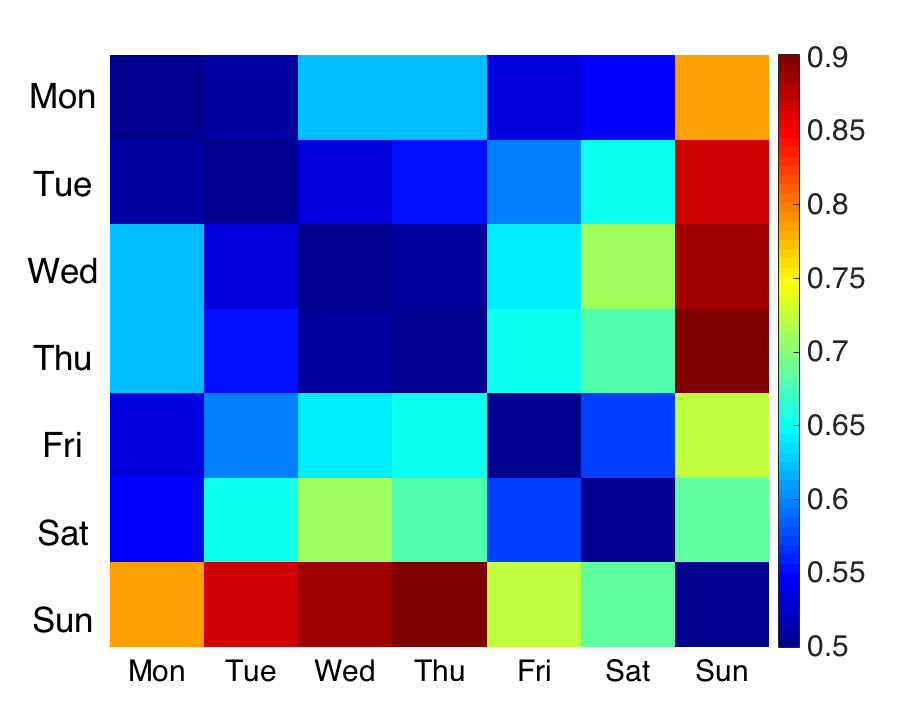} \\
      {\small (a) Classification.} &  {\small (b) Pariwise classification.}
    \end{tabular}
  \end{center}
  \vspace{-3mm}
  \caption{\label{fig:week_classification_cfm} (a) Classification and
    (b) pairwise confusion matrices using January graph signals as
    training data showing that it is relatively easy to distinguish
    weekdays from weekends.}
\end{figure}

We now fix January graph signals as training data and the graph
signals in the remaining $11$ months as testing data;
Figure~\ref{fig:week_classification_cfm}(a) shows the classification
confusion matrix.\footnote{Given $c$ classes and confusion matrix $\Mm
  \in \R^{c \times c}$, element $\Mm_{i,j}$ is a count of samples that
  belong to class $i$ but are classified as class $j$.  Perfect
  classification yields an identity confusion matrix.}  It is
relatively easy to distinguish weekdays from weekends as well as
Saturdays from Sundays; in contrast, weekdays are easily confused with
each other indicating similar traffic patterns.

We then run a series of pairwise classifications to see whether
traffic patterns on weekdays and weekends differ. We only consider
distinguishing one weekday from another at a
time. Figure~\ref{fig:week_classification_cfm}(b) shows pairwise
classification accuracies; we can easily distinguish
Sundays from weekdays, but not weekdays from other weekdays.
% \begin{figure}[h]
%   \begin{center}
%     \includegraphics[width=0.4\columnwidth]{figures/learning/week_classification_acc_pair.png}
%   \end{center}
%   \vspace{-3mm}
%   \caption{\label{fig:week_classification_acc_pair} Pairwise
%     classifications show that it is significantly easier to
%     distinguishing Sundays from weekdays.}
%   \vspace{-3mm}
% \end{figure}

\subsubsection{Discussion} We detected events and classified days of
the week from taxi pickups in Manhattan. We were able to successfully
detect locations and times of both common and special events by using
the graph dictionary learned from a spatio-temporal taxi-pickup data
volume.  In day-of-the-week classification, we were able to
significantly outperform principal component analysis and provide
insight into the traffic patterns on weekdays and weekends by using
the same graph dictionary. These results suggest that the proposed
graph dictionary learning techniques are a promising tool for
exploring mobility patterns of spatio-temporal urban data, which may
aid in urban planning and traffic forecasting.

 \vspace{-2mm}
\section{Conclusions}
\label{sec:conclusions}
We studied three critical problems allowing piecewise-constant graph
signals to serve as a tool for mining large amounts of complex data:
graph signal localization, graph signal decomposition and graph
dictionary learning. We used piecewise-constant graph signals to model
local information in the vertex-domain and showed that decomposition
and dictionary learning are natural extensions of localization. For
each of these three problems, we proposed a specific graph signal
model, an optimization problem and an efficient solver.

We conducted extensive validation studies on diverse datasets.  The
results show that cut-based localization is good at localizing
ball-shaped classes and path-based localization is good at localizing
elongated class. We also used proposed methods to analyze
taxi-pickup activity in Manhattan in $2015$ and showed that based on
these, we can successfully detect both common and special events as
well as tell apart weekdays from weekends. These findings validate the
effectiveness of the proposed methods and suggest that graph signal
processing tools may aid in urban planning and traffic forecasting.

 \vspace{-3mm}
% -------------------------------------------------------------------------
\bibliographystyle{IEEEbib}
\bibliography{bibl_jelena}

\end{document}